\documentstyle[12pt,epsf,graphicx]{article}
\def\href#1#2{#2}   
\newif\ifdraft
\draftfalse	  
\reversemarginpar   
\makeatletter
\let\mlabel=\label
\let\adkendequation=\endequation%
\def\endequation{\adkendequation\adklabel\global\@ignoretrue}
\let\adkendeqnarray=\endeqnarray%
\def\endeqnarray{\adkendeqnarray\adklabel\global\@ignoretrue}
\newbox\marglabbox
\def\adklabel{\ifvoid\marglabbox\else\marginpar{\unhbox\marglabbox}\fi}
\def\label#1{\ifdraft\ifmmode%
  \global\setbox\marglabbox=\hbox{\hfill\fbox{\tiny\verb*~#1~}}%
  \else\ifinner\else\marginpar{\hfill\fbox{\tiny\verb*~#1~}}%
  \fi\fi\fi \mlabel{#1}}
\makeatother
\ifdraft%
\fi
\font\twelvebb=msbm12
\font\tenbb=msbm10
\font\sevenbb=msbm7
  \newfam\bbfam
  \def\bb{\fam\bbfam\twelvebb}
  \textfont\bbfam=\twelvebb
  \scriptfont\bbfam=\tenbb
  \scriptscriptfont\bbfam=\sevenbb
\font\twelveeusm=eusm10 scaled 1200
\font\teneusm=eusm10
  \newfam\eusmfam
  \def\eusm{\fam\eusmfam\twelveeusm}
  \textfont\eusmfam=\twelveeusm
  \scriptfont\eusmfam=\teneusm
  \scriptscriptfont\eusmfam=\scriptfont\eusmfam
\font\twelvefrak=eufm10 scaled 1200
\font\tenfrak=eufm10
  \newfam\frakfam
  \def\frak{\fam\frakfam\twelvefrak}
  \textfont\frakfam=\twelvefrak
  \scriptfont\frakfam=\tenfrak
  \scriptscriptfont\frakfam=\scriptfont\frakfam
\def\sqr#1#2{{\vcenter{\hrule height.#2pt
   \hbox{\vrule width.#2pt height#1pt \kern#1pt
      \vrule width.#2pt}
   \hrule height.#2pt}}}

\def\bsqr#1#2{{\vrule width #1pt height#2pt}}
\def\bsquare{{\mathchoice\bsqr66\bsqr66\bsqr33\bsqr33}}
\def\badbreak{\penalty1000}

\def\Tr{\mathop{\rm tr}}		    
\def\identity{{\bb I}}			    
\def\union{\cup}                            
\def\N{{\bb N}}				    
\def\R{{\bb R}}				    
\def\Z{{\bb Z}}				    
\def\C{{\bb C}}				    
\newcommand{\muhat}{{\hat \mu}}             
\newcommand{\chihat}{{\hat \chi}}           
\newcommand{\gfive}{\gamma_{5}}             
\newcommand{\linkset}{{\eusm L}}            
\newcommand{\rgmap}{{\eusm R}}              
\newcommand{\linkall}{\Omega_1}             
\newcommand{\cD}{{\cal D}}                  
\newcommand{\hD}{{\hat D}}                  
\newcommand{\cE}{{\cal E}}                  
\newcommand{\cO}{{\cal O}}                  
\newcommand{\cP}{{\cal P}}                  
\newcommand{\cS}{{\cal S}}                  
\newcommand{\cU}{{\cal U}}                  
\newcommand{\hU}{{\hat U}}                  
\newcommand{\tU}{{\tilde U}}                
\newcommand{\tM}{{\tilde M}}                
\newcommand{\tS}{{\tilde S}}                
\newcommand{\cM}{{\cal M}}                  
\newcommand{\sobjects}{{\frak X}}           
\newcommand{\Fmaps}{{\cal F}}               
\newcommand{\Umaps}{{\eusm F}}              
\newcommand{\ken}{E^K}                      
\newcommand{\keng}{{\bar E^K}}              
\newcommand{\sha}{{E^{Sh}}}                 
\newcommand{\phibar}{{\bar\phi}}            
\newcommand{\psibar}{{\bar\psi}}            
\newcommand{\abar}{{\bar a}}                
\newcommand{\mbar}{{\bar m}}                
\newcommand{\qbar}{{\bar q}}                
\newcommand{\Ubar}{{\bar U}}                

\begin{document}

\begin{center}
{\Large{\bf A Framework for Systematic Study of QCD}} \\
\vspace*{.1in}
{\Large{\bf Vacuum Structure I: Kolmogorov Entropy}} \\
\vspace*{.1in}
{\Large{\bf and the Principle of Chiral Ordering}}\\
\vspace*{.4in}
{\large{I.~Horv\'ath}} \\
\vspace*{.15in}
University of Kentucky, Lexington, KY 40506, USA

\vspace*{0.2in}
{\large{May 11 2006}}

\end{center}

\vspace*{0.15in}

\begin{abstract}
  \noindent
  In this series of articles we describe a systematic approach for studying QCD vacuum
  structure using the methods of lattice gauge theory. Our framework incorporates 
  four major components. (i) The recently established existence of space--time order
  at all scales (fundamental structure) observed directly in typical configurations
  of regularized path--integral ensembles. (ii) The notion of scale--dependent vacuum
  structure (effective structure) providing the means for representing and quantifying 
  the influence of fluctuations at various scales on physical observables (phenomena). 
  (iii) The unified description of gauge and fermionic aspects of the theory which 
  facilitates a high level of space--time order in the path--integral ensembles. 
  (iv) The strict ``Bottom--Up'' approach wherein the process of identifying the vacuum 
  structure proceeds inductively, using the information from valid lattice QCD 
  ensembles as the only input. In this work we first elaborate on the meaning
  of the notion of space--time order in a given configuration which is conceptually 
  at the heart of the path--integral approach to vacuum structure. It is argued that the
  {\em algorithmic complexity} of binary strings associated with coarse-grained descriptions
  of the configuration provides a relevant quantitative measure. The corresponding
  ensemble averages define the ranking of different lattice theories at given 
  cutoff by the degree of space--time order generated via their dynamics. We then introduce
  the set of local transformations of a configuration, {\em chiral orderings}, in which
  the transformed gauge connection represents an effective matrix phase acquired by chiral 
  fermion when hopping over a given link. It is proposed that chiral orderings facilitate 
  the evolution in the set of actions which increases the degree of space--time order 
  while preserving the physical content of the theory, and should thus be used in the 
  search for the fundamental QCD vacuum structure. The relation to renormalization group 
  ideas is discussed, and the first step in general formulation of {\em effective lattice QCD} 
  realizing the notion of scale--dependent vacuum structure is given.  
\end{abstract}

\vfill\eject

\section{Introduction}

Understanding the vacuum structure of non-abelian gauge theories is the subject of  
active theoretical research. While no definitive answers have been obtained over
a number of years, the subject continues to captivate the interest of many researchers 
since the stakes are high. Indeed, the potential fruits of such effort include,
among other things, identifying the mechanism leading to confinement and spontaneous 
chiral symmetry breaking (SChSB) -- phenomena believed to be central for the dynamics 
of strong interactions. 

While the vacuum of quantum field theory is a state in the appropriate Hilbert space, this 
is not necessarily the representation of choice in the search for vacuum structure. 
Indeed, the discussion of related issues in QCD is frequently carried out in the Euclidean 
path-integral formalism. Here one does not work with states in the Hilbert space, 
but rather with the statistical {\em ensemble} of gauge configurations, i.e. the set 
of pairs $\cE^{QCD} \equiv \{\, (\,A, \,P[A]\,\cD A\,) \;\}$,
where $A \equiv \{\, A_\mu(x) \,\}$ collectively denotes an arbitrary space-time 
configuration of gauge potentials, and $P[A]\,\cD A \propto \exp(-S^{QCD}[A])\,\cD A$ is 
its weight (``probability'') in the path integral. Since the vacuum expectation value 
of any operator is replaced by the corresponding ensemble average, the information stored 
in the quantum vacuum state is equivalent to the information encoded in the ensemble. 
In this sense, one can view the ensemble as a particular representation of the vacuum.
  
One advantage of working with classical ensembles is that the notion of ``vacuum structure'' 
acquires a natural intuitive meaning here. Indeed, the attempts to {\em understand} QCD 
vacuum in this language aim at finding a related ensemble $\cE^{STR}$ (ensemble 
with ``structure''), 
which only involves a certain definite subset of configurations that share a specific kind 
of space-time order. The notion of ``vacuum structure'' is then loosely associated with 
the common space-time structure exhibited by these special configurations. Ideally, one 
should be able to describe $\cE^{STR}$ analytically in terms of collective degrees 
of freedom (objects in the gauge field) that encode this order. The (usually vaguely stated) 
criterion for identifying the ensemble $\cE^{STR}$ is that the ``QCD physics is reproduced'' 
while, at the same time, this physics should be naturally understood in terms of 
the associated order and/or the corresponding collective degrees of freedom. There are two 
highly interrelated problems with the above program which currently appear to prevent 
it from becoming a developed and systematic area of research.

\begin{description}

   \item[{\bf P1:}] {\em The lack of well-posed goals consistent with the fact that 
                         $\cE^{QCD}$ describes quantum field theory.}

   \item[{\bf P2:}] {\em The lack of guiding principles and the systematic framework for 
                         constructing $\cE^{STR}$ once such goals are given.}
   
\end{description}

\noindent The aim of this series of papers is to discuss a set of ideas that have a bearing
on the above issues. Most of these ideas are original, but some represent a significant 
expansion of the line of thought that already appeared in our previous work. In the remaining
parts of this section, we will discuss in more detail why we view {\bf P1} and {\bf P2} as 
pressing problems, which will allow us to naturally set the stage for our approach.

\subsection{Lattice QCD}

One of the important reasons for using the Euclidean path integral formalism in the context of 
vacuum structure is that gauge theory can be conveniently regularized in this framework. Indeed, 
the considerations of the previous section are only meaningful when viewed as a result 
of the limiting procedure involving regularized versions of the theory. Gauge invariant 
regularization on the Euclidean space-time lattice~\cite{Wil74A} provides an appropriate framework 
for both the non-perturbative definition of gauge theory, and the study of the vacuum structure 
along the lines described above. 

The power of lattice QCD in this context resides with the fact that the vacuum of a finite 
lattice system, namely the ensemble $\cE^{LQCD}$, is at our disposal in the following sense.
The ensemble $\cE^{LQCD}= \{\, (\,U, \,P(U)dU\,) \;\}$ can be fully represented by an infinite 
probabilistic chain of configurations $\{\ldots ,U^{i-1},U^{i},U^{i+1}, \ldots \}$, 
such that the probability of encountering the configuration 
$U\equiv \{\, U_{x,\mu} \in \mbox{SU(3)} \,\}$ is proportional to its weight 
$P(U) dU \propto \exp(-S^{LQCD}(U))\,dU$ in the lattice path integral (a finite-dimensional 
integral). At the same time, finite sections of such chains are generated explicitly in 
Monte Carlo simulations of lattice gauge theory on digital computers~\cite{Cr79A,Cr79B}. 
In other words, a numerical simulation gives us direct access to the vacuum in the form of 
configurations dominating the evaluation of the regularized QCD path integral which ultimately 
defines the theory. 

Given the above, one naturally expects lattice QCD (LQCD) to be very relevant in the quest for 
understanding the QCD vacuum structure. One of the points that we will emphasize throughout 
this series of papers is that the role of LQCD should in fact be elevated from auxiliary
and secondary to primary and decisive.

\subsection{The ``Top--Down'' Approach}
\label{sec:int-topdown}

For a number of years, the research in this area proceeded mainly along the line which we 
refer to as a ``Top--Down'' approach. The characteristic feature of Top--Down is that the 
{\em starting point} of an investigation is an immediate proposal for relevant collective 
degrees of freedom (e.g. an instantons, magnetic monopoles, center vortices). Such a proposal 
is made in the continuum and is usually supplemented by a ``picture'' of how confinement or 
SChSB could be understood if the collective degrees of freedom in question would indeed turn 
out to be the most relevant.

The role of LQCD is rather auxiliary in this approach. The primary aspect amounts to calculations 
supplying details needed for an eventual analytic representation of $\cE^{STR}$ in terms 
of the collective degrees of freedom chosen, i.e. determining the densities, sizes and 
other geometric characteristics of the objects relevant for the description of $\cE^{STR}$. 
Such an inquiry typically proceeds via certain procedure that assigns to every configuration $U$ of 
the numerically-generated ensemble $\cE^{LQCD}= \{\ldots ,U^{i-1},U^{i},U^{i+1}, \ldots \}$ 
an associated configuration $U_C \equiv F_C(U)$ in which the corresponding degrees of freedom 
$C$ are ``visible''. This produces a numerical ensemble with ``structure'' 
$\cE^{LSTR}= \{\ldots U^{i-1}_C,U^{i}_C,U^{i+1}_C, \ldots \}$ on which the needed 
characteristics of $\cE^{STR}$ can be measured. Another use of $\cE^{LSTR}$ could be to quantify 
(by measuring and comparing variety of relevant observables in $\cE^{LQCD}$ and $\cE^{LSTR}$) how 
``close'' $\cE^{STR}$ is to $\cE^{QCD}$, but this is usually not done beyond the string tension.

There are numerous problematic issues associated with the Top--Down approach some of which
we list below.

\noindent (1) The proposal for the collective degrees of freedom is always {\em ad hoc} in 
the sense that it is motivated by certain interesting properties (mainly of topological nature)
but not taking into account at all the global structure of quantum QCD ensemble. In that 
sense it amounts to little more than guessing (see {\bf P2}). Nevertheless, it is certainly 
possible (but maybe not likely) to get to the root of a problem via a well-motivated guess.
\smallskip

\noindent (2) The procedure for selecting $U \rightarrow U_C=F_C(U)$ is not unique in existing 
examples, and is sometimes gauge dependent. Moreover, the aspect of non-uniqueness of $F_C$ 
appears to be generic.~\footnote{There is no satisfactory mathematical meaning to the task of 
constructing the map $F_C$ for any reasonable class of $C$. In other words, the attempt
to define the ``$C$-content'' $U_C$ of configuration $U$ appears to be ill-founded.}  
The finality of the results and arguments in favor of a given picture arrived at in a Top--Down 
manner thus appears to always remain in question (see {\bf P2}). 
\smallskip

\noindent (3) It is not clear how to interpret the meaning and the scope of $\cE^{STR}$ 
arrived at in a Top--Down manner in the context of quantum field theory defined by $\cE^{QCD}$. 
Indeed, at the face value $\cE^{STR}=\cE^{STR}(\cE^{QCD},C,F_C)$ is a function of $\cE^{QCD}$, 
the collective degrees of freedom $C$, and the procedure $U \rightarrow F_C(U)$ which
makes $C$ explicit in the configuration. It is hard to find a general notion in field theory 
which would give well-defined meaning to such generic $\cE^{STR}$. To illustrate this further,
it is customary in the Top--Down approach to interpret $\cE^{STR}$ in terms of a phenomenon 
(physics) one wants to understand. For example, the ongoing attempts frequently start as an 
effort to understand confinement, i.e. to construct $\cE^{STR}$ in terms of correctly chosen $C$ 
such that string tension ($\sigma$) would be reproduced and its origin understood. There are 
similar attempts to understand chiral symmetry breaking and reproducing chiral condensate. 
However, this approach carries ambiguities that are difficult to resolve conceptually. 
For example, having constructed $\cE^{STR}_{conf}$ for confinement, do we expect that 
it will also naturally explain chiral symmetry breaking and reproduce the condensate 
(and vice versa)? Such a coincidence might not be very likely since the construction 
of $\cE^{STR}_{conf}$ is almost certainly not unique. Indeed, it appears more reasonable to 
expect (and different coexisting pictures of confinement support this expectation) that there 
are many ways to construct $\cE^{STR}_{conf}$ reproducing {\em just} $\sigma$. Similarly, there 
are probably many ways to construct $\cE^{STR}_{chi}$ which reproduce {\em just} the chiral 
condensate, thus supplying multitude of different ``pictures'' for SChSB in QCD. 
Moreover, the argument doesn't stop there. One can legitimately ask how vacuum encodes (and 
explains) the mass of the lightest scalar glueball or, for that matter, the mass and properties 
of any of the other bound states. The point of these remarks is that organizing the search 
for vacuum structure ``by phenomenon'' is probably not the appropriate way to pose the problem 
(see {\bf P1}). This simply reflects the fact that field theory is not naturally organized in 
terms of ``phenomena''.

\subsection{New Input from Lattice QCD}
\label{sec:newinput}

There were recent developments in this area that will serve as an important part of 
the conceptual input for our framework that we hope can overcome the issues related 
to {\bf P1} and {\bf P2}. To fix the notation and the language, let us denote 
by $\cE^{LQCD}_{\phibar}$ the ensemble of configurations $\phi = \phibar(U)$
(local composite field) associated with given $\cE^{LQCD}$.
\footnote{Note that the
symbol for configuration $\phi$ is distinguished from the symbol for the local map 
$\phibar$ relating it to the configuration of gauge field $U$.} 
In terms of a probabilistic chain we have
\begin{eqnarray}
   \cE^{LQCD} &=& \{\ldots U^{i-1},U^{i},U^{i+1}, \ldots \} 
   \quad \longrightarrow \quad 
   \{\ldots \phibar(U^{i-1}),\phibar(U^{i}),\phibar(U^{i+1}), \ldots \} \equiv 
   \nonumber \\ 
   &\equiv&  \{\ldots \phi^{i-1},\phi^{i},\phi^{i+1}, \ldots \} 
   \equiv \cE^{LQCD}_{\phibar}
   \label{eq:5}  
\end{eqnarray}
The probability distribution (and the associated action) governing the ensemble 
$\cE^{LQCD}_{\phibar}$ is related to the fundamental LQCD distribution P(U) via
\begin{equation}
    P_{\phibar}(\phi) \equiv e^{-S_{\phibar}(\phi)}/Z =
    \int dU \,P(U) \, \delta(\phi - \phibar(U)) 
    \label{eq:10}
\end{equation}

We now wish to highlight two aspects put forward in recent 
works~\cite{Hor02B,Hor03A,Hor05A,Ale05A} dedicated to study of a particular 
composite field, namely the topological charge density~\cite{Has98A,NarNeu95} 
associated with Ginsparg-Wilson operator~\cite{Gin82A}, i.e. 
$q_x = \qbar_x(U) = -\mbox{\rm tr} \,\gamma_5 \, (1 - D_{x,x}(U)/2)$, if
$D$ satisfies the canonical Ginsparg--Wilson relation. 

\begin{description}

\item[{\bf I1:}] {\em The configurations $q$ appearing in probabilistic chains 
representing $\cE^{LQCD}_\qbar$ exhibit an \underline{observable space-time order} 
(structure)}~\cite{Hor03A,Hor05A}.~\footnote{The nature of the ordered structure 
present in these configurations is quite intriguing (see~\cite{Hor03A,Hor05A}) but 
we will not discuss it here.} {\em This order disappears after randomization of 
space-time coordinates of the underlying gauge field}~\cite{Ale05A} {\em , and is 
thus a manifestation of space-time order in $\cE^{LQCD}$.}
\footnote{Strictly speaking, the corresponding studies have so far been 
performed in pure-glue LQCD defined by Wilson and Iwasaki gauge actions, and using 
the Wilson-based overlap operator~\cite{Neu98BA} to define the topological field. 
However, it is expected that the existence of the structure extends to generic LQCD 
and to generic Ginsparg-Wilson operator.}

\end{description}

\noindent The conceptual content of the above result is non-trivial. 
Indeed, this is the first time that a definite space-time order has been observed 
in the physically relevant composite field evaluated directly on {\em unmodified} 
configurations that appear in probabilistic chains representing $\cE^{LQCD}$. 
This fact gives substance to the (previously quite unfounded) expectation that QCD 
path integral is dominated by certain definite class of configurations exhibiting 
a common structure. We emphasize that this didn't necessarily have to be the case. 
While the ensemble averages build up coherence over distances of order 1 fm in QCD 
(as measured via properly defined physical correlators), it was not a priori 
obvious that this coherence would manifest itself at the configuration level via 
explicitly detectable space-time order. Nevertheless, it now appears that this is 
indeed so. 

\begin{description}

\item[{\bf I2:}] {\em It was proposed}~\cite{Hor02B,Hor05A} {\em that the patterns 
of structure in topological charge fluctuations described by $\qbar$ associated with 
Ginsparg-Wilson fermions can be meaningfully studied as a function of the fermionic 
response scale $\Lambda^F$. The underlying proposition is that all aspects of QCD 
vacuum structure should be viewed in a \underline{scale-dependent manner}.}
   
\end{description}

\noindent To be more explicit, the scale dependent ensemble 
$\cE^{LQCD}_{\qbar,\Lambda^F}$ for $\qbar$ was defined as
\begin{equation}
   \cE^{LQCD}_\qbar = \{\ldots q^{i-1},q^{i},q^{i+1}, \ldots \} 
   \quad \longrightarrow \quad 
   \{\ldots q^{i-1,\Lambda_F}, q^{i,\Lambda^F}, q^{i+1,\Lambda^F}, \ldots \} 
   \equiv \cE^{LQCD}_{\qbar,\Lambda^F}
   \label{eq:15}  
\end{equation}
Here for arbitrary configuration $q$, the {\em effective topological field} 
$q^{\Lambda^F}$ is defined via~\cite{Hor02B,Hor05A}
\begin{equation}
   q^{\Lambda^F}_x \;=\;
   -\sum_{\lambda \le a\Lambda^F} (1 - \frac{\lambda}{2})\,
   (\psi_x^\lambda)^\dagger \,\gamma_5\, \psi_x^\lambda
   \label{eq:20}  
\end{equation}
where $\psi^{\lambda}$ is the eigenmode of $D$ with eigenvalue $\lambda$ and $a$ 
is the lattice spacing. This formalism allows to study the patterns in topological
charge fluctuations as a function of scale.

\subsection{Resolutions: ``Bottom-Up'' and the Fundamental Structure}
\label{sec:intro_bottomup}

Given {\bf I1} and the host of problems with the Top--Down methodology described 
above, it appears highly desirable to dramatically change the point of view when 
approaching the problem of QCD vacuum structure in the path integral formalism.
Since now there exists evidence for a directly observable order in typical 
representatives of $\cE^{LQCD}$, it might no longer be necessary to rely on guessing 
associated with Top--Down strategy. One can instead take the attitude that a viable 
candidate for relevant collective degrees of freedom $C$ has to have the space-time 
attributes consistent with the structure observed in typical LQCD configurations. Thus, 
instead of making fixed $C$ a starting point of the investigation, the approach 
built here views determining $C$ as one of the final goals in a systematic study. 
We refer to this as a {\em ``Bottom--Up''} approach to QCD vacuum structure. There 
are two crucial aspects that we wish to emphasize and to associate with this 
approach.

\begin{description}

  \item[{\bf A1:}] {\em The sole input used in the Bottom--Up search for $\cE^{STR}$
  comes from various valid ensembles $\cE^{LQCD}$ defining $\cE^{QCD}$ in 
  the continuum limit. At the technical level, this can also enter via associated 
  ensembles $\cE^{LQCD}_\phibar$ of various local composite fields} 
  ({\em see Eq.}~(\ref{eq:5})).

\end{description}

\noindent Indeed, the intention in the Bottom--Up approach is that its only source 
of information are typical configurations from $\cE^{LQCD}$ (for various lattice 
formulations), and that it starts from a white page, i.e. using no assumptions at all 
about the nature of collective degrees of freedom $C$ in terms of which $\cE^{STR}$ 
will ultimately be described. Consequently, this research will naturally proceed via 
intermediate phases involving the creation of suitable concepts and characteristics 
whose behavior in $\cE^{LQCD}$ will provide hints about the nature of $\cE^{STR}$. 
In particular, one envisions the following. 
\medskip

\noindent {\bf (i) Geometry.} The Bottom--Up approach to QCD vacuum structure should start
by exploring the {\em geometric} patterns occurring in the fields typical of 
$\cE^{LQCD}$~\cite{Hor03A,Hor04A}. The simplest beginning is to study the structure 
in the basic scalar and pseudoscalar gauge invariant composite fields, i.e. action density 
and topological charge density. More complex studies of other composite fields, and perhaps
also of fundamental gauge fields, can follow. These investigations should lead to 
{\em precise geometric statements} about $\cE^{LQCD}$, which can be supported or disproved 
by numerical simulation.   
\smallskip 

\noindent {\bf (ii) Picture.} The information from geometric studies will serve as 
a sole guide and input for an emerging picture of the relevant collective degrees of freedom. 
The underlying expectation is that the requirement that such picture be consistent with 
accumulated knowledge about geometric behavior of fields will significantly restrict 
the available possibilities.     
\smallskip

\noindent {\bf (iii) Physics.} The viability of any picture consistent with geometry 
will ultimately have to be assessed on physical grounds. In other words, some relevant 
physical implications (or predictions specific for that picture) have to be compared with 
physics extracted directly from $\cE^{LQCD}$. It is worth emphasizing that stages {\bf (ii)} 
and {\bf (iii)} are naturally intertwined in that physics considerations immediately
come into focus when considering any particular picture of a collective variable.
\smallskip

\noindent {\bf (iv) Definition of $\cE^{STR}$.} When the qualitative and quantitative 
knowledge about the relevant collective variables in QCD reaches a sufficient level of 
completeness via the above systematic process, it might be possible to define the
degrees of freedom $C$ analytically. A precise definition of $\cE^{STR}$ in terms of
$C$ then might become possible and the understanding of QCD vacuum could be achieved.

\begin{description}

  \item[{\bf A2:}] {\em The vacuum structure obtained from $\cE^{LQCD}$ in a Bottom-Up 
  manner is \underline{fundamental}}~\cite{Hor03A} {\em in the sense that it involves 
  fluctuations of elementary fields at all scales upon taking the continuum 
  limit, and should thus be relevant for all aspects of QCD physics.}

\end{description}

\noindent Indeed, the fact that the structure (embodied in collective degrees of freedom
$C$) is extracted from behavior in full unmodified 
ensembles $\cE^{LQCD}$ implies that, in principle, no aspect of physics should be completely
lost in the process of identifying the structure and the subsequent construction of 
$\cE^{STR}$. We emphasize that this fact represents a non-trivial shift in the focus and 
the scope of the QCD vacuum research. Indeed, the line of thinking adopted in 
Refs.~\cite{Hor03A,Hor05A} together with the results obtained indicate that there might 
exist a structure that encodes the origin of chiral symmetry breaking, confinement, the mass 
of the lightest pseudoscalar glueball \footnote{The masses of bound states are not always 
thought of as associated with vacuum structure but in very real sense they are. After all, 
the correlation functions encoding these masses are vacuum expectation values of appropriate 
operators. In fact, all the observables arise as averages over $\cE^{QCD}$ and thus one can 
attempt to ``understand'' their origin in the vacuum structure.}, the mass of the $10$-th excited 
scalar glueball and so on -- all in one fundamental vacuum structure.

At the current stage, there are several missing pieces (of both conceptual and practical 
nature) that need to be resolved before the Bottom-Up approach to fundamental QCD vacuum 
structure can evolve into a developed area of research. (a) So far the practical framework 
only exists for a single (albeit highly relevant) composite field, namely topological
charge density. While this might eventually be sufficient to uncover the nature of $C$,
there should be a general framework that would allow to study an arbitrary composite field
and the gauge field itself. (b) There is no accepted conceptual notion for degree of
``structure'' (order) in a given individual configuration of the gauge field (or a composite 
field for that matter). 
While the intuitive notion might be sufficient to proceed, the systematic framework should 
contain at least a concept that defines what we are attempting to do. (c) Once such notion
is in place, there should exist a mechanism that guides the process of identifying 
$C$ via maximizing the degree of order in $\cE^{LQCD}$ ensembles. This might require concepts
that do not belong to the existing framework of quantum field theory, but should nevertheless
be meaningfully integrated into it and, hopefully, enrich it. Addressing these issues is 
the primary goal for the first two papers of this series.

\subsection{Resolutions: Scale-Dependent Vacuum (Effective Structure)}
\label{sec:intro_effective}

We have emphasized in Sec.~\ref{sec:int-topdown} that associating $\cE^{STR}$ with
a particular {\em phenomenon} is rather problematic and unnatural in the general
framework of quantum field theory. On the other hand, it has proved very useful 
to organize field theory calculations (and thinking) in terms of a {\em scale}. 
Given this and {\bf I2}, it is thus reasonable to expect that this will also be 
a productive way to deal with the QCD vacuum. In other words, we suggest the 
``change of basis'' in which the questions about QCD vacuum are asked. 
In particular, instead of seeking $\cE^{STR}_{phenomenon}$ for various different 
phenomena, we suggest to seek $\cE^{STR}_\Lambda$, i.e. the vacuum structure relevant 
for physics at arbitrary typical scale $\Lambda$. Another way of saying this is that 
the process involving momenta up to typical value $\Lambda$ will ``excite'' or ``see'' 
the vacuum structure involving the corresponding momenta. Thus if we wish to understand 
the role of vacuum in such phenomenon, it is $\cE^{STR}_\Lambda$ that is relevant for 
such understanding. 

Similarly to the case of fundamental structure, we wish to emphasize two aspects
associated with the Bottom--Up program applied to uncovering the scale-dependent 
structure in QCD vacuum.

\begin{description}

  \item[{\bf A3:}] {\em The sole input used in the Bottom--Up search for 
  $\cE^{STR}_\Lambda$ comes from various ensembles $\cE^{LQCD}_\Lambda$ based 
  on $\cE^{LQCD}$. Alternatively, this can also enter via associated ensembles 
  $\cE^{LQCD}_{\phibar,\Lambda}$ of composite fields $\phibar$} 
  ({\em see e.g. Eq.}~(\ref{eq:15})).

  \item[{\bf A4:}] {\em The structure obtained from $\cE^{LQCD}_\Lambda$ in 
  a Bottom--Up manner is an \underline{effective structure}} \cite{Hor02B,Hor03A} 
  {\em exhibited by elementary fields dominated by fluctuations at the momentum
  scale $\Lambda$. The scale--dependence introduced in this manner allows 
  to associate a typical scale (and the corresponding effective structure) with 
  various phenomena.}

\end{description}

\noindent We emphasize that the notion of $\cE^{STR}_\Lambda$ is essential for 
the interpretation of the {\em fundamental structure} represented in $\cE^{STR}$. 
Indeed, the existence of such an all--encompassing structure raises a conceptual 
issue namely how does one ``decode'' the structure relevant for a particular 
physical phenomenon. The concept of $\cE^{STR}_\Lambda$ resolves this and 
represents an ``unfolding'' of the fundamental structure into an effective 
structure relevant for physics at scale $\Lambda$. Note that the sought for
collective degrees of freedom $C$ characterizing the fundamental structure
can acquire a scale-dependence $C=C(\Lambda)$ in this process. 

While the discussion above summarizes our scale--dependent approach, there are 
clearly some crucial gaps to be filled before this can be viewed as a general 
theoretical/practical framework. Indeed, so far such framework (partially) exists 
only for the study of topological charge density. How does one generalize it to 
other operators and to the ensemble level? In other words, how are 
$\cE^{LQCD}_\Lambda$ and $\cE^{LQCD}_{\phibar\,,\Lambda}$ defined? Can they be 
defined for arbitrary $\cE^{LQCD}$ and arbitrary $\phibar$? Also, even in case 
of topological charge density one needs to address the question of how is 
the fermion--response scale $\Lambda^F$ (see Eq.~(\ref{eq:15})) related to the 
usual momentum scale $\Lambda$. We will start elaborating on some of 
these issues in the present work, and the problem will be addressed fully 
in the third paper of this series.

\subsection{Final Introductory Remarks}

We wish to make three sets of comments before we start.
\medskip

\noindent {\em (i)} The ideas presented in this series of papers (and the manuscripts 
themselves) have been developed over the period of several years, and certain
aspects of it were discussed by the author in less unified manner at various 
meetings. In this extended period (and before that), there appeared works that,
at least to some extent and at least in some regards, align with the points of view 
expressed in this work. Being such, we would like to mention some of them.

The idea of using low--lying fermionic modes as an unbiased probe of vacuum
structure can be viewed as a precursor to Bottom--Up.  The beginnings of the 
``mode'' approach can be traced to the set of early works~\cite{fermion_glob}, where 
they were used to assess the global properties of the underlying configuration. 
The proposal for using fermionic modes for systematic study of {\em local behavior} 
in the configuration has been made in Ref.~\cite{Hor01A}. The related early papers in 
this category include~\cite{deG01A,Followup,Hor02A,Gat02A}, and later~\cite{Other}.
\footnote{Note that even though the ``tool'' is in principle unbiased, it can be
used in a biased way. Here we will not attempt to sort out which claims contained
in all these works actually correspond to reality or can be interpreted as unbiased
statements.} Recently, there also appeared related studies that use the eigenmodes 
of the covariant lattice Laplacian for similar purposes~\cite{Gre05A,Bru05A}. 

As emphasized in~\cite{Hor02B}, the problem with studying individual eigenmodes is 
that they do not provide a direct interpretation of vacuum properties in terms of 
gauge fields or gauge invariant composite fields. As such, they have to rely on
the model assumptions (although these might be testable a posteriori). Focusing on 
studying the properties of gauge invariant composite fields expressible in terms of 
fermion eigenmodes is the natural way to proceed~\cite{Hor02B,Hor03A,Hor05A} with 
Bottom--Up. This approach has recently been taken up also in 
Refs.~\cite{follow_q}, which also confirmed the existence of the fundamental
structure of Ref.~\cite{Hor03A}. Similar fundamental structure was observed in case 
of $CP^{N-1}$ models 
in Ref.~\cite{Tha05A}. Using a certain non--standard non--local topological density
operator, vacuum properties have been very recently investigated in a largely 
Bottom Up manner in Ref.~\cite{quaternion}. The recent work on using the Laplacian 
modes for filtering the gauge fields~\cite{Bru05A} is an interesting step in the 
Bottom--Up direction as well. 
\medskip

\noindent {\em (ii)} Throughout this series of papers we will use a rather formal 
mathematical language when stating certain important conclusions.
\footnote{We have used this approach in some of our previous works and hope to
continue this practice to the largest extent possible also in our future ``Bottom--Up'' 
investigations.}
We do this even though in most cases these statements are not rigorously proved
at the time (referring to them as {\em conjectures} instead of theorems), and even 
though it is very likely that in majority of cases they might never be 
proved. The reason for this is not to do mathematics, but rather to benefit from 
borrowing its language. Indeed,the rationale for such attitude is the belief that 
stating intended conclusions precisely helps the process of obtaining meaningful results 
in case of investigating the QCD vacuum structure via path integral. 
The following points offer some clarification for our point of view.
\medskip

\noindent $\bullet\,$
If a precise statement is eventually invalidated (numerically or otherwise),
we are still left with a non--trivial positive piece of information, namely that
a particular definite possibility is excluded. 
\smallskip

\noindent $\bullet\,$
A precise statement that turns out to be false can frequently be shown to be 
approximately true in a well-defined exact sense. The information value
of such proposition can thus obviously be very large. Moreover, a false
statement of this kind can frequently be modified in such a way (after additional 
information is obtained) that it captures the truth better, or it can even evolve 
into an exactly valid proposition. Indeed, we are interested in a set of precise
statements that get better and better constrained as more numerical data is
accumulated. Such a process has a chance to converge to meaningful definitive 
results. 
\smallskip

\noindent $\bullet\,$
In a lattice theory with finite degrees of freedom (where continuum results are 
obtained via a limiting process) everything is well defined. Moreover, as emphasized in 
Sec.~\ref{sec:intro_bottomup}, the crucial part of the vacuum structure research is of 
geometric nature. There is just no good reason not to be as specific as possible in 
such setting.
\medskip

\noindent {\em (iii)} Since this work has many different aspects to discuss and our
goal is to present it in a manner that is as coherent as possible, we decided 
not to include any numerical data in this series of papers. Rather, there will be
several accompanying papers where certain conjectures proposed here will be 
supported (or possibly invalidated). Our emphasis here is on building of the
framework indeed.

\section{The Set of Lattice Actions} 
\label{sec:actions}

As indicated by {\bf A1}, our attention will be focused on the ensembles 
$\cE^{LQCD}$ corresponding to various lattice theories $S^{LQCD}$ defining 
QCD in the continuum limit. The underlying idea is to build a conceptual
framework which will guide us in the process of selecting theories judiciously
in a manner that will maximize our chances of identifying the relevant 
collective degrees of freedom in QCD. To pursue this, we first need to specify 
the set of acceptable lattice actions. For most part, this discussion will 
be carried out in the context of gluodynamics (i.e.\ pure-glue QCD). However, 
we stress from the beginning that the full power of the approach proposed here 
only appears in full QCD. Indeed, certain crucial aspects will later be 
discussed in QCD with dynamical quarks. With that understanding, we will skip 
the ``QCD'' superscripts when denoting the actions, ensembles and related 
objects. Also, since we will be mostly working with lattice definitions of 
a continuum theory, we will mostly skip the superscript $L$ signifying 
the distinction. Depending on the context, we might distinguish the gluonic 
($S^{G}$) and fermionic ($S^{F}$) parts of full QCD action ($S$).

Identifying proper lattice actions usually starts from classical 
expressions in the continuum and, to that end, we will use conventions 
where the gauge coupling only enters via the multiplicative factor of 
the gluonic part. More precisely, if $A_\mu(x) \in \mbox{\rm su(3)}$ 
(smooth vector field of traceless anti-Hermitian matrices), then the field 
strength is defined via
\begin{equation}
      F_{\mu \nu}(x) \,\equiv\, 
      \partial_\mu A_\nu - \partial_\nu A_\mu + [\, A_\mu, A_\nu \,] 
   \label{eq:25}  
\end{equation} 
while the expression for the gauge action is
\begin{equation}
     S^G \,=\, \int d^4 x\, s^G(x)     \qquad\qquad
     s^G(x) \,=\, -\frac{1}{2g^2} \Tr F_{\mu \nu}(x) F_{\mu \nu}(x)
     \label{eq:30}
\end{equation}
If $\psi(x)$ is a smooth Dirac field ($\psi(x) \in \C^{12}$ and 
$\psibar \equiv \psi^\dagger \gamma_0$),\footnote{Note that at this level we treat
fermionic field as the complex-valued field and the ``fermionic action'' 
as the functional yielding the Dirac equation via the variational principle.
We will thus view the transition to Grassmann variables as a part of 
``quantization''.}  then the part of the action involving fermions reads
\begin{equation}
  S^F \,=\, \int d^4 x\, s^F(x)     \qquad\qquad
  s^F(x) \,=\, \psibar(x) ( \gamma_\mu D_\mu + m) \psi(x)
  \label{eq:35}
\end{equation}
where we assumed a single flavor for simplicity, and 
$D_\mu \psi(x) = (\partial_\mu + A_\mu(x)) \psi(x)$.

The standard construction of lattice gauge theory proceeds via introducing
the hypercubic lattice of points $n \in \Z^4$, and associating the gauge  
degrees of freedom $U_{n,\mu} \in \mbox{\rm SU(N)}$ with links (emanating
from $n$ and ending at $n+\mu$) while the fermionic degrees of freedom
$\psi_n \in \C^{12}$ (before quantization) are associated with sites of 
the lattice. We then write the candidate action for latticized theory with 
massless quarks in the form
\begin{equation}
   S^{G} = \beta\sum_n O_n(U,g)      \qquad\qquad 
   S^{F} = \sum_{n,m} \psibar_n D_{n,m}(U,g) \psi_m
  \label{eq:40}
\end{equation}
where the bare gauge coupling in the conventional form $\beta=2N/g^2$ has been
introduced, and we have allowed for the possible explicit $g$-dependence of 
$O$ and $D$. The criteria for selecting $S^{G}$ and
$S^{F}$ are the following.
\smallskip

\noindent {\bf ($\alpha$) Classical limit.} To establish the classical limit,
the lattice structure is embedded into $\R^4$ via introduction of the 
``naive'' lattice spacing $a$ and the relation $x = a n$. If $A_\mu(x)$ and 
$\psi(x)$ are smooth fields
\footnote{By ``smooth'' we will understand differentiable arbitrarily many 
times, even though a weaker condition would be sufficient. Arguments here 
do not depend on this detail. \label{foot:smooth}} 
defined on $\R^4$, then it is required that expressions 
(\ref{eq:30},\ref{eq:35}) are reproduced upon transcription 
of (\ref{eq:40}) to the lattice via
\begin{equation}
   U_{n,\mu} \;\leftarrow \; 
   \cP \exp\Bigl(a \int_0^1 ds\, A_\mu(an+(1-s)a\muhat)\Bigr)
   \qquad\quad
   \psi_n \;\leftarrow \; a^{3/2} \, \psi(an)
   \label{eq:45}
\end{equation}
once the naive continuum limit $a\to 0$ is taken. Here $\cP$ is the path 
ordering symbol and $\muhat$ a unit vector in direction $\mu$. 
\footnote{With the standard definition of parallel transporter and 
the conventional expression for lattice covariant derivative, the link 
variable transports the vector from position $n+\muhat$ to $n$. Hence the
path ordering prescription in Eq.~(\ref{eq:45}). Simpler transcriptions
for the gauge field, such as $U_{n,\mu} \leftarrow \exp(a A_\mu(na))$ or
$U_{n,\mu} \leftarrow \exp(a A_\mu(na + a\muhat/2))$ can be used 
as well. \label{foot:classicalU}}
\smallskip

\noindent {\bf ($\beta$) Locality.} Lattice actions are required to be local
in the sense that the influence of distant fields on the contribution from 
site $n$ decays at least exponentially in lattice units. In other words, 
$\partial O_n(U,g)/\partial U_{m,\nu} \le A\, e^{-\alpha |n-m|}$, with
constants $A$ and $\alpha$ being independent of $U$ and $g$. 
\footnote{Note that partial derivatives with respect to $U_{n,\nu}$
are rather formal here since the matrix elements of $U_{n,\nu}$ are 
constrained. What one really means here are the partial derivatives in 
the canonical representation of Sec.~\ref{sec:canonical}, i.e.
$\partial O_n(u,g)/\partial u_{m,\nu}^a$.}
Similarly, the kernel $D_{n,m}(U,g)$ should decay at least exponentially in 
$|n-m|$, and $|\partial D_{n,m}(U,g)/\partial U_{l,\nu}|$ exponentially in 
$|n-l|$ and $|m-l|$. While exponential locality might not be necessary, it is 
believed to be sufficient for ensuring universality.
\smallskip

\noindent {\bf ($\gamma$) Symmetries.} From the point of view of universality
(and convenience) it is also desirable that lattice theories preserve to the
largest extent the symmetries of the action in the continuum. 
We thus require {\em (i)} gauge invariance, i.e. invariance under
\begin{equation}
     U_{n,\mu}  \;\longrightarrow G_n \,U_{n,\mu}\,G^{-1}_{n+\muhat}
     \qquad\qquad
     \psi_n \;\longrightarrow\; G_n \,\psi_n
     \label{eq:50}  
\end{equation}
with $G_n \in \mbox{\rm SU(N)}$; {\em (ii)} the invariance under transformations 
of the hypercubic lattice structure, i.e.\ lattice translations and lattice 
rotations; {\em (iii)} the chiral invariance of the massless fermionic action 
$S^{F}$, i.e.\ invariance under infinitesimal L\"uscher~\cite{Lus98A} 
transformations
\begin{equation}
   \psi \;\longrightarrow\; \psi + i\theta\gfive ( 1 - R D) \psi
   \qquad\qquad
   \psibar \;\longrightarrow\; \psibar +
        \psibar\, i\theta ( 1 - D R)\gfive\;
   \label{eq:55}  
\end{equation}
where $R$ is local and $[R,\gfive]=0$, which leads to Ginsparg-Wilson 
fermionic kernels satisfying $\{ D,\gfive \} \;=\; 2 D R \gfive D$;
{\em (iv)} Charge conjugation. {\em (v)} $\gfive$-hermiticity of $D$, i.e. 
$D^\dagger = \gfive D \gfive$. 
\medskip

\noindent {\bf ($\delta$) No Doublers.} The fermionic action $S^{F}$ requires  
additional consideration. In particular, it has to be ensured that 
the corresponding free action ($U_{n,\mu} \equiv 1$) describes a 
{\em single species} of massless Dirac fermion.  
\footnote{It is worth pointing out here that the conditions we stated
admit a highly singular behavior in the regions of configuration 
space far from classical behavior. These might become relevant for example if 
one claims that the corresponding classical ensemble (after quantization) 
can be {\em fully} represented by the probabilistic Markov chain.
We will not deal with associated subtleties here.}
\medskip

In what follows, we will denote by $\cS^{G}$ the set of lattice gauge 
actions $S^{G}$ satisfying conditions ($\alpha$ -- $\gamma$). Similarly,
the set of fermionic actions satisfying ($\alpha$ -- $\delta$) will be
denoted as $\cS^{F}$. It is worth pointing out that each element in
Eq.~(\ref{eq:40}) actually describes a one-parameter family of actions
labeled by bare coupling $g$. While the explicit $g$-dependence of lattice
operators $O$ and $D$ can be present, the requirements 
($\alpha$ -- $\gamma$) have to be satisfied for all $g$. For example, 
the correct classical limit of $S^{G}\in \cS^{G}$ requires that
\begin{equation}
   O_0(U,g) \;\longrightarrow\; 
   -a^4\, \frac{1}{4N} \Tr F_{\mu\nu}(0)F_{\mu\nu}(0)  \;+\; \cO(a^5)
   \label{eq:60}
\end{equation}
under transcription (\ref{eq:45}) independently of $g$. 

Finally, we wish to remark on the conventions that we will follow when 
denoting lattice entities. The distinction between lattice coordinates 
$n, m,\ldots$ and the physical coordinates in $\R^4$, such as 
$x=an, y=am,\ldots$ is made explicit by the change of notation. 
However, some confusion can arise due to the fact that while everything 
in Eq.~(\ref{eq:40}) is dimensionless, we sometimes wish to write 
lattice relations in the continuum-like manner where the appropriate 
dimensions are in place. In that regard, we will follow the trend in 
the literature and not distinguish the symbols for dimensionless and 
continuum-like (possibly dimensionfull) lattice entities. However, to help 
distinguish the two cases, we will place the space-time index as a subscript 
if the entity is viewed as dimensionless, while we put it in parentheses if 
proper dimension is implied. Thus, for example, we have 
$D_{n,m} \equiv D_{x,y} \equiv a D(x,y) \equiv a D(n,m)$, with the relations 
$x=an$, $y=am$ implicitly understood. Similarly, the action $S^{F}$ of 
Eq.~(\ref{eq:40}) can be equivalently written as 
$\sum_{x,y} \psibar_x D_{x,y}(U,g) \psi_y$ or
$a^4 \sum_{n,m} \psibar(n) D(n,m;U,g) \psi(m)$ or 
$a^4 \sum_{x,y} \psibar(x) D(x,y;U,g) \psi(y)$. Note also that the link
variable (being dimensionless) can take all possible forms, i.e.
$U_{n,\mu} \equiv U_{x,\mu} \equiv U(n,\mu) \equiv U(x,\mu)$.

\section{Kolmogorov Entropy}
\label{sec:kentropy}

In this section we start building the framework for a systematic exploration
of a fundamental vacuum structure using lattice QCD in a Bottom-Up manner. 
Relying on {\bf I1} we will accept the proposition that even though 
the definition of QCD in the path integral formalism involves inherent 
randomness, there exists a non-trivial space-time order in configurations 
that dominate the evaluation of the associated regularized path integral. 
In an attempt to uncover the nature of this order, and thus the nature of 
presumably existing underlying collective degrees of freedom $C$, we wish 
to study lattice definitions of QCD which tend to maximize the order and 
minimize the randomness. The goal of this section is to associate the amount 
of structure (order) in typical configurations with their {\em Kolmogorov
entropy}, which we define as the Kolmogorov complexity (algorithmic 
complexity), of binary strings representing coarse-grained configurations.
\footnote{We should point out that the term ``Kolmogorov entropy'' is used, 
with different meaning, also in the theory of dynamical systems (metric 
entropy). However, since the measure discussed here is indeed entropy--like, 
we find it appropriate to use it as well.}

The need for such conceptual step is quite clear. Indeed, while the notion 
of order (structure) is apparently central to the study of QCD vacuum via 
Euclidean path integrals (certainly so in the Bottom-Up strategy), its precise 
meaning is usually left unspecified. There are at least two reasons for this. 
First, it might be assumed that, intuitively, it is quite clear what we mean 
by saying that some configuration is more ordered than other. This is certainly 
true in extreme cases, e.g. when comparing the configuration with constant field 
strength to a configuration with randomly generated links, but the question 
becomes fuzzy in less extreme cases. Secondly, the notion is highly non-trivial 
conceptually, and touches upon the involved issues related to proper definition 
of ``randomness''. We wish to emphasize in this regard, that our interest
in these issues at this point is strictly theoretical. We will point out
that there {\em exists} a well-defined measure of space-time order 
in any configuration of lattice gauge field (and various composite fields), 
which allows to quantify the degree of space-time order for any lattice 
theory. In other words, we will discuss in what sense it is meaningful to say 
that theory $S_1$ exhibits more space time order than theory $S_2$, but will
not attempt to make these concepts into a computational scheme.

In statistical mechanics, the discussion of ``order'' usually goes hand in 
hand with the concept of entropy. Consider, for example, two classical 
statistical systems consisting of identical degrees of freedom, but involving
different interactions among them. We are inclined to say that the system 
with lower entropy is more ordered than the system with higher entropy. 
In the same spirit, consider the pure-glue lattice gauge theory
$S(g)\in \cS^{G}$. Upon quantization, it becomes a classical statistical
system with the associated canonical ensemble $\cE(g)$, where the action
plays the role of ``energy''. We could thus use the corresponding Gibbsian 
entropy of $\cE(g)$ as a quantitative measure of ``order'' associated with
that theory (at any fixed $g$ and fixed lattice volume). The problem with 
such definition is that it obscures (and avoids) the notion of the order 
at the {\em configuration level} -- the concept we are primarily interested 
in when we try to identify the collective degrees of freedom of QCD in 
a Bottom-Up manner. In other words, Gibbsian entropy is not assigned to 
a given configuration independently of an ensemble we are studying: Gibbsian
entropy is strictly an ensemble concept.

Interestingly, the issues with similar ingredients arose in the information
theory few decades ago. In particular, the Shannon's classical theory 
of communication~\cite{Sha48} treats the objects to be communicated
as outcomes of a random source with associated probability distribution. 
The central notion here is the concept of Shannon entropy which only 
depends on the distribution itself, and does not explicitly refer to the 
``amount of information'' associated with an individual object. The dual 
point of view was developed later by Solomonoff~\cite{Sol64}, 
Kolmogorov~\cite{Kol65} and Chaitin~\cite{Cha69}, and is known as the 
algorithmic information theory, Kolmogorov complexity, or algorithmic 
complexity. Here the focus is on the object itself and it defines the 
information (complexity) associated with an object with the smallest number 
of bits required to describe it via {\em effective computation}.
The standard textbook exposition to this subject is Ref.~\cite{Vitanyi},
and the earliest work we are aware of on utilizing this framework 
in physics is Ref.~\cite{Zur89}.   
Bellow we discuss in what sense we can carry these concepts into the study
of QCD vacuum structure. Since the notions of information theory needed
are not standard in this area, we will describe them in some detail before
dealing with field theories.

\subsection{Shannon Entropy}
\label{sec:shannon}

Suppose we have a finite or countable set of objects ({\em source words}) 
$x \in \sobjects$, which are intended to be communicated to the receiver by 
the means of a binary string. Enumerating these objects in an arbitrary fixed 
manner, we associate each $x_n$ with its order number $n$ (for the purposes 
of communication) which, in turn, can be mapped surjectively to the set of 
finite binary strings 
${\frak B}^\star \equiv \{0,1\}^\star$, e.g. via canonical lexicographic 
assignment 
\begin{equation}
    0 \leftrightarrow \epsilon,\; 1 \leftrightarrow 0,\;
    2 \leftrightarrow 1,\;        3 \leftrightarrow 00,\;
    4 \leftrightarrow 01,\;       5 \leftrightarrow 10,\;\ldots
    \label{eq:65}
\end{equation}
where $\epsilon$ is an ``empty word''. Given the maps, we will interchangeably
denote by $x$ the object, the integer or the string.
~\footnote{We emphasize that especially in the case of integer versus 
associated binary string, we will mostly make no distinction (verbal
or otherwise) whatsoever, and use e.g. $x$ in the context of an integer 
even if it was previously referred to as a string.}

For a meaningful communication, the sender and the receiver have to agree 
both on the above maps, and on the method which reproduces $x$ from its code 
$y$. This is characterized by a definite {\em decoding function}
$D:\, {\frak B}^\star \rightarrow {\frak B}^\star$ such that $D(y)=x$. While 
$D$ does not necessarily have to have an inverse (there could be several 
{\em code words} representing the same source word), in this discussion we 
assume that the encoding substitution $E(x) = \{\,y\, : D(y)=x \,\}$ contains 
a unique element $y$. If $l(z)$ denotes the length (in bits) of arbitrary 
string $z$, then the efficiency of coding a particular $x=D(y)$ is given by 
$l(y)\equiv L_D(x)$.

When communicating a stream of source words, it is advantageous to consider 
only codes for which recognizing the end of the current code word in the stream
does not require observation of the subsequent words. Such codes are referred 
to as instantaneous codes or {\em prefix codes}. Formally, the subset 
$A \subset {\frak B}^\star$ is prefix--free, if $xy\notin A$ (concatenation
of $x$ and $y$ is not in $A$) for any non-empty $x,y \in A$. The code 
specified by $D$ is then called a prefix code if its domain is prefix-free.

Assume that the source words are generated by a random source specified by
the probability distribution $P(x)$ on $\sobjects$.~\footnote{This can clearly
be thought of as a distribution on ${\frak B}^\star$ with the corresponding
mappings assumed, and with the probability of ``unused'' strings occuring 
being zero.} In other words, we are interested in the problem of communicating 
the class of messages, such that the source words are distributed according 
to $P(x)$. The effectiveness of the communication accomplished via particular 
decoding function $D$ is then measured by the average code--word length, i.e. 
${\bar L}_D \equiv \sum_x P(x)\, L_D(x)$.     
A natural problem in this setting is to ask what is the minimal possible
average code--word length, i.e. ${\bar L} \equiv \min_D \{{\bar L}_D\,\}$, 
where the minimization goes over all one-to-one prefix codes $D$. The result 
is the Shannon's {\em Noiseless Coding Theorem}~\cite{Sha48,Vitanyi}.

\medskip
\noindent{\bf Theorem 1} {\em If $H(P)=-\sum_{x} P(x) \ln P(x)$ 
is the Shannon entropy associated with $P$, then}
\begin{equation}
   H(P) \,\le \, {\bar L} \le H(P) \,+\, 1
   \label{eq:70}
\end{equation}

\noindent The optimal average code--word length is thus (practically) 
equal to the Shannon entropy. The above discussion attempted to highlight
two facts. (1) The concept of Shannon entropy is completely independent of 
the nature of the objects from $\sobjects$ that source-words represent. 
In ``information'' terms, $H(P)$ is the amount of information the observer 
(aware of $P$) gains when he witnesses an outcome of $P$. It is a measure 
of uncertainty in $P$. (2) We are led to Shannon entropy via the minimization 
of the ``{\em cost of description}'' of the ensemble $\{\,(x,P(x))\,\}$. 
Indeed, the message (binary string) $y$ such that $D(y)=x$ can be viewed as 
a description of the binary string $x$ representing some real object 
from \sobjects. However, the optimal coding of $x$ (its optimal description) 
is completely determined by $P(x)$ and not $x$ itself.
 
\subsection{Kolmogorov Complexity}
\label{sec:kolmogorov}

The notion of Kolmogorov complexity represents an attempt to quantify the amount 
of information associated with any particular binary string $x$. If such string
completely characterizes some real object from $\sobjects$ (i.e. it is not just
a string assigned to it by {\em ad hoc} enumeration procedure, but faithfully
reflects all of its attributes and no more), then it would provide an absolute 
amount of information stored in that object. 

It turns out that a concept can indeed be defined in such a way that it will 
satisfy the above features for asymptotically long strings. To explain the basic
idea, we consider the same setup as in the previous section in that we 
consider the set of decoding functions $D$: $\N^0 \rightarrow \N^0$ 
(or equivalently ${\frak B}^\star \rightarrow {\frak B}^\star$), and call 
$y$ such that $D(y)=x$ a description of $x$ under $D$. 
\footnote{We will follow the convention where the set of positive integers 
(natural numbers) is denoted as $\N$, while the set of non-negative integers
as $\N^0$. The associated finite subsets of consecutive integers will 
be denoted as $\{1,2,\ldots,m\} \equiv \N_m$ and 
$\{0,1,\ldots,m-1\} \equiv \N^0_m$. \label{foot:intconv}}
We do not need to assume that $D$ is one-to-one 
(there can be several descriptions of the same object after all) and, to have
maximal generality, we initially do not require the domain of $D$ to be prefix-free.
In what follows, we will refer to this general set of functions as {\em partial} 
functions, where the term emphasizes that a particular function can be undefined 
for some arguments $y$. Even though our goal is to define the amount of information 
in the {\em individual} string $x$ (quantified by the length of the minimal 
description $y$), it is obvious that we need to consider $x$ in the context of 
all other strings from ${\frak B}^\star$. Indeed, in the opposite case we 
could simply choose any decoding function $D$ such that $D(1)=x$, and thus 
associate a single bit of information with any particular $x$, which is clearly 
not satisfactory. Thus, a natural way to proceed would appear to try to identify 
a function $D_o$ such that the associated minimal description is optimal for 
each $x$. More precisely,
\begin{equation}
   L_{D_o}(x) \,\le\, L_D(x)  \quad \forall x \,, \forall D   \qquad\qquad
   L_{D}(x) \equiv \min_y \{\,l(y) \,:\, D(y)=x \,\}
   \label{eq:90} 
\end{equation}
where we assume that if for given $D$ there is no $y$ such that $D(y)=x$, 
then $L_D(x) \equiv \infty$.
However, it is easy to see that the partial function $D_o$ with above
properties doesn't exist. Indeed, assume the opposite and choose $x_1$, $x_2$
such that $L_{D_o}(x_1) < L_{D_o}(x_2)$, with the corresponding minimal 
descriptions $x_1^\star$, $x_2^\star$ 
(i.e. $D_o(x_1^\star)=x_1$ and $D_o(x_2^\star)=x_2$). Define
the function $D_o^\prime(y) = D_o(y)$ for all $y\neq x_1^\star, x_2^\star$, 
and $D_o^\prime(x_1^\star) = x_2$, $D_o^\prime(x_2^\star) = x_1$. This obviously 
leads to shortening of the minimal description for $x_2$ and, in particular,
$L_{D_o^\prime}(x_2) = L_{D_o}(x_1) < L_{D_o}(x_2)$ thus producing 
a contradiction with the premise that $D_o$ is optimal.

The above argument shows that, in order to get a meaningful concept, it might 
be fruitful to relax the notion of an ``optimal'' function.~\footnote{Indeed, 
it is easy to see that definition (\ref{eq:90}) would be meaningless even if we 
were interested in describing $x$ in the context of finite number of other 
strings, let alone in the context of ${\frak B^\star}$.} A natural proposal
in this direction is to consider {\em additive optimality}. In particular,
the partial function $\Phi$ is additively optimal (universal) for the set of 
partial functions if for every $D$ there is a constant $\gamma(\Phi,D)$ 
(independent of $x$), such that
\begin{equation}
   L_\Phi(x) \,\le\, L_D(x) \,+\, \gamma (\Phi,D)  \qquad\, \forall x\,, \forall D 
   \label{eq:95} 
\end{equation}
Assuming that a universal function exists, there is no reason for it to be 
unique. However, the optimal lengths under two universal descriptions clearly 
differ from each other at most by an $x$-independent constant, i.e. 
$|L_{\Phi_1}(x) - L_{\Phi_2}(x)| \le \gamma(\Phi_1,\Phi_2)$. While this means 
that $L_\Phi$ will not define a strictly absolute amount of information in 
a finite string, there is universality here which applies for asymptotically 
large ones. 
To see that, we first point out that any $L_D(x)$ assigning finite value 
to all finite $x$ ($L_\Phi(x)$ obviously has to be of this kind), has an 
unbounded lower envelope $L_D^{min}(x) \equiv \min_{z\ge x} \{L_D(z)\}$.
Indeed, let $\bar{\frak B}(l)$ denote the set of all binary strings up to
length $l$. This is a finite set, and thus there exists
$\tilde x(x) \equiv \max \{D(y), y\in \bar{\frak B}(L_D^{min}(x))\}$.   
By construction, for all $z>\tilde x(x)$, we have $L_D^{min}(z)> L_D^{min}(x)$,
since all possible encodings of length $L_D^{min}(x)$ are simply exhausted
for such $z$. Repeating this argument for $\tilde x(x)+1$ and recursively
shows that $L_D^{min}(x)$ (and thus $L_D(x)$) is unbounded.
For two universal functions $\Phi_1$, $\Phi_2$ we then have 
\begin{equation}
    \frac{|L_{\Phi_1}(x) - L_{\Phi_2}(x)|}{L_{\Phi_1}(x)} \;\le\;
    \frac{\gamma(\Phi_1,\Phi_2)}{L_{\Phi_1}(x)} \;
    \longrightarrow 0 \quad
    \mbox{\rm for} \quad
    l(x)\longrightarrow \infty
    \label{eq:100}
\end{equation}
In this sense the additive optimality is universal for asymptotically large
strings.

The above discussion suggests that one could define the amount of information
in $x$ as $L_\Phi(x)$ using arbitrary (but fixed) $\Phi$ from the equivalence 
class of universal partial functions. However, it turns out that there is still 
no additively optimal element $\Phi$ on the set of partial functions. 
Indeed, assume the opposite and consider arbitrary increasing sequence
of universal minimal encodings $x_1^\star < x_2^\star < x_3^\star \ldots$ 
(i.e. $\Phi(x_1^\star)=x_1$, $\Phi(x_2^\star)=x_2$, $\ldots$).
\footnote{Note that the minimal encoding substitution 
$E_D(x)=\{y: \,D(y)=x,\, l(y)=L_D(x)\}$ can be made into a one-to-one 
function by fixing $y\equiv x^\star$ that appears first in the lexicographic 
ordering of binary strings.} 
Choose an arbitrary subsequence $z_i^\star \equiv x_{j(i)}^\star$, such that
$\log_2 x_i^\star < (\log_2 x_{j(i)}^\star)/2$, and define the new function 
$\Phi^\prime$ via $\Phi^\prime(y)\equiv \Phi(y)$ for $y\ne x_i^\star$, and 
$\Phi^\prime(x_i^\star)=\Phi(x_{j(i)}^\star)$. By construction, the encodings
under $\Phi^\prime$ for all $x_{j(i)}$ will be shorter than encodings under $\Phi$ 
and, in particular, $L_{\Phi^\prime}(x_{j(i)})=l(x_i^\star) \le l(x_{j(i)}^\star)/2 
= L_\Phi(x_{j(i)})/2$. Here the inequality follows from the fact that in the canonical
correspondence (\ref{eq:65}) between integers and binary strings we have 
$l(x)=\lfloor \log_2(x+1) \rfloor$, where $\lfloor . \rfloor$ denotes the floor
function. We thus arrived at the contradiction since
$L_{\Phi^\prime}$ is not related to $L_\Phi$ via $x$-independent constant. 
Consequently, the set of partial functions does not have the lowest element under 
additive optimality.

While this appears to be a serious blow for the development of the concept 
we are interested in, it actually turns out to be a virtue in the end (see 
bellow). To expose the basic trick, let us consider arbitrary enumerable subset 
$A=\{\,D_1,\, D_2,\, D_3,\ldots\}$ of partial functions. We can construct 
the partial function (specification method) $D_o(y)$ that minorizes all elements
of $A$ under additive optimality in the following way. Fix certain
one-to-one correspondence 
$(n,y) \longleftrightarrow z \equiv \langle n,y\rangle$ 
(i.e. bijection $\N^0 \times \N^0 \longleftrightarrow \N^0$). If string 
(integer) $z$ represents the pair $(n,y)$, i.e. $z=\langle n,y\rangle$,
then $D_o(z)\equiv D_n(y)$. By construction, the function $D_o$ can emulate all 
the functions $D_n$ by properly selecting its arguments $z$. Let us consider the 
``description length'' function associated with $D_o$, i.e.
\begin{equation}
   L_{D_o}(x) \equiv \min_z \{\,l(z) \,:\, D_o(z)=x \,\} \;=\;
                     \min_{n,y} \{\,l(\langle n,y\rangle) \,:\, D_n(y)=x \,\}
   \label{eq:105} 
\end{equation}
and compare it to $L_{D_k}$ for arbitrary (but fixed) $k$. Denoting by $x_D^\star$
the canonical minimal encoding of $x$ under $D$, there exists $x$-dependent $i$ 
such that
\begin{equation}
   L_{D_o}(x) \,=\, l(\langle i, x^\star_{D_i}\rangle ) \,\le\,
                    l(\langle k, x^\star_{D_k}\rangle ) \,=\, 
                    l(x^\star_{D_k}) + c_k \,=\, L_{D_k}(x) + c_k
   \label{eq:110}  
\end{equation}
where $c_k\equiv l(\langle k, x^\star_{D_k}\rangle )-l(x^\star_{D_k})$ does not
depend on $x$ if the pairing bijection 
$(n,y) \longleftrightarrow z \equiv \langle n,y\rangle$ is chosen suitably.
For example, with one of the standard prescriptions 
$\langle n,y\rangle = \bar n y$ (string $\bar n$ concatenated with string
corresponding to $y$), where the string $\bar n = 1^n0n$ (meaning
a string of $n$ symbols $1$ followed by $0$ followed by binary string 
corresponding to integer $n$), we have
$c_k = 2l(k)+1$. Thus, indeed, the function $D_o$ minorizes all $D_k \in A$. 
Assuming that things can be arranged so that $D_o$ itself is an element 
of $A$, then $D_o\equiv \Phi$ is a universal element for $A$ under additive 
optimality. 

The notion of Kolmogorov complexity is built on two basic results from the theory 
of computation, namely the fact that the set of {\em recursive (computable)} 
functions is enumerable, and that the minorizing functions constructed according 
to (\ref{eq:105}) are themselves recursive, thus representing universal 
elements. The fact that the set of recursive functions allows for a meaningful 
definition of minimal description length based on additive optimality 
lends the feeling of inherent ``rightness'' for the concept attempting to quantify 
the absolute amount of information stored in a string. Indeed, the use of a decoding
function $D$ in communicating an information on object $x$ might be viewed as 
pointless unless the message can be effectively decoded by computing machines. 
At the same time, it is generally accepted that the notion of ``computability'' 
(or effective computation) is equivalent with the notion of recursive functions, 
i.e. functions that can be implemented on a Turing machine.
\footnote{More precisely, the partial function $D(y)=x$ (from $\N^0$ to $\N^0$) 
is recursive if there is a Turing machine $T$ such that for inputs $y$
for which $D(y)$ is defined, $T$ outputs $D(y)$ in a finite number of steps,
while it runs forever on inputs for which $D(y)$ is not defined.}  

Focusing on recursive functions, the language of information theory starts to mesh 
with the language of computation theory. In the former case the string $y$ such that
$D(y)=x$ is referred to as a code for $x$ under $D$ (or simply an argument
of function $D$), while in the latter case one refers to the input string $y$ 
placed on a tape of a Turing machine $T$ implementing $D$ as a {\em program} 
producing the output $x$. The enumerability of all Turing machines (and thus of 
partial recursive functions) can be easily seen from the fact that one can enumerate 
all possible finite {\em lists of rules} that define the function of a valid Turing 
machine. In fact, each list of rules can be mapped into a unique binary string. 
We will assume the canonical construction of this type, which is straightforward but 
tedious, and can be found e.g. in Ref.~\cite{Vitanyi}. Searching in the lexicographic 
order through all strings, let $n(i)$ be the $i$-th string encoding a possible Turing 
machine, which we denote as $T_i$. This enumeration of machines induces an associated 
enumeration of recursive functions $D_i$ implemented by $T_i$. 
\footnote{It should be emphasized that while $T_i$, $T_j$ always represent distinct 
machines for $i \ne j$, they might be implementing the same function. The set of 
recursive functions is defined as the set of all distinct elements in the sequence
$\{D_i,i=1,2,\ldots\}$.}
The crucial point is that the canonical enumeration is recursive in the sense 
that there is a Turing machine which can decide whether a given string $n$ encodes
a valid machine. From this it can be seen rather easily that applying a construction
(\ref{eq:105}) to sequence $\{D_i,i=1,2,\ldots\}$ yields a minorizing $D_o$ which 
is itself recursive. Indeed, one can construct the machine $T^u$ which takes the pair 
$(i,y)\leftrightarrow \langle i,y\rangle$ as its input, and will simulate 
(on its own tape) machine $T_i$ running program $y$. Given $i$, this machine will 
start generating binary strings in lexicographic order and check whether they represent 
the list of rules for a Turing machine. When encountering $i$-th such string $n(i)$, 
it will run $y$ using the set of rules specified by $n(i)$ it found. By construction 
the machine $T^u$ computes $D_o$.
\footnote{We note that the above description of $T^u$-s function is not the proof
of its existence. Such proof has to demonstrate the existence of the set of rules 
(list of quadruples of integers) defining the machine that will function as described.}

To summarize, what we have described above are the basic logical ingredients leading 
to the fact that the set of partial recursive functions contains a universal element 
$\Phi \equiv D_o$ under additive optimality. This element has been identified as a
function computed by the {\em universal} Turing machine $T^u$, i.e. machine that can 
simulate the actions of any other Turing machine. Thus, the length of the shortest 
description under $\Phi$ (namely $L_\Phi (x)$) provides an absolute measure for 
information stored in an asymptotically long string $x$. It is the number of bits 
that have to be transmitted to the receiver who can decode it in an effective manner 
(using a Turing machine). In the language of computation theory it is the length of 
a shortest program that, without any additional input, outputs $x$ and halts when 
run on $T^u$. 

\smallskip
\noindent We wish to make few comments at this point:
\smallskip

\noindent (1) There are infinitely many universal Turing machines and infinitely
many universal functions. Due to the universal nature of the concept discussed above,
we can fix the universal machine and the universal function to be the canonical 
elements described above. Then the {\em plain Kolmogorov complexity} of string
(natural number) $x$ is defined as
\begin{equation}
    C(x) \;\equiv\; L_\Phi (x)   \qquad\qquad 
    \Phi\, \mbox{\rm computed by reference machine} \,T^u
    \label{eq:115}
\end{equation}  

\noindent (2) Kolmogorov complexity formalizes the intuitive notion that strings 
with lots of regular patterns can be described succinctly (e.g. the string
{\tt 11111...1} ), while strings without any appreciable regularities are not 
amenable to short description no matter what method is used. Thus, apart from 
information-theoretic meaning of the unavoidable amount of information to be 
transmitted,\footnote{We note in this context that strings for which 
$C(x) < l(x)$ (i.e. they allow shorter description than themselves)
are referred to as {\em compressible}, while the rest of them as 
{\em incompressible}.}
it is also an absolute measure of {\em order} present in the asymptotically 
long string. In particular, when considering strings $x_1$ and $x_2$ of the same 
length, then we view $x_1$ as more ordered than $x_2$ if $C(x_1)<C(x_2)$.
\smallskip

\noindent (3) Contrary to the discussion of Shannon entropy, we have not restricted
ourselves to the prefix (instantaneous) codes $D(y)$ so far. However, it turns out 
that the whole logical construction described above can be repeated if one considers 
only partial recursive functions with prefix-free domains. Such functions are 
implemented by {\em prefix Turing machines}, which only yield a halting calculation 
on the elements of such domain. Fixing a canonical enumeration of prefix machines
\footnote{One way to proceed is to use a standard procedure of transforming any 
given Turing machine into a prefix machine. This way the canonical enumeration of 
Turing machines directly induces the enumeration of prefix machines.} 
one can construct the universal prefix machine $T^u$ according to recipe described 
above and use it to define the universal prefix function $\Psi$ and the 
corresponding {\em prefix Kolmogorov complexity}   
\begin{equation}
    K(x) \;\equiv\; L_\Psi (x)   \qquad\qquad 
    \Psi\, \mbox{\rm computed by reference prefix machine} \;T^u
    \label{eq:120}
\end{equation}
The functions $C(x)$ and $K(x)$ are asymptotically equal. However, $K(x)$ has several
desirable properties which $C(x)$ does not.
\footnote{One relevant difference is that, unlike plain complexity, the prefix version 
is {\em subadditive}, i.e. $K(xy) \le K(x) + K(y) + \cO(1)$ (here $xy$ means 
concatenation of strings $x$ and $y$).} Because of this (and also to continue
the comparison to Shannon's theory) we will use the prefix version of Kolmogorov 
complexity in what follows.
\smallskip

\noindent (4) While the concept of Kolmogorov complexity is built on the notion of 
effective computation, the complexity functions $C(x)$, $K(x)$ are themselves not 
effectively computable (recursive). Thus, the significance of the framework 
is mainly conceptual and theoretical. Nevertheless, many of its aspects have been 
successfully utilized in many varied areas. Also, the basic (and apparently deep) 
idea of minimal description length frequently serves as a guide for devising less 
general (and typically less elegant) notions that can be applied in practice in a 
more quantitative manner.
\smallskip

\noindent (5) An elementary result from the theory of Kolmogorov complexity that will 
be useful for us is that, if $f(x)$ is an invertible recursive function, then
the complexities $C(x)$ and $C(f(x))$ differ by at most a constant independent
of $x$. This is a simple consequence of the fact that to arrange for evaluation
of $f(x)$ on an appropriate universal machine, one only needs to specify (input)
the ordering number $i$ of the machine $T_i$ that computes $f$. Analogous
statement obviously holds for $K(x)$ as well.
\smallskip

\noindent (6) It is sometimes instructive and useful to think about Kolmogorov
complexity in terms of programs written in high-level universal languages 
(such as C$^{++}$, LISP or Java). While the whole approach can be formalized
and even used for practical purposes, we will not be concerned with this
aspect here, and view it only as a tool for developing intuition.

\subsection{Kolmogorov Entropy: $\Z_2$ Gauge Theories}
\label{sec:z2ken}

The preliminaries in the previous two subsections allow us to come back to the 
question of how to quantify the level of space-time order in lattice gauge 
theories in an interesting manner. Before dealing with SU(N) theories, it is 
first useful to take a simpler case. For example, consider the set of pure 
gauge theories $\cS^{G}$ with N=2. To every $S^{G} \in \cS^{G}$ assign
a $\Z_2$ gauge theory $S^{GZ}$ by replacing $SU(2)$ variables by $\Z_2$ 
variables, and the path integral by a corresponding discrete sum. This defines 
the set $\cS^{GZ}$ of $\Z_2$ gauge theories. Our goal is to answer (in principle) 
the following generic question. Take two theories $S^{GZ}_1(\beta_1)$ and 
$S^{GZ}_2(\beta_2)$ ($\beta$-s fixed) both defined on a finite discretized 
symmetric torus, i.e. both involving $L^4$ lattice sites. Which theory possesses 
more space-time order?

Our aim is to approach this issue via concept that allows for even more 
detailed question. In particular, given two $\Z_2$ configurations $U_1$, $U_2$,
which one exhibits more space-time order? We propose to answer this question
using the notion of Kolmogorov complexity. Indeed, for fixed $L$, the set 
{\sobjects} of objects we are interested in is the set of all possible $\Z_2$ 
configurations $U$. It is clear that every $U \in \sobjects$ can be 
{\em completely described} by a binary string $z(U)$ of length $4L^4$.
Indeed, the two possible values of the field $U(n,\mu)$ associated with
link $(n,\mu)$ can be represented by the two binary values
$b \in {\frak B} \equiv \{0,1\}$. Moreover, if we fix some one-to-one 
function $F(n,\mu) = k \in \{1,2,\ldots ,4L^4\}$, then we can form a 
corresponding binary string via
\begin{equation}
    z_F(U) \,\equiv b_1 b_2 \ldots b_{4L^4} \qquad\qquad
    b_k \,=\, U(F^{-1}(k))
    \label{eq:125}
\end{equation} 
We need to be more specific about the nature of the function $F$, which 
has to be defined for all $L$. In other words, $L$ has to be one of the 
arguments of $F$, and also (a dummy) component of the image so that 
$F$ is formally invertible. To do that, let us first agree that the 
coordinates $n \equiv (n_1,n_2,n_3,n_4)$ on the torus of size $L$ take 
the values $n_i \in \{0,1,\ldots ,L-1\} = \N^0_L$,
and remind ourselves that $\mu \in \{1,2,3,4\} = \N_4$. Then we consider
invertible partial functions 
\begin{equation}
  F: \; \N \times (\N^0)^4 \times \N_4  \,\mapsto\, \N\times \N   \quad
  \mbox{\rm such that}  \quad
        (\N^0_L)^4 \times \N_4  
        \, \mapsto \, 
        \N_{4L^4} 
        \quad \mbox{\rm under} \quad F(L,n,\mu)
    \label{eq:130}
\end{equation}
for arbitrary fixed $L$, where we skipped the $L$-component of the 
domain/range in the second part of the expression (see also equation 
(\ref{eq:140}) bellow). We then define the {\em Kolmogorov entropy} of 
configuration $U$ relative to $F$ via 
\begin{equation}
     \ken_F(U)  \,\equiv\, K(z_F(U))
     \label{eq:135}
\end{equation} 
i.e. we associate it with prefix Kolmogorov complexity of $z_F(U)$. 

The degree of order detected in $z^F(U)$ via its prefix complexity is a 
result of two factors. The first is the level of inherent space-time 
order present in configuration $U$, while the second is the influence 
of map $F$ which, although fixed for all $U$, appears rather arbitrary 
at this point. However, we will argue bellow that this arbitrariness
is innocuous for asymptotically large configurations $U$ if we 
restrict ourselves to recursive functions $F$. Indeed, denoting the set 
of invertible recursive maps satisfying (\ref{eq:130}) as $\Fmaps$, 
let us consider $F_1, F_2 \in \Fmaps$, and compare the complexities 
$K(z_{F_1}(U))$, $K(z_{F_2}(U))$. Since $F_1$, $F_2$ are recursive, 
the binary strings $z_{F_1}(U)$ and $z_{F_2}(U)$ are recursively related 
via invertible map
\begin{equation}
   z_{F_2}(U) \,=\, G^{F_1F_2}(z_{F_1}(U))  \quad \forall \,U 
   \qquad
   \mbox{\rm where} 
   \qquad
   (G^{F_1F_2}(x))_i \,\equiv\, x_{F_1(F_2^{-1}(i))}
   \label{eq:140}
\end{equation}
and we have denoted the $i$-th bit of string $x$ by $x_i$. Consequently 
(see comment (5) in previous section), the complexities $K(z_{F_1}(U))$, 
$K(z_{F_2}(U))$ differ at most by an $U$-independent constant $\gamma(F_1,F_2)$. 
We can thus conclude that Kolmogorov entropy (\ref{eq:135}) is additively 
optimal with respect to $\Fmaps$, with arbitrary $F \in \Fmaps$ being 
a universal element.
 
It should be remarked that since additive optimality is inherent in 
the notion of Kolmogorov complexity already, it is only natural that 
the Kolmogorov entropy (\ref{eq:135}) has an additive arbitrariness 
related to the choice of $F$. While we can choose and fix any 
$F \in \Fmaps$, it is relevant here that the complexity of $F$ can affect 
significantly the value of Kolmogorov entropy in small configurations, 
and possibly mask the true level of their space-time order. Thus, contrary 
to the selection of universal machine, there is a natural general preference 
for the choice of maps $F$, namely for those with low complexity. In fact, 
given $T^u$, the appropriate choice $F^u$ can be constructed as follows. 
Let $n(i)$ be the sequence of natural numbers such that $T_{n(i)}$ is 
the $i$-th Turing machine computing function $F_i \in \Fmaps$. The set
$\{K(n(i))\}$ is bounded from bellow and thus has a minimum $K_{min}$.
We define
\begin{equation}
   F^u \,\equiv\, F_k  \qquad \mbox{\rm such that} \qquad
   k \,=\, \min \{ \, i : K(n(i))=K_{min} \, \}
   \label{eq:145}
\end{equation}
In other words, $F^u$ is specified by lexicographically first Turing machine 
computing an element of $\Fmaps$, whose complexity is equal to $K_{min}$.
\footnote{\label{foot:kenf} It is worth emphasizing that one defines 
the complexity of a general recursive function $f$ as 
$K(f) \equiv \min_i \{K(i): T_i$ computes $f \}$. The reason is that,
assuming a canonical order of machines, specifying $i$ is all that
is needed to specify the map.} 
Similarly to fixing universal machine $T^u$, we now fix the map $F$ 
to $F^u$ and thus define the Kolmogorov entropy of configuration $U$ as 
\begin{equation}
     \ken (U)  \,\equiv\, K(z(U))  \qquad\qquad
     z(U) \;\, \mbox{\rm computed via} \;\,F^u 
     \label{eq:150}
\end{equation} 
Given two configurations $U_1$, $U_2$ on the same space-time lattice,
we define $U_1$ to exhibit more space-time order than $U_2$ if
$\ken(U_1) < \ken(U_2)$.

We are now in the position to quantify the degree of space-time order 
associated with any theory $S \equiv S(\beta) \in \cS^{GZ}$. In particular,
we identify this measure with the average Kolmogorov entropy in 
ensemble $\cE$ associated with $S$, namely 
\begin{equation}
      \ken [S] \,\equiv\,  \langle \, \ken (U) \,\rangle_{S} 
      \,=\, \sum_{U} \ken (U) \, P_S(U)
      \qquad\qquad
      P_S(U) = e^{-S(U)}/Z 
     \label{eq:155}
\end{equation}
The meaning of $\ken [S]$ is quite clear. If the ensemble $\cE$ is dominated
by configurations with high level of space-time order, then its Kolmogorov
entropy is low. Equivalently, if the individual configurations typically 
appearing in probabilistic chains representing $\cE$ exhibit a high degree 
of space-time order, then the Kolmogorov entropy of the theory is low.
\smallskip

\noindent We wish to make few important remarks at this point:
\smallskip

\noindent
(1) For fixed finite $L$, Kolmogorov entropy $\ken [S,L]$ induces an order
    relation on the set of actions $\cS^{GZ}$.
    \footnote{More precisely, it induces a linear order on the set 
    of equivalence classes, where $S_1 \sim S_2$ if $\ken [S_1]=\ken [S_2]$.}
    In particular, we say that $S_1 < S_2$ if $\ken [S_1,L] < \ken [S_2,L]$.
    Since the elements of $\cS^{GZ}$ are local, it is expected that any two 
    actions can be consistently compared in the infinite volume limit, i.e.
    that there exists $L_{12}$ such that 
    $\ken [S_1,L] < \ken [S_2,L]$ for all $L>L_{12}$.
    As discussed extensively above, the nature of Kolmogorov entropy is such
    that there is an inherent arbitrariness by a constant (related to the 
    choice of universal machine and the choice of map $F$), and the notion 
    acquires an absolute meaning in the limit $L\rightarrow \infty$.
    At the same time, the above ordering relation is expected to be affected 
    very little by this arbitrariness even at finite $L$ since $T^u$ and
    $F^u$ are fixed for all $U$ and all theories. Nevertheless, if one wants
    to ensure strict universality in the infinite--volume limit, then bounded 
    differences of entropies need to be considered as irrelevant. In particular, 
    if there is an $L$-independent constant $C_{12}$ such that 
\begin{equation}
    |\, \ken [S_1,L] - \ken [S_2,L] \,| \;<\; C_{12} \qquad \forall \,L
    \label{eq:160}
\end{equation}     
    then $S_1$ and $S_2$ should be viewed as possessing the same amount
    of space-time order. In other words, it is natural in the 
    $L \rightarrow \infty$ limit to consider the equivalence classes 
    of theories, with the equivalence relation defined by 
    (\ref{eq:160}).
\smallskip

\noindent
(2) In definition (\ref{eq:150}) we took into account that it is
    desirable (but not necessary) to minimize the influence of the map $F$ 
    on the complexity $K(z(U))$. This can be done alternatively (and in some
    ways more elegantly) by considering the {\em conditional} 
    Kolmogorov complexity $K(z(U)|F)$. Roughly speaking, this means 
    that information (and necessary input into $T^u$) about $F$ doesn't  
    count as a contribution to the complexity of $z(U)$. We will not discuss
    this possibility here in detail.
\smallskip

\noindent
(3) Since we are dealing with gauge theories, it is natural to ask how gauge 
    freedom enters the issue of inherent space-time order. In particular, given 
    gauge transformation $g$, the Kolmogorov entropies $\ken (U)$ and $\ken (U^g)$ 
    can be quite different. In other words, $\ken (U)$ is not gauge invariant in 
    the usual sense (at the configuration level). We can see three acceptable 
    views concerning this issue. 
    (i) Gauge invariance is not really necessary for the concept characterizing 
    the space-time order in the configuration.
    \footnote{Note that considerations related to Elitzur's theorem 
    do not play a role here since $\ken (U)$ is a global quantity for
    which no sensible local definition exists.}
    Indeed, it is simply a fact that gauge freedom can introduce an unphysical 
    space-time noise. Since this noise is integrated over in the path integral 
    for all theories in the same manner, the ordering of theories via $\ken [S]$ 
    preserves its intended meaning even if one is interested in gauge-invariant 
    quantities only. 
    (ii) We can make $\ken (U)$ gauge invariant and define 
    the associated Kolmogorov entropy for the theory via
    \begin{equation}
     \keng(U)  \,\equiv\, \min_g K(z(U^g))  \qquad\qquad
      \keng[S] \,\equiv\,  \langle \, \keng(U) \,\rangle_{S} 
     \label{eq:165}       
    \end{equation}
    In other words, the gauge invariant Kolmogorov entropy
    of $U$ is taken to be the minimal Kolmogorov complexity of $z(U^g)$ over 
    the gauge orbit $\{U^g\}$ of $U$. Since $K(x)$ is bounded from bellow, 
    the minimum indeed exists for arbitrary $L$. 
    (iii) Instead of taking descriptional complexities of $z(U)$
    as a measure, we could use the complexities of strings 
    $z(\phibar(U))$ describing some fixed reference gauge invariant 
    composite field $\phibar(U)$. This possibility is less general, but can certainly 
    be exploited for comparing the theories (via $\ken_\phibar(U)$) with respect to 
    a given relevant composite field (i.e. topological charge density in case of QCD).

\subsubsection{Information Issues and the Information Gauge}
\label{sec:infogauge}

It is interesting to elaborate on the information theory viewpoint 
of the definitions discussed above. Assume that the sender wishes 
to communicate a complete information on $\Z_2$ configuration 
$U$ to a receiver via message decodable using effective computation. 
Then the (additively) optimal amount of information necessary to 
transmit is given by $\ken (U)$. However, if we are interested (as 
usually is the case) in the gauge-invariant content of the 
configuration, then we can further minimize the cost of transmission 
by communicating only $\keng(U)$ bits to the receiver. In effect, 
instead of using $U$, we are sending a description of a particular 
configuration $V=V(U)$ from the associated gauge orbit, which possesses 
the highest level of space-time order. Such $V(U)$ may not be unique, 
but we can make it unique e.g. by selecting a configuration which 
appears first in the lexicographic order of associated binary strings. 
More precisely, if ${\cal V}(U)$ is the ordered set of such
configurations, namely
\begin{equation}
   {\cal V}(U) \,\equiv\, \{\, U^{g_i} \;:\;  
                               \ken(U^{g_i}) = \keng(U), \;
                               z(U^{g_{i+1}}) > z(U^{g_{i}}) \, \}   
                               \qquad
                               \mbox{\rm then}
                               \qquad
                               V(U) \,\equiv\, U^{g_1}
    \label{eq:170}       
    \end{equation}
where the ordering of binary strings is defined by the canonical 
correspondence (\ref{eq:65}). The above prescription assigns
a unique configuration to a gauge orbit of arbitrary $U$, and thus
can be viewed as a gauge--fixing prescription. 

Assume now that the sender intends to communicate to the receiver 
information about the gauge--invariant content of the $\Z_2$ gauge 
theory with action $S$, and he wishes to do that via transmitting 
the description of some relevant configurations. This sender would 
generate $N$ independent configurations $U_j$ from distribution 
$P_S(U)$ and, instead of transmitting minimal description of $U_j$, 
he would transmit the description of $V_j$ fixed by prescription 
(\ref{eq:170}) in order to minimize the cost. In other words, he 
would prefer ``fixing the gauge'' accordingly. For this reason, 
we refer to the gauge specified above as the {\em information gauge}. 

It is also important to emphasize at this point that the role of 
information theory in our construction is purely to assess the level
of space-time order exhibited by the configuration (theory), and
we are not attempting to connect it to the absolute amount 
of ``relevant information'' stored in it. The basic concepts 
characterizing the latter, though of utmost interest, are yet to be 
developed. In fact, one should readily appreciate the apparent 
contradictory tendencies in that regard. Indeed, the configurations 
with high Kolmogorov entropy require large number of bits to 
describe, and in that sense one can view them as storing a lot
of information. However, by the very nature of our approach, 
we expect their relevance for physics to be rather limited. On the 
other hand, we hope that configurations with low Kolmogorov entropy,
exhibiting a specific kind of space-time order and allowing for
a short description, will contain large amount of the physically
relevant information.

\subsubsection{Shannon Entropy and Kolmogorov Entropy}
\label{sec:sha_kol}

Let us recall that we started our discussion from the fact 
that Gibbsian entropy, though intuitively acceptable, does not 
provide a measure of space-time order derived from a notion
directly applicable to an individual configuration. At the same 
time, in the path--integral approach it is the order at the 
configuration level we are seeking when trying to identify 
the collective variables relevant in QCD. This led us to 
the information theory where the analog of Gibbsian entropy is 
represented by Shannon entropy, which explicitly relates to 
``ensemble'' properties, and does not address the information 
content of individual objects from the ensemble. 
\footnote{\label{foot:kolsha} In fact, in case of $\Z_2$ gauge 
theories which we discuss, Shannon entropy is identical to Gibbs 
entropy. However, this is not true for probability distributions 
with continuous sample spaces (e.g. lattice QCD).} 
However, on the information theory side there exists a notion 
of Kolmogorov 
complexity which we used to define the ``Kolmogorov entropy'' 
of a configuration, and proposed that this is the appropriate 
measure of its space-time order. 
One can then define the corresponding measure for the $\Z_2$
theory $S$ via the ensemble average specified by $P_S$. We thus
ended up with two distinct ensemble concepts for theory $S$, 
namely its Kolmogorov entropy and the Shannon entropy of the 
associated $P_S$. What differences can one expect?

It turns out that information theory provides a partial answer 
to this question for theories $S$ that lead to {\em computable} 
probability distributions. The theorem bellow is a direct 
transcription of the relevant result for probability
distributions on the set of binary strings 
(see e.g. Ref.~\cite{Vitanyi}) to the framework developed here.

\medskip
\noindent{\bf Theorem 2} {\em Let $S \in \cS^{GZ}$ (for fixed
size $L$ of a symmetric torus) leads
to a computable probability distribution $P_S(U)$, and
$\sha[S] \equiv -\sum_U P_S(U) \ln P_S(U)$. Then
\begin{equation}
     0 \;\le\; \ken[S] - \sha[S] \;\le \; K({\bar P}_S) + \cO(1)
   \label{eq:175}
\end{equation}
where ${\bar P}_S(z(U))\equiv P_S(U)$ is the induced probability
distribution on ${\frak B}^\star$.}
\medskip

\noindent It needs to be emphasized here that the distribution
${\bar P}_S$ (and its computability) has to be viewed as
a function of the lattice size as well, i.e. 
${\bar P}_S={\bar P}_S(x,L)$. We also recall that $K({\bar P}_S)$ 
is the Kolmogorov complexity of ${\bar P}_S$ (see discussion
in section~\ref{sec:z2ken} and footnote~\ref{foot:kenf}). 
\medskip

\noindent In view of the above theorem, we need to explicitly 
emphasize few points:
\medskip

\noindent
(1) {\em Theorem 2} implies that if $\Z_2$ theories $S_1$, $S_2$ 
    have computable probability distributions, then their order 
    relation induced by $\ken[S]$ is the same as the one induced
    by $\sha[S]$ in $L \rightarrow \infty$ limit. Indeed, this
    follows from the equivalence relation (\ref{eq:160}) and the fact 
    that $\ken[S]$ and $\sha[S]$ only differ by an $L$--independent
    constant for a theory with computable $P_S$. Thus, if it could
    be established that all theories in $\cS^{GZ}$ lead to computable
    distributions, one could also use Shannon entropy to distinguish 
    different levels of space-time order in the infinite volume limit. 
    In the specific case of $\Z_2$ gauge theories this would justify 
    the use of Gibbsian entropy (see footnote~\ref{foot:kolsha}) 
    as a measure of space--time order in the theory. However, 
    the corresponding analysis of $\cS^{GZ}$ is not available yet, 
    and one also needs to realize that the potential extension of such 
    result would only be valid for theories with discrete local variables.
\medskip

\noindent
(2) In connection with the above remark we wish to emphasize that 
    we view $\ken[S]$ as the fundamental measure of space-time order 
    in the theory, while $\sha[S]$ is an auxiliary concept that can be 
    used, in some cases, to estimate $\ken[S]$. For certain issues the 
    use of $\ken[S]$ is simply insufficient. For example, consider 
    the problem of minimizing $\ken[S]$ on some subset of $\cS^{GZ}$ 
    with computable distributions. Unless $K(P_S)$ is bounded on such 
    subset (which is not necessary the case if the subset has infinite 
    number of elements), then minimizing $\sha[S]$ will not help us at 
    all even in the infinite volume limit. The recurring problem that 
    appears in similar considerations is that Kolmogorov entropy and 
    Shannon entropy are related by a ``theory--dependent'' constant 
    ($K({\bar P}_S)$).

\subsection{Kolmogorov Entropy: SU(N) Gauge Theories}

The extension of the above discussion to SU(N) gauge theories is 
straightforward in certain regards and problematic in others. From
the point of view of numerical lattice QCD it is natural to view
a configuration (to a given numerical accuracy) as a binary string
since that's how it is in fact stored in the memory of a computer.
While the basic idea remains to apply the Kolmogorov complexity 
measure to such strings and thus define the Kolmogorov entropy of 
a configuration/theory, the whole procedure needs to be formalized 
properly. Moreover, there are new conceptual problems arising due to 
the fact that local variables are continuous and thus involve infinitely 
many bits even on a finite lattice. The issues related to taking 
the appropriate limits are very relevant and need to be addressed.

\subsubsection{Canonical Coarse-Graining of the Gauge Group}
\label{sec:canonical}

In this section we discuss a systematic way of discretizing the gauge
group which is convenient for associating the SU(N) gauge configurations
with binary strings. A particular way we propose is motivated by the
issues of space--time order discussed here, but also by the anticipation
of studying gauge theories from the point of view of information theory
later. After discretizing the space-time coordinates via introduction
of a lattice, it is quite natural to discretize all continuous aspects
of the theory, including the continuous degrees of freedom. However,
this is usually not done since it is not necessary for proper 
regularization of a continuum system. At the same time, continuous gauge 
variables allow for incorporation of exact gauge invariance in essentially 
continuum--like manner. In reality though, exact gauge invariance carries
mainly an aspect of ``elegance'' in this case since it is well under control
even with a discretized gauge group, and its violations boil down to 
a numerical precision issue (See the discussion later in this section.).
Moreover, for our purposes, the role of the gauge coarse--grained 
approximations is to define the measure of space--time order in 
the original lattice theory with continuous gauge variables, and not to 
supply a different lattice construction of continuum QCD.

Discretization of gauge variables is made easier by the fact that SU(N)
groups are compact and each element can be specified by N$^2$-1 real
numbers chosen from bounded intervals. To provide a sufficient level of 
detail, let us focus on the group SU(3) which is most relevant for our
purposes. It is convenient to start from a particular trigonometric 
parametrization for matrix elements of $g\in$ SU(3) in a fundamental
representation, namely~\cite{Bron88}
\begin{eqnarray}
   g_{11} & = &  \cos\theta_1 \cos\theta_2\, e^{i\phi_1}  \qquad
   g_{13}   =    \cos\theta_1 \sin\theta_2\, e^{i\phi_4}  \qquad
   g_{12}   =    \sin\theta_1\, e^{i\phi_3}               \nonumber  \\
   g_{22} & = &  \cos\theta_1 \cos\theta_3\, e^{i\phi_2}  \qquad
   g_{32}   =    \cos\theta_1 \sin\theta_3\, e^{i\phi_5}  \nonumber  \\   
   g_{21} & = &  \sin\theta_2 \sin\theta_3\, e^{-i(\phi_4+\phi_5)} \,-\, 
                 \sin\theta_1 \cos\theta_2 \cos\theta_3\, 
                 e^{i(\phi_1+\phi_2-\phi_3)}              \nonumber   \\ 
   g_{33} & = &  \cos\theta_2 \cos\theta_3\, e^{-i(\phi_1+\phi_2)} \,-\, 
                 \sin\theta_1 \sin\theta_2 \sin\theta_3\, 
                 e^{-i(\phi_3-\phi_4-\phi_5)}    \label{eq:176}       \\ 
   g_{23} & = &  -\cos\theta_2 \sin\theta_3\, e^{-i(\phi_1+\phi_5)} \,-\, 
                 \sin\theta_1 \sin\theta_2 \cos\theta_3\, 
                 e^{i(\phi_2-\phi_3+\phi_4)}              \nonumber   \\
   g_{31} & = &  -\sin\theta_2 \cos\theta_3\, e^{-i(\phi_2+\phi_4)}  \,-\, 
                 \sin\theta_1 \cos\theta_2 \sin\theta_3\, 
                 e^{i(\phi_1-\phi_3+\phi_5)}              \nonumber  \\
   0 & \le & \theta_1, \, \theta_2, \, \theta_3 \,\le \, \pi/2  \qquad
                 0 \,\le \, \phi_1, \ldots ,\phi_5 < 2\pi \nonumber
\end{eqnarray}
with the invariant group measure given by
\begin{equation}
   d g \;=\; \frac{1}{2 \pi^5} 
             \sin\theta_1 \cos^3\theta_1 d\theta_1 \,
             \sin\theta_2 \cos\theta_2 d\theta_2 \,
             \sin\theta_3 \cos\theta_3 d\theta_3 \,
             d \phi_1 \ldots d \phi_5
             \label{eq:177}
\end{equation} 
Using the above expressions we can now proceed to change the variables and 
define a canonical parametrization of the group where all eight parameters 
are chosen from a unit interval and the group measure becomes uniform (flat).
In particular, we have 
\begin{equation}
   u_1 \equiv 1-\cos^4 \theta_1    \qquad
   u_2 \equiv 1-\cos^2 \theta_2    \qquad
   u_3 \equiv 1-\cos^2 \theta_3    \qquad
   u_{3+a} \equiv \frac{1}{2\pi} \phi_a
   \label{eq:178}
\end{equation}
where
\begin{equation}
   u_a \in \cases{[0,1]\,,\;&a=1,2,3  \cr 
                  [0,1)\,,\;&a=4,\,\ldots ,8 \cr}
   \qquad\quad
   dg \,=\, \prod_{a=1}^{8} \, d u_a
   \label{eq:179}
\end{equation}

Since group elements are uniformly distributed in the space of 
parameters $u_a$, we can proceed with the discretization in the usual 
manner. Instead of representing a value of $u_a$ exactly, we choose 
to keep only a ``coarse-grained'' information on it. In particular, 
to store such information in $k$ bits, we can partition the associated 
unit interval into $2^k$ equal-sized subintervals and specify only
which subinterval $u_a$ belongs to. Such arrangement obviously assumes 
the existence of a bijection between the set of subintervals and 
the set ${\frak B}^\star_k$ of binary strings with length $k$. To set 
up such bijection let us first define a set ${\eusm V}_k$ of standard 
``lattice values'' in the unit interval, and the ``lattice spacing''
$\epsilon_k$ for arbitrary $k$, namely
\begin{equation}
   {\eusm V}_k \,\equiv\, \{\, r^{(k)}_i \equiv \frac{i}{2^k} , \;  
                          i=0,1,2,\ldots,2^k-1 \,\} \qquad\quad 
    \epsilon_k \,\equiv\, \frac{1}{2^k} \,=\, r^{(k)}_1 
   \label{eq:180}
\end{equation}
Note that there is a natural bijection between ${\frak B}^\star_k$ 
and ${\eusm V}_k$. Indeed, every $i\in \N_{2^k}^0$ has a unique 
binary representation of the form 
$i = b_1\,2^{k-1} + b_2\,2^{k-2} + \ldots + b_{k-1}\,2 + b_k$
and then
\begin{equation}
   {\eusm V}_k \ni r^{(k)}_i \,=\, 0.b_1b_2 \ldots b_k  
   \; \longleftrightarrow \;    
        b_1b_2 \ldots b_k \,\equiv\, y^{(k)}_i \in {\frak B}^\star_k  
   \label{eq:185}
\end{equation}
We will use the above map to define a bijection between binary strings 
and partitions of a unit interval by associating the values $r^{(k)}_i$
with individual subintervals. In doing so, we wish to treat periodic 
parameters ($u_4,\ldots,u_8$) differently from non-periodic parameters 
($u_1,u_2,u_3$) because of their different geometric meaning. In particular, 
we wish to end up with a discretization for which the immediate vicinity 
of an identity matrix corresponds to a single lattice cell rather than being 
split over several different ones. This requires the values of periodic 
parameters close to $0$ and $1$ being associated with the same subinterval. 
We thus define 
\begin{equation}
   {\frak B}^\star_k \ni y^{(k)}_i  \;\longleftrightarrow\;
   \cases{\;[\,r^{(k)}_i,r^{(k)}_i + \epsilon_k\,) \;,\;i=0,1,\ldots,2^k-2 \cr
          \;[\,r^{(k)}_i,r^{(k)}_i + \epsilon_k\,] \;,\;i=2^k-1 \cr}
   \label{eq:190} 
\end{equation}
for non-periodic parameters and
\begin{equation}
   {\frak B}^\star_k \ni y^{(k)}_i  \;\longleftrightarrow\;
   \cases{\;[\, r^{(k)}_i-\frac{\epsilon_k}{2},r^{(k)}_i+\frac{\epsilon_k}{2} \,) 
            \;,\quad i=1,2,\ldots,2^k-1 \cr
          \;[\,r^{(k)}_i,\frac{\epsilon_k}{2}\,] \union [\, 1-\frac{\epsilon_k}{2},1 \,) 
            \;,\;\, i=0 \cr}
   \label{eq:195}
\end{equation}
for periodic ones. Let us denote the subintervals defined above as ${\eusm I}^{(k)}_i$, 
with an implicit understanding of different definition for periodic and non-periodic 
cases. 

The bijections (\ref{eq:185}), (\ref{eq:190},\ref{eq:195}) imply
that we can interchangeably specify a value ($r^{(k)}_i$), an interval 
(${\eusm I}^{(k)}_i$) or a binary string ($y^{(k)}_i$) when speaking of 
``coarse-graining'' a parameter $u_a$ into $k$ bits. For future reference, let us 
define the operation of coarse-graining explicitly. In terms of a value
of the parameter we have
\begin{equation}
    u_a  \; \longrightarrow \; u_a^{(k)} \,\equiv\, 
    r^{(k)}_i \in {\eusm V}_k  \qquad \mbox{\rm such that} \qquad
    u_a \in {\eusm I}^{(k)}_i
    \label{eq:200}
\end{equation}
At the level of binary strings, we first point out that we can associate any possible 
value of $u_a$ with an infinite binary string $z_a \equiv b_1 b_2 b_3 \ldots$ 
formed by digits of its binary representation $u_a=0.b_1b_3b_3 \ldots$. 
If $u_a$ has two binary representations  (e.g. $u_a=0.10000\ldots = 0.01111\ldots$) 
we choose the option with infinitely many zeros. Note that we include $u_a=1$ 
via $u_a=0.11111\ldots$. With these definitions the operation of coarse-graining into 
$k$ bits translates at the level of binary strings into
\begin{equation}
    z_a  \; \longrightarrow \; z_a^{(k)} \,\equiv\, 
    y^{(k)}_i \in {\frak B}^\star_k   \qquad \mbox{\rm such that} \qquad
    u_a \in {\eusm I}^{(k)}_i
    \label{eq:205}
\end{equation}
Note that in the non-periodic case coarse-graining simply comes down to keeping first 
$k$ bits of a binary representation for $u_a$. In the periodic case it is the same 
after applying an appropriate periodic shift.

\subsubsection{Coarse--Grained Configurations and Binary Strings}

Consider an SU(3) configuration $U \equiv \{ U(n,\mu) \}$ on a discretized 
hypercubic symmetric torus of size $L$. Emphasizing the canonical representation
of the previous section, let us denote the same object as
$u \equiv \{ u(n,\mu) \}$, where 
$u(n,\mu) \equiv (u_1(n,\mu),u_2(n,\mu), \ldots ,u_8(n,\mu))$. 
We associate with $u$ (and hence with $U$) a sequence of $k$-th level coarse-grained 
configurations $u^{(k)}$ defined via
\begin{equation}
   u \;\longrightarrow \; u^{(k)} \equiv \{\,u^{(k)}(n,\mu)\,\} \equiv
   \{\, (u_1^{(k)}(n,\mu),u_2^{(k)}(n,\mu), \ldots ,u_8^{(k)}(n,\mu)) \,\} 
   \label{eq:210}
\end{equation}
where the coarse-grained gauge parameters are defined in Eq.~(\ref{eq:200}), and 
the ``$k$-th level'' refers to the fact that each parameter is represented by $k$ 
bits. By construction $\lim_{k \to \infty} u^{(k)} = u$ in the standard Euclidean 
norm.

In an analogous manner we now proceed with defining a sequence of $k$-th level binary 
strings corresponding to $u^{(k)}$. To that effect we first define a binary 
string $z^{(k)}(n,\mu)$ of length $8k$ assigned to a given link $(n,\mu)$ via
\begin{equation}
    z^{(k)}(n,\mu) \,\equiv\, z^{(k)}_1(n,\mu)\, z^{(k)}_2(n,\mu) \ldots
                              z^{(k)}_8(n,\mu)
    \label{eq:215}
\end{equation} 
where $z^{(k)}_a(n,\mu)$ are defined in (\ref{eq:205}) and the operation on 
the right--hand side means a simple concatenation. To form a string representing
the whole configuration we have to fix the recursive invertible map $F \in \Fmaps$ 
of Eqs.~(\ref{eq:130}), relating a linear order on the string to underlying 
space-time geometry. For arbitrary such map we then have 
\begin{equation}
     z^{(k)}_F(U)  \,\equiv\, z^{(k)}(F^{-1}(1))\,  z^{(k)}(F^{-1}(2)) \ldots
                              z^{(k)}(F^{-1}(4L^4))
     \label{eq:220}
\end{equation}
In other words, $z^{(k)}_F$ is the concatenation of local binary strings 
$z^{(k)}(n,\mu)$ in the order provided by the map $F$. By construction, the length
of this string is $l(z^{(k)}_F)=32 L^4 k$. 
\smallskip

\noindent Let us finally make few remarks pertaining to these issues.
\smallskip

\noindent (1) Note that in addition to the choice of map $F$, there is additional 
freedom in assembling the binary string representing the coarse--grained configuration 
$u^{(k)}$. For example, instead of placing different bits of $z^{(k)}_a(n,\mu)$
into a localized part of the string, we could concatenate $k$ blocks of size
$32 L^4$, where the $j$-th block would be formed by the $j$-th bit of
$z^{(k)}_a(n,\mu)$. However, these freedoms do not affect asymptotic considerations. 
\smallskip

\noindent (2) We emphasize that we will interchangeably use the notation $U^{(k)}$ and 
$u^{(k)}$ to denote the same object. In the former case one should realize that 
$U^{(k)}(x,\mu)$ is a matrix constructed from $u^{(k)}(x,\mu)$ via (\ref{eq:176}) rather 
than matrix of coarse-grained matrix elements.  
\smallskip

\noindent (3) For future reference, we will denote the set of all configurations $U$
on the symmetric torus with $L$ lattice sites on the side as $\cU^L$. Similarly,
the finite set of $k$-th level coarse grained configurations $U^{(k)}$ will be 
denoted as $\cU^{L,k}\subset \cU^L$. 

\subsubsection{Kolmogorov Entropy}

With the above preparations, we can now turn to the issue of quantifying the degree 
of space--time order for SU(3) (and analogously for SU(N)) configurations/theories. 
Following the same path as in the $\Z_2$ case, the underlying idea is to apply 
the Kolmogorov complexity measure to binary strings describing the configurations. 
However, there are several problems with carrying out such a program. 
Among others, an immediate issue is that, even for finite $L$, the set {\sobjects} 
of objects we wish to describe (set of all SU(3) configurations) is not countable. 
This has several undesirable consequences, the most obvious one being that we cannot 
form a one-to-one correspondence between binary strings from ${\frak B}^\star$ 
and configurations.
\footnote{Note that it is not a problem to form a bijection between the sets of 
{\em infinite strings} and configurations at given fixed $L$. However, we need
a bijection that includes all possible values of $L$.}

In order to proceed, we will abandon the notion that the binary string associated
with $U$ has to encode this configuration completely. 
\footnote{By ``complete description'' we mean such that any property of a configuration 
can be inferred from properties of the associated string and vice-versa.} 
Instead, we consider a sequence of coarse--grained configurations
$\{\,U^{(k)}, \; k=1,2,\ldots\}$, each of which approximates $U$ in an increasingly
fine detail. Clearly, the ability to analyze $U^{(k)}$ for arbitrary $k$ is
equivalent to analyzing $U$ fully. At the same time, as we discussed in the previous 
section, each $U^{(k)}$ can be completely described by a binary string $z^{(k)}_F(U)$ 
of length $32 L^4 k$. For given fixed $k$, we can thus proceed and define the $k$-th 
level Kolmogorov entropy of $U$ corresponding to map $F$ via
\begin{equation}
     \ken_F(U,k)  \,\equiv\, K(z^{(k)}_F(U))
     \label{eq:225}
\end{equation} 
As discussed in Section \ref{sec:z2ken}, this measure is additively optimal with respect
to set $\Fmaps$ with arbitrary $F$ being an universal element. Following the same logic 
as in $\Z_2$ case we fix $F^u$ of Eq.~(\ref{eq:145}) and define
\begin{equation}
     \ken(U,k)  \,\equiv\, K(z^{(k)}(U))  \qquad\qquad
     z^{(k)}(U) \;\, \mbox{\rm computed via} \;\,F^u 
     \label{eq:230}
\end{equation} 
If $S \equiv S(\beta)$ is a theory from $\cS^{G}$, we define its $k$-th level 
Kolmogorov entropy as
\begin{equation}
      \ken [S,k] \,\equiv\,  \langle \, \ken (U,k) \,\rangle_{S} 
      \,=\, \int dU \ken (U,k) \, P_S(U)
      \qquad\qquad
      P_S(U) = e^{-S(U)}/Z 
     \label{eq:235}
\end{equation}
i.e. as an ensemble average of Kolmogorov entropies assigned to individual configurations.

In case of $\Z_2$ gauge theories, we could directly associate Kolmogorov entropy with
the measure indicating the degree of space--time order dynamically generated by the 
theory in its typical configurations. For QCD the situation is slightly more involved.
Indeed, it is important to realize that binary strings $z^{(k)}(U)$ can become 
asymptotically large by either increasing the lattice size $L$, or by increasing the level 
of coarse--graining $k$. Naively, one is tempted to take $k \to \infty$ while keeping
$L$ fixed in order to completely characterize the theory in a finite lattice volume.
While this can certainly be done, the Kolmogorov entropy in such limit will not be 
a measure of space--time order at all. Rather, it will reflect how ordered are the digits
of a typical local gauge parameter $u_a(n,\mu)$. Thus, it turns out that in order to
obtain the construct we are interested in, we have to do just the opposite. In particular,
we will say that $\ken(U,k,L)$ ($\ken [S,k,L]$) is a measure of space--time order
in the configuration (theory) if and only if
\begin{equation}
   k \,\ll \, L^4
   \label{eq:240}
\end{equation}
where we have made the $L$-dependence of $\ken$ explicit. 

The above considerations indicate that in dealing with issues of space--time order, we 
are simply required to approach any particular question at fixed coarse-graining level
$k$. For example, the Kolmogorov entropy $\ken [S,k,L]$ induces a ``$k$-th level'' order
relation on the set $\cS^{G}$ of theories defined on the torus with lattice
size $L$, namely
\begin{equation}
      S_1 \; \mathrel{\mathop<\!{\scriptstyle k}} \;S_2
      \qquad \mbox{\rm if} \qquad
      \ken [S_1,k,L] \,<\, \ken [S_2,k,L] 
      \label{eq:245}    
\end{equation}
The underlying interpretation is that $S_1$ possesses more space--time order than 
$S_2$ if $S_1 \; \mathrel{\mathop<\!{\scriptstyle k}} \;S_2$ (for $k \ll L^4$). 
Given that interactions are local, we expect that the above order relation 
becomes $L$--independent at sufficiently large $L$ for any two theories, thus
characterizing the interactions themselves. In the $L \rightarrow \infty$ limit 
it is natural to consider an induced order relation on the set of equivalence classes,
where theories whose Kolmogorov entropies differ at most by a constant belong to
the same class (see Eq.~(\ref{eq:160})).

Finally, let us point out that in complete analogy with the $\Z_2$ case we can make 
the definition of $k$-th level Kolmogorov entropy gauge invariant by minimizing
$\ken (U,k,L)$ on the gauge orbit of $U$ (see Eq.~(\ref{eq:165})). We can also 
gauge--fix to the information gauge via an obvious modification of Eq.~(\ref{eq:170}).

\subsubsection{The Conjecture of Hierarchic Structure}

One of the main reasons behind introducing the measure of space--time order based on 
Kolmogorov complexity is its universality in the sense that it can capture the ordered
space-time behavior in any possible form. This is important because our geometric
intuition for what the domain--correlated behavior of a function can be is very 
limited. Even taking just a scalar function of two space--time coordinates as 
an example that human brain can handle relatively comfortably visually, we have 
absolutely no intuition on possible degrees of space--time coherence beyond extreme 
cases. Indeed, we tend to deal with such functions via their graphs which we can picture 
mentally if they are nice smooth surfaces in 3-d space. The common intuition then  
is that smoother graphs (functions) are ``more ordered'' than rougher 
graphs (functions). However, even in the extreme case of very high order our 
intuition is incomplete since there exist very singular functions whose behavior 
is maximally ordered in the sense that local patterns determine their behavior 
everywhere. 

Returning back to lattice configurations, the above--mentioned generality of
the approach based on Kolmogorov complexity, while highly desirable, has certain 
implications that have to be dealt with. In particular, the point that needs
to be emphasized at this stage is that the linear order on the set of configurations 
induced by $\ken(U,k,L)$ is not necessarily preserved if the ``resolution'' $k$ changes.
More precisely, if $U_1,\,U_2 \in \cU^{L}$, and $k_1 < k_2 \ll L^4$ with 
$U_1 \mathrel{\mathop<\!{\scriptstyle k_1}} U_2$, then it does not necessarily follow 
that $U_1 \mathrel{\mathop<\!{\scriptstyle k_2}} U_2$. Indeed, we can easily imagine 
a situation where first $k_1$ bits of gauge parameters in $U_1$ exhibit high 
space--time coherence, while being disordered in $U_2$. However, in the trailing 
$k_2-k_1$ bits the situation can reverse in favor of $U_2$ if supplied with highly 
ordered digits. The point is that while we intuitively tend to think that leading digits 
are ``more important'' for the structure, the Kolmogorov complexity measure 
does not distinguish such hierarchy and the order around 1-st digit counts in principle 
as much as the order around the 100-th digit. We emphasize again that this is a 
{\em desirable} feature of Kolmogorov entropy measure since our goal is to be completely
general. Thus, if there is a secret message encoded into configuration $U$ via
inserting a highly space-time ordered binary string into various digits of $u$, it can 
still be detected by analyzing $\ken(U,k,L)$ for sufficiently large $k$.

The point we wish to convey in the above reasoning is that at the {\em configuration level}, 
the ranking introduced by Kolmogorov complexity has to be viewed strictly as a 
resolution--dependent concept in this case. In other words, it is meaningful to say that 
configuration $U_1$ is more ordered than configuration $U_2$ only when they are both analyzed 
with fixed resolution $k$. However, when we characterize the theory $S(g)\in \cS^{G}$,
via $\ken [S,k]$ things are expected to be much more universal in this regard. 
Indeed, since our theories are analytic (at least in certain corner of the
configuration space), we expect that with other things being equal
(such as multiplicities), the space--time order in leading digits has a more
decisive influence on probability of occurrence than possibly equivalent
order in higher digits. This is simply because leading digits influence the value of 
the action (and thus a probability of occurrence) more significantly than higher digits. 
In other words, it is natural to assume that a theory in $\cS^{G}$ will feature 
typical configurations with {\em hierarchical order}. By this we mean that the relative 
degree of order will not increase when higher digits are included, and will in fact
typically decrease. Put still differently, the higher digits in typical configurations
are expected to exhibit increasingly more space--time randomness. 

We wish to formalize the above observations into a definite statement concerning 
the measure $\ken [S,k]$ of space--time order on the set $\cS^{G}$. While at present 
we do not have an apparatus to attempt the proof of the following statement,
we propose it for future considerations because of its relevance.
\medskip

\noindent {\bf Conjecture CI1.} {\em Let $k$ be a fixed positive integer and 
let ``$\mathrel{\mathop<\!{\scriptstyle k}}$'' be the linear order relation on
$\cS^{G}$ induced by $\ken [S,k,L]$ in the $L \rightarrow \infty$ limit.
Then ``$\mathrel{\mathop<\!{\scriptstyle k}}$'' is universal with respect to $k$,
i.e. if $S_1 \mathrel{\mathop<\!{\scriptstyle k_1}} S_2$, then 
$S_1 \mathrel{\mathop<\!{\scriptstyle k_2}} S_2$, for arbitrary $k_1,k_2$ 
and arbitrary $S_1,S_2 \in \cS^{G}$.} 
\smallskip

\noindent Few points need to be emphasized in connection with the above conjecture.
\smallskip

\noindent (1) The utility of Kolmogorov entropy as a measure quantifying the degree 
of space--time order in different theories does not rely on the validity of {\bf CI1}. 
Indeed, in our quest for uncovering the relevant collective degrees of freedom 
we can use $\ken [S,k]$ at fixed resolution $k$ as a general measure distinguishing 
different theories. Obtaining a complete picture will require changing the resolution
$k$ in either case. Nevertheless, from a theoretical standpoint it would be very
appealing if {\bf CI1} could be demonstrated at least in some simplified framework.
In what follows we will implicitly assume its validity.
\smallskip

\noindent (2) A weaker statement with similar consequences (namely existence
of an absolute ranking of theories by the degree of space--time order they generate)
can be obtained if one postulates the existence of $k_0[S_1,S_2]$ such that 
the relative ordering of $S_1$ and $S_2$ is preserved for all $k>k_0$.  
\smallskip

\noindent (3) We remind the reader again that it is crucial in the formulation 
of {\bf CI1} that the infinite--volume limit is taken at fixed $k$ to define 
the ordering. This is to nullify the influence of various non-universal constants
implicitly present in the definition of $\ken [S,k,L]$. Also, it is implicitly
understood in {\bf CI1} that in $L\rightarrow \infty$ limit the linear order
is defined on the equivalence classes of theories whose Kolmogorov entropies
are bounded by a constant (see Eq.~(\ref{eq:160})).

\subsubsection{The Associated Coarse-Grained Theory and Shannon Entropy}

In section \ref{sec:sha_kol} we have argued that in the case of $\Z_2$ gauge 
theories it is useful to consider the Shannon entropy of the corresponding ensembles. 
Even though this is an auxiliary concept from our point of view, its usefulness 
stems from the fact that Kolmogorov entropy is a non-computable notion. 
Using {\em Theorem 2}, Shannon entropy gives us at least a possibility to calculate
which of the arbitrary two $\Z_2$ gauge theories with computable probability 
distributions generates more space--time order in the infinite--volume limit. 
This raises a natural question at this point, namely how to define this auxiliary 
concept in case of SU(N) lattice gauge theories. 
 
As we argued extensively above, we carry out the analysis of space--time structure 
at a given coarse--grained level $k$, wherein the configuration $U$ is associated 
with its coarse--grained counterpart $U^{(k)}$ (or $u^{(k)}$), completely represented 
by a binary string $z^{(k)}$. Given $S \in \cS^{G}$ and fixed $L$, our goal is to 
define a probability distribution on a finite set $\cU^{L,k}$ of coarse--grained 
configurations $U^{(k)}$, such that its Shannon entropy is related to Kolmogorov 
entropy $\ken [S,k,L]$ in the same manner as in {\em Theorem 2} 
(see Eq.~(\ref{eq:175})). To do that, it is useful to think about the theory $S$ in 
terms of the corresponding ensemble $\cE[S]$, which is fully represented by an 
appropriately chosen infinite probabilistic chain. The coarse--graining procedure 
corresponds to the replacement 
\begin{equation}
  \cE[S] \equiv
  \{\ldots ,U_{i-1},U_{i},U_{i+1}, \ldots \} 
  \;\,\longrightarrow\;\, 
  \cE^{(k)}[S] \equiv 
  \{\ldots ,U^{(k)}_{i-1},U^{(k)}_{i},U^{(k)}_{i+1}, \ldots \} 
  \label{eq:250}
\end{equation}
The ensemble $\cE^{(k)}[S]$ defines a lattice theory of discrete variables 
$u_a^{(k)}$, which we denote as $S^{(k)}$, so that we can write formally
$\cE^{(k)}[S] \equiv \cE[S^{(k)}]$.
\footnote{Definition of action $S^{(k)}$ is more than formal though
in the sense that its value for arbitrary $U^{(k)}\!\in\cU^{L,k}$ can be 
explicitly written down (see bellow).} 
The probability density of encountering a configuration $U\!\in\cU^L$ in
$\cE[S]$ is given by $P_S(U) = \exp(-S(U))/Z$, while the probability of 
encountering $U^{(k)}\!\in\cU^{L,k}$ in $\cE^{(k)}[S]$ is by construction
\begin{equation}
    P_{S^{(k)}} \Bigl( U^{(k)} \Bigr) \,\equiv \,
    \int dV \,\delta \Bigl( U^{(k)} - V^{(k)} \Bigr) \, 
    P_S\Bigl( V \Bigr)
    \label{eq:255}
\end{equation}
where $V^{(k)}$ is a $k$-th level coarse-grained configuration associated
with $V$. Note that using our canonical representation
of gauge variables, we can write the above equation in a precise (rather
than formal) manner as
\begin{equation}
    P_{S^{(k)}} \Bigl( U^{(k)} \Bigr) \,=\,
    \int \prod_{x,\mu,a} d v_a(x,\mu) \,
    \prod_{x,\mu,a} \delta \Bigl( u^{(k)}_a(x,\mu) - 
                                  v^{(k)}_a(x,\mu) \Bigr) \, 
    P_S ( v )
    \label{eq:260}
\end{equation}
Moreover, for actual evaluation of this probability one could use an 
explicit form without the $\delta$--functions. To do that, we define 
the $k$-th level displacement variables 
$\tau^{(k)}\equiv \{\, \tau^{(k)}_a(x,\mu) \,\}$ with values in the intervals
\begin{equation}
   \tau^{(k)}_a(x,\mu) \in \cases{\;
        [0,\epsilon_k]\,,\;&a=1,2,3  \cr 
        \;[-\epsilon_k/2,\epsilon_k/2]\,,\;&a=4,\,\ldots ,8 \cr}
   \label{eq:265}
\end{equation}
in terms of which we can write
\begin{equation}
    P_{S^{(k)}} \Bigl( U^{(k)} \Bigr) \,=\,
    \int d \tau^{(k)} \,
    P_S\Bigl( u^{(k)} + \tau^{(k)} \Bigr)   \qquad\quad
    d \tau^{(k)} \equiv \prod_{x,\mu,a} d \tau^{(k)}_a(x,\mu)
    \label{eq:270}
\end{equation}
where the periodicity is implicitly assumed to be taken into account 
for arguments of $P_S$ with $a=4,\,\ldots ,8$.
      Equation (\ref{eq:270}) defines a theory $S^{(k)}$ of discrete
variables $u_a^{(k)}$ such that its Kolmogorov entropy is equal to
$k$-th level Kolmogorov entropy of $S$, namely
\begin{equation}
      \ken [S^{(k)}] \,\equiv\, 
      \sum_{U^{(k)}} \ken (U^{(k)}) \, P_{S^{(k)}}(U^{(k)})
      \;=\;
      \ken [S,k] \,\equiv\, 
      \int dU \ken (U,k) \, P_S(U)
     \label{eq:275}
\end{equation}
where in the context of $S^{(k)}$ (left hand side; discrete variables) we 
have $\ken (U^{(k)}) \equiv K(z^{(k)})$, which is equal to $\ken (U,k)$
in the context of $S$ (right hand side; continuous variables).

For theory $S^{(k)}$ with computable probability distribution, there is 
a relation between its Kolmogorov and Shannon entropies analogous 
to {\em Theorem 2} in $\Z_2$ case. This then induces the same relation 
between Shannon entropy of $S^{(k)}$ and the $k$-th level Kolmogorov
entropy of $S$ which is summarized in the following theorem.
\medskip

\noindent{\bf Theorem 3} {\em For $L$, $k$ fixed, let $S \in \cS^{G}$ 
leads to a discrete theory $S^{(k)}$ with computable probability distribution 
$P_{S^{(k)}}(U^{(k)})$, and $\sha[S^{(k)}] \equiv -\sum_{U^{(k)}} 
 P_{S^{(k)}}(U^{(k)}) \ln P_{S^{(k)}}(U^{(k)})$. 
Then
\begin{equation}
     0 \;\le\; \ken[S,k] - \sha[S^{(k)}] \;\le \; K({\bar P}_{S^{(k)}}) 
          + \cO(1)
     \label{eq:280}
\end{equation}
where ${\bar P}_{S^{(k)}}(z=z^{(k)})\equiv P_{S^{(k)}}(U^{(k)})$ is the induced 
probability distribution on ${\frak B}^\star$.}
\medskip

\noindent To close this section, we wish to emphasize two points.

\smallskip
\noindent (1) The remarks given at the end of section \ref{sec:sha_kol} apply
in the case discussed here as well. 
\smallskip

\noindent (2) The associated coarse--grained theory $S^{(k)}$ is not exactly
gauge--invariant with violations at the $k$-th bit level. This
does not have any undesirable implications in this context since the role of 
$S^{(k)}$ is to assist us with quantifying the degree of space--time order 
in $S$.

\section{The Principle of Chiral Ordering}
\label{sec:pco_main}

In the previous section we have introduced the notion of Kolmogorov entropy
for SU(N) gauge theories and proposed it as a measure quantifying the degree
of space--time order generated by the theory at the configuration level. 
Since the concept is based on Kolmogorov complexity of binary strings describing 
coarse--grained configurations, it is completely general in the sense that 
it reflects space--time order of coarse--grained configurations/theories in any 
form. Moreover, if we accept the hypothesis of {\em hierarchic order} 
(see conjecture {\bf CI1}), then we have a measure providing {\em universal}
ranking of theories in $\cS^{G}$ by the degree of space--time order generated
in their typical configurations. 

Before proceeding further, let us note that we included a detailed elaboration 
on Kolmogorov entropy in the previous section simply because we view the recent 
findings expressed in {\bf I1} (see section \ref{sec:newinput}) as an important 
new aspect that needs to be properly incorporated into the framework of quantum 
field theory. Equipped with the notion of Kolmogorov entropy we can, at least 
at the conceptual level, attempt to formulate problems that are relevant 
for the program of uncovering the {\em fundamental structure} of QCD vacuum in 
the path integral formalism. In that regard, it we would certainly be of crucial
importance if we had means to achieve the following major goal.
\smallskip

\noindent
{\em \underline{Goal 1}: Map out the landscape of Kolmogorov entropy in the set
of actions $\cS^{G}$ and understand its behavior.} 

\smallskip\noindent
Indeed, in order to uncover the nature of relevant collective variables $C$, we 
would eventually prefer to deal with theories possessing low Kolmogorov entropy
at finite lattice spacing. In such theories the distorting influence of 
randomness on $C$ is expected to be low thus giving us best chance of correctly 
identifying it. 

To start speaking in these terms, let us clear up some terminology first. 
While $\cS^{G}$ is the set of individual lattice actions, we tend to speak of 
lattice ``theories'' belonging there in two different ways. Firstly, we view
the quantum dynamics of finite (countable) degrees of freedom defined by any 
individual element $S(g)$ (g and other parameters fixed) of $\cS^{G}$ as a 
``theory''. Note that we talked about it in this sense when discussing 
the Kolmogorov entropy. For example, we can compare the space--time order of 
two theories differing only by values of $g$, 
e.g. $S(g_1) \, \mathrel{\mathop<\!{\scriptstyle k}} \,S(g_2)$.
\footnote{In fact, we naturally expect that typical configurations will become
increasingly spatially ordered as $g$ decreases. In other words, we expect
that $S(g_1) \, \mathrel{\mathop<\!{\scriptstyle k}} \,S(g_2)$ for any
$g_1 < g_2$.}
On the other hand, it is sometimes useful to think of lattice theory as 
a one-parameter family of interactions $S(g)$ parametrized by bare coupling. 
In this case the ``theory'' prescribes a lattice interaction at any value of 
$g$ thus parametrizing the approach to the continuum limit.

Thinking now in terms of this second interpretation, consider the set of pure-glue 
lattice gauge theories contained in $\cS^{G}$. Upon quantization, each theory 
$S(g)\in \cS^{G}$ is associated with ensembles $\cE(g)$ which determine the 
lattice quantum averages for various observables. To obtain values in physical 
units, one usually fixes the appropriate quantity to its physical value,
independently of $g$, which introduces the lattice spacing $a$.
From the physical point of view it makes clearly more sense to compare 
space--time order of two theories at the same lattice spacing rather than
the same bare coupling, i.e. to view the theory $S$ as being parametrized 
by lattice spacing rather than bare coupling. Since this is what we will do,
let us describe the procedure of introducing the lattice spacing in more 
detail. Assume, for the sake of argument, that the mass of the lightest scalar
glueball has been fixed to its value $\mbar$ for all theories $S(g)$, 
in each of which its dimensionless value is $m^S(g)$. For given $S(g)$
this defines the lattice spacing $a^S(g)$ via
\begin{equation}
    a^S(g) \,\mbar \;\equiv\; m^S(g)
    \label{eq:80}
\end{equation}
For simplicity, assume that the set of other interesting physical observables 
consists of masses $\mbar_i$, of all the other stable particles in the theory,
which we want to predict. In this restricted setting the statement that all 
actions $S(g) \in \cS^{G}$ define the same continuum theory (universality)
means that 
\begin{equation}
    \lim_{g \to 0} \frac{m_i^S(g)}{a^S(g)} \;=\; 
    \mbar \lim_{g \to 0} \frac{m_i^S(g)}{m^S(g)} \;=\;     
    \mbar_i \qquad\qquad
    \forall i, \quad \mbox{\rm independent of} \; S 
    \label{eq:85}
\end{equation}
In other words, all lattice theories $S \in \cS^{G}$ will predict the same 
ratios of masses of stable particles in the continuum ($g \to 0$) limit.
Note that in the above considerations we have implicitly assumed that all 
theories in $\cS^{G}$ have a second order phase transition at $g=0$. If we 
further assume that the lattice correlation lengths 
$\xi^S_i(g) \equiv 1/m^S_i(g)$ are monotonically increasing as $g \to 0^+$ 
(at least for $g < \hat g$, where $\hat g$ is independent of $S$), then we 
can invert the functions $a^S(g)$ and talk about lattice actions parametrized 
by lattice spacing $a$. In other words, we set 
$S^\prime (a) \equiv S(g = g^S(a))$, but in what follows we will keep 
the same symbol ($S$) for both dependences to simplify the notation. 

With all the above pieces in place, it is tempting to conclude that, at least 
at the conceptual level, it is now clear how to proceed with the program of 
identifying the fundamental structure of QCD vacuum (collective variables $C$). 
Being restricted to some finite range of non-zero lattice spacings $a$ by 
practical considerations, we could minimize the Kolmogorov entropy 
$\ken[S(a),k]$ in the set of actions at fixed $a$ to identify the action
$S^o(a)$ which maximizes the space--time order. We would then study 
the corresponding ensemble $\cE^o(a)$ in a Bottom-Up manner to extract
as much information as possible about the nature of space--time order present. 
The lattice spacing could then be decreased and the procedure repeated in 
an attempt to gain the information about the sought-for collective variables 
$C$ in the continuum limit.

As simple and appealing this might sound, it is most likely a wrong way
to pose the relevant task (even at the level of principle, leaving out 
the practical considerations). Indeed, apart from the issues of existence 
and uniqueness 
of $S^o(a)$ (which might not be serious in this case~\footnote{We note 
that such considerations were actually part of the reason why we chose 
to parametrize theories in terms of $a$ rather than $g$. Indeed, in the latter 
case, the set $\cS^{G}$ restricted to actions at fixed $g$ most likely does not 
have a bottom element with respect to $\ken$. The other reason obviously is that 
when we compare Kolmogorov entropies for two actions with the same number of 
lattice degrees of freedom and fixed $a$, we are comparing the degrees of 
space--time order in the same physical volume.}), there is a fatal flaw in the
program as formulated above. Indeed, assume that we start the minimization 
to determine $S^o(a)$ from the action $S^{ini}(a)$ such that theory $S^{ini}$
is well in the scaling region for lattice spacing $a$ considered. Thus, for
$S^{ini}$ we have that $m^{S^{ini}}_i(a)/a \approx \mbar_i$ for physical 
masses $\mbar_i$. Since the minimization of $\ken$ makes no reference 
to correlation lengths other than 
$\xi^S(a) \equiv 1/m^S(a) = 1/a\mbar$ (recall that lattice spacing 
is kept fixed), there is no reason to expect that the scaling behavior will be 
preserved during the minimization. In other words, the masses of stable
particles (other than the one fixing the lattice spacing) would most likely
drift away from their physical values. Thus, even though the Kolmogorov 
entropy might be low in $S^o(a)$, we would certainly not want to use it to 
make conclusions about the nature of space--time order in the continuum
limit, no matter how small $a$ might be. These considerations suggest that in 
an attempt to develop a systematic program for studying space--time structure 
in QCD vacuum, we would ideally (in principle) like to be able to do 
the following.

\smallskip
\noindent  {\em \underline{Goal 2}: Starting from some action 
$S^{ini}(a)$, we would like to move in the set $\cS^{G}$ toward actions with
lower Kolmogorov entropy such that if $S^{ini}(a)$ is in the scaling
region, then the theories reached will remain in the scaling region, and 
if $S^{ini}(a)$ is not in the scaling region, then the theories reached 
will not diverge further away from it.}

\smallskip\noindent
As we emphasized along the way, the above goal specifies the ``in principle''
strategy for identifying a fundamental structure in QCD vacuum. Indeed,
it appears unlikely that at the practical level we could proceed via direct 
constrained minimization of Kolmogorov entropy in the set of actions. 
Instead, we will view {\em Goal 2} as a conceptual guide against which 
our attempts should measure their potential merit. The second part of this 
article deals with a proposal in that direction.

To start describing our approach, it is instructive to first highlight 
the crucial obstacles to practical realization of {\em Goal 2} which need
to be addressed. 
($\alpha$) There is no standard explicit way of parametrizing the elements 
of $\cS^{G}$ and thus no standard practical way of ``moving in the set 
of actions''. 
($\beta$) There is no clear underlying principle that would guide us to move 
in $\cS^{G}$ while ``preserving physics'' as demanded by {\em Goal 2}. 
One could view the renormalization group (RG) blocking as a procedure of such 
kind, except that the decimation of the lattice poses a problem. At the heart 
of our method is the proposal for such principle. 
($\gamma$) Since Kolmogorov entropy is not computable, we cannot directly 
minimize it or at least check aposteriori by explicit calculation that it became 
lower during the procedure. One thus needs to discuss why is the chosen evolution 
in the set of actions believed to lower the Kolmogorov entropy. In the sections 
that follow we will focus on the above issues in turn.

\subsection{Configuration-Based Deformation of the Action}
\label{sec:deformation}

In this section we describe a generic way of ``moving'' in the set of actions
based on the local deformation of individual configurations. We will use 
the fact that it is not important for us to know analytically (or otherwise)
the exact form of the action we work with. Indeed, we only need to be able to 
inspect the {\em ensemble} associated with the action and to make sure that it 
represents a valid element of $\cS^{G}$. Thus, consider a map
\begin{equation}
   \cM : \, \cU^L \longrightarrow \cU^L \qquad\quad   
   U_{n,\mu} \;\longrightarrow\; \cM_{n,\mu}(U)
   \label{eq:285}
\end{equation}
Let $S\in \cS^{G}$ and $\cE_S$ be the associated ensemble. The map $\cM$ then
induces the new ensemble
\begin{equation}
   \cE_S \equiv \{\ldots U^{i-1},U^{i},U^{i+1}, \ldots \} 
          \quad \longrightarrow \quad 
          \{\ldots \cM(U^{i-1}),\cM(U^{i}),\cM(U^{i+1}), \ldots \} 
          \equiv \cE_{S^{\cM}} 
   \label{eq:290}  
\end{equation}
where the old probability density $P_S(U)=e^{-S(U)}/Z_S$ is modified to 
\begin{equation}
   P_{S^\cM}(U) \,=\, \int dV \,P_S(V) \,\, \delta \Bigl( U- \cM(V) \Bigr)  
   \label{eq:295}  
\end{equation}
and is associated with the ``action'' $S^{\cM}$ given by
 \begin{equation}
   e^{-S^\cM(U)} \,\equiv\, P_{S^\cM}(U) \,Z_{S^\cM}   \qquad\qquad
   Z_{S^\cM} = Z_S  
   \label{eq:300}  
\end{equation}
We now impose the following requirements on the map $\cM$:
\smallskip

\noindent {\em (i) Symmetries.} We require that the symmetries of the original
action $S(U) \in \cS^{G}$ are also present in new action $S^\cM(U)$. One can
easily inspect that this is guaranteed if $\cM_{n,\mu}$ transforms in the same
way as $U_{n,\mu}$ under respective symmetry transformations. Thus, for example
\begin{equation}
     \cM_{n,\mu}(U)  \;\longrightarrow G_n \,\cM_{n,\mu}(U)\,G^{-1}_{n+\muhat}
     \qquad \mbox{\rm under} \qquad 
     U_{n,\mu}  \;\longrightarrow G_n \, U_{n,\mu} \, G^{-1}_{n+\muhat}
     \label{eq:305}  
\end{equation}
and similarly for the symmetries of the hypercubic lattice structure. 
\smallskip

\noindent {\em (ii) Locality.} Considering the configuration in the 
canonical representation (see Sec.~\ref{sec:canonical}), i.e.
$U = \{ U_{n,\mu} \} \equiv \{ u_{n,\mu} \} = u$
we require the existence of constants $C$, $\gamma$ such that
\begin{equation}
     \frac{\partial \cM_{n,\mu}^a(u)}{\partial u_{m,\nu}^b} 
     \,\le\, C e^{-\gamma |n-m|}
     \label{eq:310}
\end{equation}
independently of $u$. 
\smallskip

\noindent {\em (iii) Classical Limit.} If $A_\mu(x)$ is a classical (smooth)
field (see footnote \ref{foot:smooth}), and $U_{n,\mu}(a)$ is its 
discretization (\ref{eq:45}) with classical lattice spacing $a$, then we
require that $\cM_{n,\mu}(U,a)$ is also its valid discretization. In other
words, that 
\begin{equation}
      U_{n,\mu}(a) - \cM_{n,\mu}(U,a) \;=\; \cO(a^2)
      \label{eq:315}
\end{equation} 

\noindent Note that the nature of the above conditions (especially {\em (ii)}
and {\em (iii)}) is such that it is implicitly assumed that $\cM$ is defined
for arbitrary $L$. In the following, we will denote the set of maps satisfying
the conditions {\em (i)--(iii)} as \Umaps. 

We propose that the action $S^\cM$ defined via $\cM \in \Umaps$ satisfying 
{\em (i)--(iii)} will generically be a valid element of $\cS^{G}$. It needs
to be emphasized that this conclusion is not the result of a rigorous proof
at this stage. Indeed, while the symmetries are directly guaranteed by {\em (i)}, 
it is not straightforward to precisely formulate and prove the relevant statement
regarding locality of $S^\cM$ based on (\ref{eq:310}), or the statement on
the classical limit based on (\ref{eq:315}).
\footnote{This is not to say that it is impossible to formulate and prove
such statements (with appropriate restrictive conditions stipulated). However, 
we will not deal with these rather involved issues rigorously here.}
On the other hand, from the physical point of view it is quite clear that the 
transition from $S$ to $S^\cM$ will not cause any major pathologies if 
{\em (i)--(iii)} are satisfied. Indeed, consider arbitrary set of local operators 
$\{O_\alpha \}$ and the related set $\{O_\alpha^\cM \}$, where
\begin{equation}
    O_\alpha^\cM(U) \,\equiv\, O_\alpha(\cM(U))
    \label{eq:320}
\end{equation}
The conditions {\em (i)--(iii)} guarantee that $O_\alpha^\cM$ will inherit 
the transformation properties, locality and classical limit of $O_\alpha$,
while at the same time
\begin{equation}
     \langle \, O_\alpha(U) \,\rangle_{S^\cM} \,=\,
     \langle \, O_\alpha^\cM(U) \,\rangle_S
     \label{eq:325}
\end{equation} 
Thus the expectation values of arbitrary local operators (and their correlation
functions) in $S^\cM$ correspond to values in original theory $S$ (element of 
$\cS^{G}$) for manifestly ``valid'' discretizations of the same continuum operators.
Roughly speaking, our overall level of rigor here is analogous to that of 
Wilson's renormalization group (RG) approach~\cite{RG}, wherein one moves in the
set of actions via RG transformation.

\subsection{The Principle of Chiral Ordering}
\label{sec:pco_sub}

Consider a configuration-based deformation $S^\cM$ of action $S$, given by an
ultralocal map $\cM\in\Umaps$. By ultralocality we mean that $\cM_{n,\mu}(U)$ 
does not depend on $U_{m,\nu}$ for $|n-m|$ larger than some configuration--independent
positive constant $\kappa$. Using the same arguments as those invoked in the case 
of RG transformations, we can conclude that the behavior of $S^\cM$ at large lattice
distances is the same as that of $S$. In other words, the bound--state masses 
$\mbar_i^S$ in lattice units are expected to be the same (on an infinite lattice) 
as $\mbar_i^{S^\cM}$. However, there are several setbacks associated with 
such generic approach to moving in the set of actions. {\em (a)} Transition 
from $S$ to $S^\cM$ can strongly affect observables for which the gauge field 
fluctuations at the scale of $\kappa$ lattice spacings are crucial (i.e. observables
that are not determined entirely by long--distance behavior of correlators), even 
on an infinite lattice. {\em (b)} On a finite lattice, even the estimates of masses
in $S^\cM$ can change from the estimates in $S$. Moreover, after applying the 
deformation many times, all observables in $S^{\cM^n}$ on a finite lattice can 
change uncontrollably (both in lattice and physical units) relative to $S$. 
There is nothing to protect them from changing. {\em (c)} For the generic choice
of $\cM$, (which typically can be interpreted as a {\em ``smoothing''} operation),
the repeated application of the map on given configuration ($\cM^k(U)$) is not 
expected to converge to a non-trivial limit. Indeed, one rather expects the
convergence to a trivial configuration with gauge potentials vanishing 
($U_{n,\mu}=\identity$ (identity matrix)), or a diverging behavior. Thus, even 
if Kolmogorov entropy of $\cM(U)$ decreases relative to $U$ for all relevant 
configurations, (which also means that $\ken[S^\cM,L]<\ken[S,L]$), we cannot expect 
a unique/interesting result of minimizing procedure based on the repeated application 
of $\cM$. Taken together, the above points serve to illustrate that to approach 
{\em Goal 2} via configuration--based deformation of the action with {\em generic} 
$\cM$ is not expected to succeed. We also note that these issues become slightly 
more complicated in the non-ultralocal case since lattice correlation lengths in 
$S^\cM$ can in principle change (by at most the range of non-ultralocality) relative 
to $S$, and thus the masses (and the lattice spacing) can change accordingly even 
on an infinite lattice.   

In this section we wish to propose a class of maps $\cM$ which, we believe, can
potentially alleviate most of the problems encountered in generic case. It is quite 
obvious from the above discussion that such maps must be a very special subset of
all possibilities and, in an attempt to identify them, we will proceed quite 
heuristically. Indeed, our only initial guide will be the striking success
of linking the topological density operator to chiral symmetry, which resulted
in {\bf I1} of Sec.~\ref{sec:newinput}, i.e. the observation of space--time order 
at the fundamental level~\cite{Hor03A}. It will become clear soon in the process
however, that the approach potentially carries much deeper connotations. Indeed,
we will use chiral symmetry here in different and more general manner. Consider 
a fermionic action $S^F$ specified by lattice Dirac kernel $D$. Apart from all 
conditions imposed on $D(U)$ by the fact that $S^F$ is an element of $\cS^F$ (which 
includes requirement of chiral symmetry), we impose an additional and highly 
non-trivial requirement. In particular,
\begin{equation}
    D_{n,n+\mu}(U) \,=\, D^f_\mu \times \Ubar_{n,\mu}
    \qquad\quad
    \forall\, n,\,\mu \,,\, \forall\, U
    \qquad\quad
    \Ubar_{n,\mu} \in \mbox{\rm SU(N)}
    \label{eq:330}
\end{equation}
where $D^f_m$ is a free fermion matrix (no color indices) representing hopping by vector 
$m$, namely
\begin{equation}
   (D^f_m)_{\alpha,\beta} = 
   D_{n,n+m}(\identity)_{\scriptstyle \alpha,\beta \atop 
                \scriptstyle a,a} 
   \qquad\quad
   \mbox{\rm color index} \;a\; \mbox{\rm fixed.}
   \label{eq:335} 
\end{equation}
In other words we require that the color-spin matrix $D_{n,n+\mu}(U)$ is a direct 
product of a free fermion matrix $D^f_{\mu}$ and an SU(N) ``matrix phase'' 
$\Ubar_{n,\mu}$. Consequently, the fermionic action $S^F$ specified by $D$ satisfying 
(\ref{eq:330}) defines a map $\cM^D$ which associates with every configuration $U$ 
a new configuration $\Ubar$ via 
\begin{equation}
   U \,\longrightarrow\, \cM^D(U) \qquad\qquad 
   \cM^D_{n,\mu}(U) \,\equiv\, \Ubar_{n,\mu}
   \label{eq:340}
\end{equation}
We will call $\Ubar=\cM^D(U)$ a {\em chirally ordered} version of $U$ according to $D$. 
One can easily check that the map $\cM^D$ satisfies the conditions {\em (i)--(iii)} of 
Sec.~\ref{sec:deformation}. 
\footnote{Indeed, the locality and the symmetries follow from the locality and 
the symmetries of $S^F$. The classical limit is also preserved as one can see from 
the expansion of $D_{n,n+\mu}(U)$ in terms of gauge--covariant paths.} 
Consequently, $\cM^D$ is expected to induce a valid deformation of the gauge 
action $S \,\longrightarrow\, S^{\cM^D} \in \cS^G$. Action $S^{\cM^D}$ will be referred
to as a {\em chirally ordered} version of $S$ according to $D$.
\smallskip

\noindent Few crucial points pertaining to the above definition of chiral ordering need to be 
made clear at this point.
\smallskip

\noindent {\em (1)} While it is not straightforward to prove that the elements of $\cS^F$ 
satisfying (\ref{eq:330}) exist, their existence does not appear to contradict any 
requirements imposed on valid fermionic actions.
\footnote{The non-trivial part in constructing such elements obviously comes from reconciling
the requirement of chiral symmetry (Ginsparg--Wilson relation) with equation
(\ref{eq:330}). See also point {\em (3)}.}
Moreover, later in this paper, we will claim that there exist converging sequences of elements 
in $\cS^F$ (with the limit that could be outside the set) that will satisfy (\ref{eq:330}) to
arbitrary precision. Accepting conditionally the possibility, let us denote the subset 
of such special fermionic actions as $\cS^{F,C}$. Note also that from now on we will frequently 
identify the action $S^F$ with its defining kernel $D$ and so we will interchangeably
write e.g. $S^F \in \cS^F$ and $D \in \cS^F$.  
\smallskip

\noindent {\em (2)} According to Eq.~(\ref{eq:330}) the operation of chiral ordering replaces
the link $U_{n,\mu}$ with the interaction matrix phase $\Ubar_{n,\mu}$ involved in  
``hopping'' of {\em chiral fermion} (quark) from site $n+\mu$ to site $n$. In other words, 
it is a measure of how much is the elementary hopping of chiral fermion affected by the presence 
of non-zero gauge field relative to the dynamics of free fermion. Thus, as far as chiral
fermion is concerned, the physics content of configuration $U$ is very much embodied 
in configuration $\Ubar=\cM^D(U)$. This is our rationale for the expectation that the relevant
physics encoded in gauge action $S$ is very close to physics in $S^{\cM^D}$. We take this 
as a basis for the proposition that moving in the set of actions via chiral ordering can 
potentially solve the problem $(\beta)$ stipulated in the introduction to 
Sec.~\ref{sec:pco_main}.
\footnote{See also points {\em (a)-(c)} in the beginning of this section.} 
\smallskip

\noindent {\em (3)} We emphasize that the condition (\ref{eq:330}) can be easily satisfied by 
lattice fermionic actions that are not chirally symmetric. For example, the Wilson--Dirac 
operator $D_W$ (which is not an element of $\cS^F$) satisfies it with $\Ubar_{n,\mu}=U_{n,\mu}$, 
i.e. with $\cM^{D_W}$ being an identity map. It is thus the {\em chiral symmetry} of the probing
fermion that turns the chiral ordering condition into a highly non-trivial requirement. This is 
natural since we view the chirality of lattice fermion as a necessary requirement for properly 
regularizing the continuum physics. At the technical level, this comes as a result of 
non-ultralocality~\cite{nonultr} of $D \in \cS^F$ in gauge variables. Indeed, as a consequence
of Ginsparg--Wilson relation, the matrix $D_{n,n+\mu}(U)$ is expected to receive contributions
from all possible gauge paths starting at $n$ and ending at $n+\mu$. Condition 
(\ref{eq:330}) demands that these contributions add up to an effective SU(N) phase. This is 
clearly a highly complex requirement, and the associated $\cM^D$ is a highly complex map.
\smallskip

\noindent {\em (4)} Viewed as a requirement for the fermionic action, Eq.~(\ref{eq:330})
represents a condition that is independent from the usual ones imposed on elements 
of $\cS^F$. In particular, it is not implied by standard symmetries.
\footnote{This does not necessarily mean that the condition cannot be phrased in the language
of symmetry. However, if it does, then the associated symmetry is different from the standard 
ones.}  
This is why we sometimes use the term ``principle'' when referring to it. However, our 
present aim in imposing Eq.~(\ref{eq:330}) is not to define a better lattice fermionic theory, 
but to provide a guide for evolution in the set of gauge actions that suits our goal of 
studying the QCD vacuum structure. In that regard, we view the {\em principle of chiral ordering}
as the following generic statement.
\smallskip

\noindent{\bf The Principle of Chiral Ordering:} {\em The configuration--based deformation     
of the gauge action defined by map $\cM$ representing the effective SU(N) interaction phase 
associated with elementary hopping of lattice chiral fermion has a very mild effect on 
lattice physics. This applies to both long and short--distance behavior (in lattice units).}
\smallskip

\noindent Thus, what we mean by the principle of chiral ordering is the broad statement according
to which using the maps $\cM_{n,\mu}$ with tendency to capture the effective interaction phase 
of chiral fermion are particularly suitable for exploring the nature of space--time order
in QCD. We emphasize that the potential usefulness of this observation does not crucially depend 
on the fact that the set $\cS^{F,C} \subset \cS^{F}$ is not empty. Indeed, we view the condition 
(\ref{eq:330}) as a clearest expression for the underlying idea, which we can use as a guide 
for extracting the approximate matrix phase corresponding to {\em arbitrary}
chiral lattice fermion. This will be discussed more fully in Sec.~\ref{sec:aco}.

\subsection{The ``Ordering'' Conjecture}
\label{sec:pco_ordering}

In the previous two subsections we have addressed the points $(\alpha)$, $(\beta)$ put forward
in the opening part of Sec.~\ref{sec:pco_main}. In particular, we have suggested how to move
in the set of actions (via configuration--based deformation) while minimally changing the
lattice physics (via principle of chiral ordering). In these considerations we have already
implicitly assumed, as the name ``chiral ordering'' suggests, that this will also lead
to lowering of the associated Kolmogorov entropy, i.e. that the point $(\gamma)$ is 
automatically satisfied. To prove the relevant rigorous statement within the framework of 
Kolmogorov complexity is a difficult task that we will not attempt here. However, we need 
to make it clear that our considerations assume the validity of the following conjecture.
\smallskip

\noindent {\bf Conjecture CI2.} {\em Let $k$ be an arbitrary but fixed positive integer.
Then $\ken[S^{\cM^D},k,L] \,<\, \ken[S,k,L]$ for sufficiently large $L$.}     
\smallskip

\noindent In other words, we assert that chiral ordering leads to a gauge theory with lower 
Kolmogorov entropy. Note that the only restriction on lattice size $L$ in {\bf CI2} 
is related to applicability of Kolmogorov complexity concepts. Also, we do not restrict 
the validity of {\bf CI2} to chiral orderings defined by $D\in \cS^{F,C}$, i.e. 
we propose that it is valid also for the approximate orderings performed with arbitrary 
$D \in \cS^F$, which we will discuss later in this paper. It should also be emphasized that 
we do not claim that the conjecture is strictly valid at the configuration level, i.e.
it is not necessary true that $\ken(\cM^D(U),k) \,<\, \ken(U,k)$ for arbitrary
$U\in \cU^L$, even though it is reasonable to expect that it is true for most configurations.

The heuristic argument for validity of the ``ordering'' conjecture is tied to the principle
of chiral ordering and is as follows. The underlying idea behind the principle of chiral 
ordering is that the physical content of a configuration mostly resides in the SU(N) interaction
phases associated with elementary hopping of a lattice chiral fermion. At the same time, the 
path integral approach to QCD vacuum implicitly relies on the assumption that the fluctuations 
associated  with ``physics'' exhibit high degree of space--time order (and will lead 
to identification of collective variables C), while the physically irrelevant fluctuations 
are mostly disordered. Thus, by identifying and keeping the physically relevant part of 
the fluctuations, we should increase the order on average, and thus to lower the Kolmogorov 
entropy. Needless to say, it would be very illuminating if the statement analogous to {\bf CI2} 
could be proved rigorously at least in some model situation.

\subsection{Repeated Chiral Ordering and Perfectly Chirally Ordered \\ Configurations}
\label{sec:pco_repeated}

It is interesting (and in the case of approximate chiral ordering very important) to think about 
repeating the chiral ordering procedure iteratively. To start speaking in these terms, let us 
denote the configuration obtained from $U$ in $j$-th step as $U^{(j)}$. More precisely
\begin{equation}
   U^{(0)} \equiv U  \qquad\qquad
   U^{(j)} \equiv (\cM^D)^j(U)\,, \qquad
   j=1,2,\ldots 
   \label{eq:345}
\end{equation}
At the same time, the original gauge action $S \equiv S^{\cM^D}_{(0)}$ evolves into 
$S^{\cM^D}_{(j)}$ after $j$ steps.
The configuration $U^{(1)}(=\Ubar)$ represents the interaction--phase content of $U^{(0)}(=U)$.
Is the interaction--phase content of $U^{(1)}$ equal to itself, i.e. is $U^{(1)}$
equal to $U^{(2)}$? Even though this would seem natural, it is not necessarily so. 
On the other hand, from the point of view of the principle of chiral ordering, 
the configurations stable under map $\cM^D$ are clearly special. In fact, we expect
them to be the configurations that reflect the space--time order present in the vacuum 
in the clearest possible manner. Let us thus define the corresponding set precisely.
\medskip

\noindent {\bf Definition 1.} {\em Let $D(U) \in \cS^F$ be an arbitrary chiral lattice Dirac 
operator defined on a latticized torus of size $L$ (i.e. $U\in \cU^L$). The configuration
$\hU$ satisfying the condition
\begin{equation}
    D_{n,n+\mu}(\hU) \,=\, D^f_\mu \times \hU_{n,\mu}
    \qquad\quad
    \forall\, n,\,\mu 
    \label{eq:350}
\end{equation}
is said to be perfectly chirally ordered. The subset of all such configurations will be denoted
as $\cU^{L,D}$.}

\medskip
\noindent We emphasize that in the above definition we do not restrict ourselves to Dirac operators
that are elements of $\cS^{F,C}$. Indeed, the set of perfectly chirally ordered configurations
is defined for arbitrary chirally symmetric action, and will play an important role in the
procedure of approximate chiral ordering discussed later in this paper. 
 
As is obvious from its definition, the set $\cU^{L,D}$ is not empty for arbitrary $D$, since it 
contains a trivial configuration with $U_{n,\mu}\equiv \identity^c$ 
(``free--field'' configuration; $\identity^c$ is an identity in color space). 
However, the notion of perfectly chirally ordered configuration (and the whole program of
chiral ordering) would not be of much interest if this was the only possibility. Thus, 
the following conjecture is to be viewed as one of the cornerstones for the approach to vacuum
structure proposed in this paper.

\medskip
\noindent {\bf Conjecture C1.} {\em The set $\cU^{L,D}$ contains non--trivial configurations
for arbitrary $D\in \cS^F$ for sufficiently large $L$. More precisely, it contains configurations
with non--vanishing gauge invariant composite fields.}     
\smallskip

\noindent An important feature of the Conjecture C1 is that one can attempt its numerical 
verification for overlap fermions~\cite{Neu98BA}.

Returning now to the repeated exact chiral ordering, let us consider the sequences
$\{\,U^{(j)}, j=0,1,2,\ldots \,\}$ and $\{\,S^{\cM^D}_{(j)}, j=0,1,2,\ldots \,\}$
defined above for $D \in \cS^{F,C}$. According to the principle of chiral ordering 
and Conjecture CI2 we expect that, with increasing $j$, we increasingly reduce the randomness
present in the configuration, while changing lattice physics very mildly. Thus, in some sense,
our approach expresses the view that random fluctuations are mostly responsible for violating 
the condition (\ref{eq:330}), while the relevant fluctuations (``physics'') tend to satisfy it.
For the elements of $\cU^{L,D}$ this process ends immediately -- they are invariant 
configurations under $\cM^D$. For $U \notin \cU^{L,D}$ it is expected that the amount of 
removed noise decreases fast with $j$ and, asymptotically, it is not possible to remove 
any more randomness. In other words, the elements of $\{\,U^{(j)}, j=0,1,2,\ldots \,\}$ 
behave asymptotically as elements of $\cU^{L,D}$. This expectation is expressed in the 
following conjecture.

\medskip
\noindent {\bf Conjecture C2.} {\em Let $U \in \cU^{L}$ and $D \in \cS^{F,C}$. (a) The limit
$\lim_{j\to \infty} U^{(j)}$ exists generically and is an element of $\cU^{L,D}$. (b) There 
exist configurations $U$ for which this limit is non--trivial at sufficiently large $L$.}
\medskip

\noindent Note that by ``generically'' we mean up to some possible small set of measure zero 
which we postulate not to be relevant in the continuum limit. We wish to emphasize the following
points.

\smallskip
\noindent {\em (1)} The non--trivial part of conjecture C2 is the generic existence of the limit
and the fact that there exist configurations for which the limit is non-trivial. Indeed, if 
$\lim_{j \to \infty} U^{(j)} = \hU$ exists, then it is easy to see that it must be an element 
of $\cU^{L,D}$ since
\begin{displaymath}
   \lim_{j\to \infty} ||\, U^{(j)}_{n,\mu} - \cM^D_{n,\mu}(U^{(j)}) \,||   \,=\,
   \lim_{j\to \infty} ||\, U^{(j)}_{n,\mu} - U^{(j+1)}_{n,\mu} \,||  \,=\,0
\end{displaymath}
and so $\cM^D_{n,\mu}(\hU) = \hU_{n,\mu}$. In the above equations $||.||$ denotes an arbitrary
matrix norm. The existence of a non-trivial limit (at the configuration level) for the 
procedure of chiral ordering is one of the main features that, we believe, distinguishes 
this approach from a deformation based on a generic map $\cM$ 
(See point {\em (c)} in Sec.~\ref{sec:pco_sub}.). In a way we are suggesting that 
the deformation that preserves the interaction phases (and hence the physics) avoids converging
to trivial configurations where physics is clearly changed. Instead, the tendency to 
preserve physics is strong enough so that a convergence to non-trivial invariant 
configurations arises.

\medskip
\noindent {\em (2)} As a result of conjecture C2 the sequence  
$\{\,S^{\cM^D}_{(j)}, j=0,1,2,\ldots \,\}$ converges to a non-trivial action 
${\hat S}(S,\cM^D)$. 
By ``non-trivial'' we mean that the corresponding action assigns a non-zero probability 
density to at least some non-trivial configurations. Note that while all individual actions 
$S^{\cM^D}_{(j)}$ are expected to be valid elements of $\cS^{G}$, we do not necessarily claim 
that the limiting action  ${\hat S}$ is also such an element. Indeed, while the classical
limit and the symmetries are surely preserved, the action could in principle become 
asymptotically non-local. However, we emphasize that even if this happened in some cases, 
the relevant message is that ${\hat S}$ can be approximated to arbitrary precision 
by an element of $\cS^{G}$.

\smallskip
\noindent {\em (3)} At the operational level, the special significance of the elements of 
$\cU^{L,D}$ (invariant configurations) comes from the fact that their probability of 
occurrence for actions reached via repeated chiral ordering cannot decrease. In fact, the 
total probability of these configurations and their small deformations will grow and, 
according to the conjecture C2, they will eventually dominate the path integral. Thus, 
to the extent to which the principle of chiral ordering reflects reality (i.e. the extent 
to which the effective interaction phases affecting chiral fermion encapsulate the physics), 
we suggest that it is the elements of $\cU^{L,D}$ that should be examined in order 
to determine the nature of collective degrees of freedom relevant in QCD vacuum.  

\smallskip
\noindent {\em (4)} Note that the actions reached via the procedure of chiral ordering,
while describing pure--glue theory, become functions of lattice Dirac kernel $D$ describing 
the associated chiral fermions. This ``appearance'' of structure associated with fermions 
in gauge part of the action will be a recurring theme in this series of papers and, in fact,
it is very natural in the context of full QCD. Indeed, we will elaborate on the usefulness 
of tying the gauge and the fermion structure later in this paper and extensively in the in 
the second paper of this series.   
   
\smallskip
\noindent {\em (5)} We emphasize that everything discussed here for the case of exact chiral
ordering (Conjecture C2 in particular), also applies for various cases of approximate chiral
ordering that will be defined in Sec.~\ref{sec:aco}. This is important since it gives 
the possibility to test the ideas proposed here, namely by using the overlap Dirac operator 
as a starting point. We also remind the reader that the set of perfectly chirally ordered 
configurations $\cU^{L,D}$ is already defined for all chirally symmetric operators 
$D\in \cS^F$.

\subsection{More on the Rationale for the Principle of Chiral Ordering}
\label{sec:pco_rationale}

It is sometimes emphasized that the physical content of the gauge field is associated 
with its influence on a charged particle. This is usually expressed by saying that 
the wave function of a particle (charged under the gauge group) acquires a non-integrable 
(path-dependent) matrix phase when traveling over a specific path in 
space--time~\cite{Cre83a, Yan75a}.
Let us impose this condition in the lattice theory for the fermion in the fundamental
representation and an ``infinitesimal'' path so that we can avoid the discussion about 
restricting the particle to a particular macroscopic path and/or about the interference 
of contributions from different paths. The elementary displacement of the particle in 
the lattice theory at finite cutoff is not really ``infinitesimal'' of course. However, 
we will impose it as if it was in order to mimic the behavior in the continuum theory. 
The quantum dynamics of the particle is described by its propagator, and for Dirac particle 
of mass $m$ we obtain the condition
\footnote{It is worth noting that the above motivating discussion in the continuum uses the 
Hamiltonian language for the probing particle. This is actually problematic in the lattice
theory with Ginsparg--Wilson fermions due to non-ultralocality~\cite{nonultr}
(see also~\cite{Cre01a}) and thus the discussion cannot be carried out entirely within
the lattice context. Thus, Eq.~(\ref{eq:355}) should be viewed as a discretized version
of the relation in the continuum where the connection of the propagator formalism to 
Hamiltonian is formally made.}
\begin{equation}
    (\, D+m \,)^{-1}_{n,n+\mu}(U) \,=\, (\,D^f+m \,)^{-1}_\mu \times \tU_{n,\mu}
    \qquad\quad
    \forall\, n,\,\mu \,,\, \forall\, U
    \qquad\quad
    \tU_{n,\mu} \in \mbox{\rm SU(N)}
    \label{eq:355}
\end{equation}
where $(\,D^f+m \,)^{-1}$ is a free (translation invariant) propagator (no color indices). 
The above equation is to be compared with Eq.~(\ref{eq:330}). 

Assuming that the fermionic actions satisfying the above condition exist, or using 
its approximate version with arbitrary $D\in \cS^{F}$, the map $\cM: \,U \rightarrow \tU$ 
represents a valid deformation of the gauge action for non-zero positive $m$ (constant in 
lattice units). Indeed, the only concern is locality which is however expected to be present  
for finite lattice mass. One could thus view Eq.~(\ref{eq:355}) as a condition defining
the chiral ordering along the similar lines as we did for standard chiral 
ordering Eq.~(\ref{eq:330}). However, the above line of reasoning would not have much 
meaning if the configuration $\tU$ depended strongly on the mass $m$ chosen in the map.
In other words, it would not be natural if the intended definition of physical content of 
$U$ depended strongly on the mass of the probing particle. The consequences associated with 
demanding the independence of map $\cM$ on $m$ can be analyzed at large $m$. Indeed, 
the spectrum of $D \in \cS^F$ is bounded, and we can choose the value of $m$ larger than 
its upper bound, i.e. $m>2$ for the standard overlap operator. We then have 
\begin{displaymath}
    (\, D+m \,)^{-1} \,=\, \frac{1}{m} 
    \Bigl (\, \identity - \frac{D}{m} + \frac{D^2}{m^2} - \frac{D^3}{m^3} + \ldots \Bigr)    
\end{displaymath} 
which is a converging Taylor-like expansion in $t \equiv 1/m$. The requirement that
the configuration $\tU$ is independent of $m$ then directly leads to the set of 
conditions
\begin{equation}
    D_{n,n+\mu}(U) = D^f_\mu \times \tU_{n,\mu}\,,
    \quad
    (D^2)_{n,n+\mu}(U) = (D^f)^2_\mu \times \tU_{n,\mu}\,,
    \quad \ldots
    \label{eq:360}
\end{equation}
for arbitrary positive powers of $D$. Thus the chiral ordering condition 
Eq.~(\ref{eq:330}) represents a leading term (in $t$) which facilitates the independence
of the ``propagating phase'' on the mass of the (heavy) probing fermion. This further 
supports our proposition that the effective interaction phase associated with elementary 
hopping of chiral lattice fermion tends to preserve the lattice physics.
\medskip

\noindent We finally add two relevant remarks.
\smallskip

\noindent {\em (1)} From our discussion in this section it is tempting to conclude that 
the chiral ordering prescription given by Eq.~(\ref{eq:355}) is more fundamental 
than that of Eq.~(\ref{eq:330}), and that the principle of chiral ordering formulated
in Sec.~\ref{sec:pco_sub} should be taken with respect to this relation. This may
in fact turn out to be true. However, from Eq.~(\ref{eq:360}) we see that the two are
quite closely related. Moreover, the perfectly ordered configurations with respect
to (\ref{eq:355}) form a subset of those that are perfect under (\ref{eq:330}). We will
thus not loose any physical information by proceeding with the original principle 
of chiral ordering which can be numerically investigated with computational power 
currently available. 

\smallskip
\noindent {\em (2)} Classical configurations (see footnote~\ref{foot:smooth}) 
belong to the set of perfectly chirally ordered configurations in the classical 
continuum limit, i.e. on an asymptotically large lattice. This is true for both 
versions based on Eqs.~(\ref{eq:330}) and (\ref{eq:355}).
To make it more precise, consider a classical field $A_\mu(x)$ on a symmetric 
torus of physical size $L_p$. Introducing a ``classical'' lattice spacing via
$L_p = La$ we construct the corresponding configuration $U^c(A_\mu,a)$ on an
$L^4$ lattice via Eq.~(\ref{eq:45}) (see also footnote~\ref{foot:classicalU}). 
We then have in case of Eqs.~(\ref{eq:330}) that
\begin{equation}
    D_{n,n+\mu}\Bigl( U^c(A_\mu,a) \Bigr) - D^f_\mu \times U^c(A_\mu,a)
    \,=\, \cO(a^2)
    \label{eq:365}
\end{equation}
i.e. the linear term in $a$ vanishes. In this classical case it is a simple consequence 
of gauge invariance. We emphasize this fact to reassure ourselves that the principle 
of chiral ordering is compatible with the requirement of classical correspondence. 
It obviously doesn't imply that classical configurations necessarily dominate 
the path integral for actions $S^{\cM^D}_{(j)}$ at fixed large $j$ in the quantum 
continuum limit.

\section{Approximate Chiral Ordering} 
\label{sec:aco}

Our discussion of the principle of chiral ordering in the previous section 
was mostly based on the assumption that there exist valid fermionic actions 
(elements of $\cS^F$) for which the condition (\ref{eq:330}) is strictly satisfied
(elements of $\cS^{F,C}$). This allowed us to proceed with our reasoning in a 
straightforward manner. However, we also made it clear that the basic idea is 
meaningful (and can be pursued) without this assumption. In this section we define 
various forms of approximate chiral ordering procedure (map $\cM^D$) which can be 
realized e.g. with the standard overlap operator~\cite{Neu98BA}. It will turn out 
that our discussion will also shed some light on how should one view the possible 
existence of (non-empty) set $\cS^{F,C}$.

\subsection{Various Forms of Approximate Chiral Ordering}
\label{sec:aco_various}

The proposition that the physical content of the gauge field is largely associated 
with the interaction phase of the probing chiral fermion can be implemented in many
different ways. In an ideal case, these would all lead to equivalent behavior. 
However, it is also possible that some are superior to others in the sense that 
the degree to which the procedure tends to preserve the physical content of 
the configuration/theory can vary. We discuss some of the possibilities bellow.
If not stated explicitly, the maps defined bellow are understood to be elements
of $\Umaps$, i.e. they satisfy conditions {\em (i)--(iii)} of Sec.~\ref{sec:deformation}.

\medskip
\noindent {\bf (I)} The simplest way to proceed is to consider the formula 
(\ref{eq:330}) specifying the chirally ordered link $\Ubar_{n,\mu}$ in the exact case, 
and use it as a basis for the ordering map also in the inexact case. The resulting 
map $\cM^D$ is then a composition of three maps 
$\cM^D = \cM^{(3)} \circ \cM^{(2)} \circ \cM^{(1)}$ specified by the following steps.

\medskip
{\em (a) Projection to color space}, i.e.
\begin{eqnarray}
   \cM^{(1)}: \,\; U_{n,\mu} \,\longrightarrow\, M 
   & \equiv &
   \frac{1}{4} \mbox{\rm tr}^s \Bigl[\,( \, D_{n,n+\mu}(\identity) \,)^{-1}
   \, D_{n,n+\mu}(U) \,\Bigr] \nonumber \\
   & = &
   \frac{1}{4} \mbox{\rm tr}^s \Bigl[\,( \, D^f_\mu \times \identity^c \,)^{-1}
   \, D_{n,n+\mu}(U) \,\Bigr] \nonumber \\
   & \equiv &
   \frac{1}{4} \mbox{\rm tr}^s \Bigl[\,( \, D^f_\mu \,)^{-1}\, D_{n,n+\mu}(U) \,\Bigr] 
   \label{eq:370}
\end{eqnarray}
where tr$^s$ denotes a spinorial trace,
\footnote{If $O$ is a matrix in color--spin space, then tr$^s\,O$ is the matrix in the
color space with matrix elements 
$( \mbox{\rm tr}^s\,O)_{a,b} \equiv \sum_\alpha O_{\alpha a,\alpha b}$, where $\alpha$
denotes the spinorial and $a,b$ the color indices.} 
$\identity^c$ is an identity in the color 
space, and $\identity$ is a configuration of such identity matrices.
\footnote{Note that if $D^f_\mu$ happens to be non-invertible (which can occur e.g. 
in the case of the overlap operator for special values of negative mass and the Wilson 
parameter), the map $\cM^{(1)}$ is still well defined by Eq.~(\ref{eq:330}).}
The last form is just a shorthand for the first two in this definition. 
One can easily see that in the case of exact chiral ordering (\ref{eq:330}) we would 
have $\Ubar_{n,\mu}=M$ and thus $\cM^D=\cM^{(1)}$.

\smallskip
{\em (b) Unitary projection.} This is provided by the polar decomposition which
offers a natural and unique definition of the ``matrix phase'' for a non-singular 
square matrix in direct analogy to the phase of a non-zero complex number. 
According to the polar decomposition theorem, if matrix $M$ is non-singular
then it can be uniquely decomposed as
\begin{equation}
     M \,=\, M^u \, M^h     \;\quad \mbox{\rm such that} \;\quad
     (M^u)^\dagger M^u = \identity^c \,, \quad (M^h)^\dagger = M^h
     \label{eq:375}    
\end{equation}
and $M^h$ is positive semidefinite. This associates a ``matrix phase'' (unitary
matrix $M^u$) with matrix $M$. 
\footnote{Note that the decomposition also holds for singular matrices but in
that case it is not unique. Also, the matrix phase $M^u$ defined by 
(\ref{eq:375}) should be called the {\em left} matrix phase since one can also 
define the {\em right} phase by switching the order of matrices. Left and right 
phases are equal for normal matrices $M$.}
Thus the map $\cM^{(2)}$ is defined by
\footnote{Projection to the gauge group based on the polar decomposition 
(steps {\em (b)} and {\em (c)} here) was used by several authors in different 
contexts. Reference~\cite{Lia92A} is the earliest we are aware of.}
\begin{equation}
   \cM^{(2)}: \; M \longrightarrow M^u \equiv M \frac{1}{\sqrt{M^\dagger M}}
   \label{eq:380}
\end{equation}
where $\sqrt{\;}$ denotes a positive semidefinite branch of the square--root. 
\footnote{We implicitly assume here that the cases when $M$ is strictly singular 
will essentially never be encountered in practice. One can also take more strict 
attitude and use the polar decomposition with an additional condition to fix $M^u$ 
uniquely also in singular cases. We will not elaborate on this here.} 

\smallskip
{\em (c) SU(N) projection.} Unitary matrix $M^u$ can be mapped on the element of 
SU(N) via rescaling by the N--th root of its determinant. Selecting the N--th
root that leads to minimal modification of $M_u$ (in arbitrary matrix norm)
specifies a unique prescription, namely
\begin{equation}
   \cM^{(3)}: \; M^u \longrightarrow \Ubar_{n,\mu} \equiv 
                 e^{-i\varphi/N} \,M^u  \qquad\;
                 \det M^u \equiv e^{i\varphi}  \qquad\;
                 \varphi \in (-\pi,\pi]
   \label{eq:385}
\end{equation}  
The full map $\cM^D = \cM^{(3)} \circ \cM^{(2)} \circ \cM^{(1)}$ can be expressed 
in a single explicit formula if so desired, and is an element of $\Umaps$.

\medskip
\noindent {\bf (II)} Using the same logic as in method {\bf (I)}, we can choose
to perform the approximate chiral ordering via more ``continuum--like'' formula
than that of Eq.~(\ref{eq:370}). Indeed, consider a Clifford decomposition
of $D^f_\mu$, namely
\begin{equation}
    D_\mu^f \,=\, \sum_{a=1}^{16} \, B_\mu^a \, \Gamma^a \,, \qquad
    a=1,2,\ldots 16
    \label{eq:390}
\end{equation}
where $B_\mu^a$ are complex numbers and $\Gamma^a$ are the elements of an 
orthogonal Clifford basis normalized such that 
tr$\,\Gamma^a \Gamma^b = 4\delta_{a,b}$. Let us denote by $B_{\mu\nu}$ 
the Clifford component of $D_\mu^f$ corresponding to basis element 
$\gamma_\nu$, i.e.
\begin{equation}
    B_{\mu\nu} \,\equiv\, B_\mu^a \quad \mbox{\rm such that} \quad 
    \Gamma^a \,=\,\gamma_\nu
    \label{eq:395}
\end{equation}
We then define the projection to color space (map $\cM^{(1)}$) via
\begin{equation}
   \cM^{(1)}: \,\; U_{n,\mu} \,\longrightarrow\, M 
   \,\equiv\, 
   \frac{1}{4B_{\mu\mu}} \,\mbox{\rm tr}^s 
       \Bigl[\,( \, \gamma_\mu \times \identity^c \,)\, D_{n,n+\mu}(U) \,\Bigr] 
   \,\equiv\, 
   \frac{1}{4B_{\mu\mu}}\, \mbox{\rm tr}^s \gamma_\mu \, D_{n,n+\mu}(U)
   \label{eq:400}
\end{equation}
where the second form in the above equation is an abbreviation for the first form.
One can easily check that if $D \in \cS^{F,C}$ (exact chiral ordering) we would have 
$\Ubar_{n,\mu}=M$ and thus $\cM^D=\cM^{(1)}$. In the approximate case we define
$\cM^D = \cM^{(3)} \circ \cM^{(2)} \circ \cM^{(1)}$, where $\cM^{(2)}$ and 
$\cM^{(3)}$ are specified in method {\bf (I)} above.

\medskip
\noindent {\bf (III)} The non--trivial aspect of the chiral ordering condition 
(\ref{eq:330}) resides in the fact that it forces a factorization 
of local spin--color matrix in the respective degrees of freedom. This gives
meaning to the statement that the fermion acquires a matrix phase when it
hops from site $n$ to site $n+\mu$. In other words, it ensures the fact that 
all fermionic degrees of freedom at site $n$ get rotated by the gauge field in 
an identical manner. It is thus sensible to define the approximate chiral
ordering prescription based on the ``degree of factorizability''. For example,
we can construct the map $\cM$ in the same way as in case {\bf I} but replace 
$\cM^{(1)}$ with 
\begin{equation}
   \cM^{(1)}: \,\; U_{n,\mu} \,\longrightarrow\, M \qquad
   \mbox{\rm such that} \qquad  
   \min_{\tM \in \C^{N\times N}} \,F(\tM) \,=\, F(M)
   \label{eq:405}
\end{equation}
where $\C^{N\times N}$ is the set of $N\times N$ complex matrices and 
\begin{equation}
   F(\tM) \,\equiv\, ||\, D_{n,n+\mu}(U) - D^f_\mu \times \tM \, ||
   \label{eq:410}  
\end{equation}
with $||.||$ denoting a matrix norm. One can easily check that for non--zero
$D^f_\mu$ the solution of the above minimization problem in Frobenius norm 
(i.e. $||A||^2 \equiv \mbox{\rm tr} A^{\dagger}A$ for matrix $A$), is unique
and given by
\begin{equation}
   M \,=\, 
   \frac{1}{\mbox{\rm tr} [(D^f_\mu)^\dagger D^f_\mu]} \,   
   \mbox{\rm tr}^s \Bigl[\,(D^f_\mu)^\dagger \, D_{n,n+\mu}(U) \,\Bigr] 
   \label{eq:415}
\end{equation}  
Comparing the above expression with Eq.~(\ref{eq:370}) we see that the two 
projections to color space are identical if
\begin{equation}
   (D^f_\mu)^\dagger D^f_\mu  \,=\, 
   \frac{\mbox{\rm tr} [(D^f_\mu)^\dagger D^f_\mu]}{4} \, \identity^s 
   \label{eq:420} 
\end{equation}
i.e. when $D^f_\mu$ is a unitary matrix rescaled by a real number. This is
true e.g. for the overlap Dirac operator but not necessarily in general.
Note that if $D\in \cS^{F,C}$ (case of exact chiral ordering) then it has 
to satisfy both Eqs.~(\ref{eq:370},\ref{eq:415}) and, consequently, 
the unitarity condition (\ref{eq:420}) also has to be satisfied.

Let us also point out that when replacing $D_\mu^f$ in the minimization 
problem (\ref{eq:405},\ref{eq:410}) with arbitrary spinorial matrix $S$,
then the solution is given by (\ref{eq:415}) with $D_\mu^f$ replaced
by $S$. Consequently, one can easily check that the color--space projection 
$\cM^{(1)}$ of case {\bf II} is equivalent to minimization of $F(\tM)$ for 
$S=B_{\mu\mu} \gamma_\mu$ ($\mu$ not summed).

If one wishes to focus the definition of color projection on the spin-color 
factorizability property, then one can obtain a prescription that is different 
(but approximately equal in generic case) from those in cases {\bf I} and 
{\bf II} by relaxing the condition that the spinorial factor is fixed. 
More precisely, one can associate the projection to the color space with 
the following minimization problem.
\begin{equation}
   \cM^{(1)}: \,\; U_{n,\mu} \,\longrightarrow\, M \qquad
   \mbox{\rm such that} \qquad  
   \min_{\tM \in \C^{N\times N},\tS \in \C^{4\times 4}} \,F_1(\tM,\tS) 
   \,=\, F_1(M,S)
   \label{eq:425}
\end{equation}
where 
\begin{equation}
   F_1(\tM,\tS) \,\equiv\, ||\, D_{n,n+\mu}(U) - \tS \times \tM \, ||
   \label{eq:430}  
\end{equation}
Note that the above problem is invariant under simultaneous
rescaling $S\rightarrow \alpha S$, $M\rightarrow \alpha^{-1} M$, and we 
fix the resulting degeneracy of solutions by requiring that the 
$\gamma_\mu$-component in Clifford expansion of $S$ coincides with that 
of $D^f_\mu$, i.e.
\begin{equation}
   \mbox{\rm tr} \, \gamma_\mu S \,=\, 4 B_{\mu\mu}
   \label{eq:435}
\end{equation} 
At the numerical level the solution can be obtained e.g. iteratively by 
alternating the solutions of type (\ref{eq:415}) with spinorial/color
matrix factors fixed. The resulting map $\cM^{(1)}$ is then composed
with $\cM^{(2)}$ and $\cM^{(3)}$ of {\bf I} to obtain the full map $\cM^D$.

\medskip
\noindent {\bf (IV)} In the remaining three cases we list prescriptions that 
are based on the spin--color factorizability property of exact chiral ordering,
but avoid steps {\em (b), (c)} (i.e. unitarity projection and SU(N) projection).
In other words, they are based on a direct minimization in SU(N) (rather
than $\C^{N\times N}$). As a result, they will not be expressed in the explicit
form, but rather defined only as solutions to the corresponding minimization
problems. The uniqueness is hard to demonstrate in these cases, and will 
be assumed. The actual numerical work will reveal if this is reasonable or
if it will be necessary to impose additional conditions to make them unique.  
We thus define the map $\cM^D$ for the current case via the solution to the
following minimization problem
\begin{equation}
   \cM^D: \,\; U_{n,\mu} \,\longrightarrow\, \Ubar_{n,\mu} \qquad
   \mbox{\rm such that} \qquad  
   \min_{V \in SU(N)} \,F(V) \,=\, F(\Ubar_{n,\mu})
   \label{eq:440}
\end{equation}
where  
\begin{equation}
   F(V) \,\equiv\, ||\, D_{n,n+\mu}(U) - D^f_\mu \times V \, ||
   \label{eq:445}  
\end{equation}
The above map represents perhaps the most natural generalization of exact chiral
ordering prescription (\ref{eq:330}) to arbitrary $D\in \cS^F$.

\medskip
\noindent {\bf (V)} The prescription that formally looks more ``continuum--like'' 
and represents the analog of case {\bf II} is given by map (\ref{eq:440})
with $F(V)$ given by
\begin{equation}
   F(V) \,\equiv\, ||\, D_{n,n+\mu}(U) - B_{\mu\mu} \gamma_\mu \times V \, ||
   \label{eq:450}  
\end{equation}
 
\medskip
\noindent {\bf (VI)} Finally, the analog of case {\bf III} 
(Eqs.~(\ref{eq:425}--\ref{eq:435})) is specified by the minimization problem
\begin{equation}
   \cM^D: \,\; U_{n,\mu} \,\longrightarrow\, \Ubar_{n,\mu} \qquad
   \mbox{\rm such that} \qquad  
   \min_{V \in SU(N),\tS \in \C^{4\times 4}} \,F_1(V,\tS) \,=\, F_1(\Ubar_{n,\mu},S)
   \label{eq:455}
\end{equation}
where
\begin{equation}
   F_1(V,\tS) \,\equiv\, ||\, D_{n,n+\mu}(U) - \tS \times V \, ||
   \label{eq:460}  
\end{equation}
Note that since $V$ is constrained to be unitary, it is not necessary to impose 
the analog of condition (\ref{eq:435}) in this case for SU(3) group.

\subsection{Repeated Approximate Chiral Ordering}
\label{sec:aco_repeat}

Various versions of approximate chiral ordering defined in the previous section represent
some of the natural choices but certainly do not exhaust all possibilities. In fact,
given arbitrary $D\in \cS^F$, we have not explicitly defined the set of acceptable maps 
$\cM^D$ that we would consider approximate chiral orderings. Let us thus put forward the 
following definition.

\medskip
\noindent {\bf Definition 2.} {\em Let $D(U) \in \cS^F$ be an arbitrary chiral lattice Dirac 
operator. We call the map $\cM^D$ an approximate chiral ordering defined by $D$ if
(a) $\cM^D \in \Umaps$; (b) the elements of $\cU^{L,D}$ (perfectly ordered configurations) 
are its fixed points, namely
   \begin{equation}
      \cM^D(\hU) \,=\, \hU \;,  \qquad\quad \hU \in \cU^{L,D} 
      \label{eq:465}
   \end{equation}
(c) it approximately satisfies the chiral ordering relation for generic configurations, 
i.e.
   \begin{equation}
       \epsilon_{n,\mu} \,\equiv\, 
       \frac{\;|| D_{n,n+\mu}(U) \,-\, D^f_\mu \times \cM^D_{n,\mu}(U) ||\;}
            {|| D_{n,n+\mu}(U) ||} \ll 1 
       \qquad\quad
      \forall\, n,\,\mu 
      \label{eq:470}
   \end{equation}
}

\noindent As usual, by ``generic'' we mean all up to a possible set of measure zero which we 
assume not to be relevant.
 
The reason for introducing the approximate chiral ordering is that basically all the 
features discussed for the exact case are expected to carry over generically to 
the approximate case. In other words, the principle of chiral ordering 
(see Sec.~\ref{sec:pco_sub}) is viewed as a robust statement about the nature of 
the gauge field. In that regard, we wish to explicitly mention the following points.
\smallskip

\noindent {\em (1)} As emphasized in Sec.~\ref{sec:pco_ordering}, we assume that there 
exist approximate chiral orderings for which the ordering conjecture {\bf CI2} is valid. 
\smallskip

\noindent {\em (2)} Fixing arbitrary $D \in \cS^F$ and a particular approximate chiral ordering 
$\cM^D$, we can consider the sequences $\{\,U^{(j)},\, j=0,1,2,\ldots \,\}$ and 
$\{\,S^{\cM^D}_{(j)},\, j=0,1,2,\ldots \,\}$ defined by repeated application of $\cM^D$ as in 
the exact case. As $j$ increases, it is expected that the configurations (and the ensembles
representing the theory) will become less affected by randomness with lattice physics changing 
very mildly. Very important expectation is that the evolved configurations will 
asymptotically approach the elements of $\cU^{L,D}$ and the ensembles will approach a non-trivial
lattice theory. To express this, we modify the conjecture {\bf C2} of exact chiral ordering
as follows.

\medskip
\noindent {\bf Conjecture C2a.} {\em Let $U \in \cU^{L}$ and $D \in \cS^F$. There exist
approximate chiral orderings $\cM^D$ such that (a) $\lim_{j\to \infty} U^{(j)}$ exists
generically and is an element of $\cU^{L,D}$; (b) there exist configurations $U$ for which 
this limit is non--trivial at sufficiently large $L$.}
\medskip

Needless to say, performing numerical tests of the principle of chiral ordering 
(as well as conjectures {\bf C1} and {\bf C2a}) in the context of overlap Dirac 
operator will be a crucial step towards making the construction proposed here viable.

\subsection{Dual View of the Chiral Ordering Evolution}
\label{sec:aco_dual}

Let us recall that the evolution of arbitrary configuration $U$ under chiral ordering 
(exact or approximate) corresponding to $D\in \cS^F$ is characterized by the sequence 
$\{\,U^{(j)},\, j=0,1,2,\ldots \,\}$, which in turn induces the evolution of gauge 
theory $S$ in the set of actions, namely  
$\{\,S^{\cM^D}_{(j)},\, j=0,1,2,\ldots \,\}$ where the actions are defined via 
the associated configuration--based deformation. However, it is also sometimes useful 
to think of chiral ordering as an evolution of the underlying chiral fermionic action or 
equivalently a Dirac operator. Let us thus consider a sequence of operators 
$\{\, D^{(j)}, j=0,1,2,\ldots \,\}$ defined via
\begin{equation}
      D^{(j)}(U) \,\equiv\,  D(U^{(j)}) \,=\, 
      D\Bigl( (\cM^D)^j (U) \Bigr)  \qquad
      j=1,2, \ldots  \qquad\;
      D^{(0)}(U) \,\equiv\, D(U)  
      \label{eq:475}      
\end{equation}
Since $\cM^D$ is an element of $\Umaps$, it is easy to check that $D^{(j)} \in \cS^F$
for arbitrary $j$. Moreover, the free Dirac operators associated with $D^{(j)}$ coincide
with that of $D$, i.e.
\begin{equation}
      D^{(j)}(\identity) \,\equiv\, D(\identity)  \qquad\;
      j=1,2, \ldots  
      \label{eq:480}      
\end{equation}
The basic property of $D^{(j)}$ built in by construction is that
\begin{equation}
   U^{(j)} \,=\, (\cM^D)^j (U) \,=\, \cM^{D^{(j-1)}}(U)
   \label{eq:485}
\end{equation}
In other words, many iterations of chiral ordering using $D$ can be thought of 
as a single iteration with another element of $\cS^F$. Assuming the validity 
of conjecture {\bf C2a} ({\bf C2}) we can conclude that for generic configurations
$U$ there exist limits
\begin{equation}
    \lim_{j\to \infty} U^{(j)} \,\equiv\, \hU \qquad\quad
    \lim_{j\to \infty} D^{(j)}(U) \,\equiv\, \hD(U)
    \label{eq:490}
\end{equation}
satisfying the relations
\begin{equation}
    \hD_{n,\mu}(U) \,=\, D_{n,\mu}(\hU) \,=\, \hD_{n,\mu}(\hU) \,=\, 
    D_\mu^f \times \hU_{n,\mu}
    \label{eq:495}    
\end{equation}
We can then conclude the following. {\em (i)} The operator $\hD$ satisfies the exact 
chiral ordering property (\ref{eq:330}) regardless of whether the initial operator $D$ 
satisfies it or not. In fact, it is an operator which yields a fixed--point configuration 
(perfectly ordered configuration; element of $\cU^{L,D}$) in one step. {\em (ii)} Operators
$D$ and $\hD$ share some of their perfectly ordered configurations. 

The above considerations suggest that the validity of conjecture {\bf C2a} is sufficient 
to conclude that the set $\cS^{F,C}\subset \cS^F$ is non-empty. Indeed, every $D^{(j)}$ 
of chirally ordered sequence is an element of $\cS^F$, and one thus naturally 
expects that the limit $\hD$ belongs there as well, in which case $\hD\in \cS^{F,C}$.  
However, we caution that the limiting action could become asymptotically non-local
in some cases. Nevertheless, even if that happens, the important result of these 
considerations is that the elements of $\cS^F$ can satisfy the chiral ordering condition
(\ref{eq:330}) to arbitrary precision.

\section{Partial Chiral Ordering}
\label{sec:pco}

In our treatment of the principle of chiral ordering we required that the ordering 
condition (\ref{eq:330}) be satisfied for all lattice links. In other words, 
well-defined interaction phases were assumed to exist for all possible elementary
space--time ``hops'' of chiral fermion, and for all (generic) configurations.
However, one could also take a view (and some results of~\cite{Hor03A} could be
interpreted that way) that there is a sizable fraction of elementary space--time
filaments (links) which are not used by the propagating fermion. For such links
the interaction phase would not have to be well-defined. Turning the argument 
around, one could also say that the regions of space--time where condition
(\ref{eq:330}) is satisfied to high accuracy are highly susceptible to fermion
propagation, while the ones where (\ref{eq:330}) is satisfied poorly form
sort of an ``excluded'' space--time. 

To incorporate the above possibility into our reasoning (and to be able to put 
it eventually to the test), requires few changes in the formalism and we will 
describe them in this section. However, our overall line of logic will not change
at all. In fact, we will essentially trace the same steps as we did in case 
of full chiral ordering, and will mostly concentrate on what needs modification. 
Note also that we will not use different notation for analogous concepts in 
the two cases which will always be distinguished by the context (either ``full'' 
or ``partial'' chiral ordering) in the future.    
\medskip

\noindent{\em (i)} To build the concept of chiral ordering we started with 
the definition of the special class of fermionic actions $\cS^{F,C}$, which
in the case of full chiral ordering was defined as the subset of elements 
from $\cS^F$ satisfying condition (\ref{eq:330}). To modify this for the partial
case let $\linkall$ be the set of all link coordinates (i.e. all $(n,\mu)$ 
of a given lattice) and $\linkset \subset \linkall$ its arbitrary subset.
We say that $D\in S^{F,C}$ if $D\in S^F$ and if for generic configuration $U$
there exists a non-empty $\linkset (U)$ such that 
\begin{equation}
    D_{n,n+\mu}(U) \,=\, D^f_\mu \times \Ubar_{n,\mu}
    \qquad\;
    \forall\, (n,\,\mu) \in \linkset
    \qquad\;
    \Ubar_{n,\mu} \in \mbox{\rm SU(N)}
    \label{eq:500}
\end{equation}
The associated partial chiral ordering map $\cM^D$ is defined via
\begin{equation}
   U \,\longrightarrow\, \cM^D(U) \qquad\qquad 
   \cM^D_{n,\mu}(U) \,\equiv\, 
   \cases{\Ubar_{n,\mu} \,,\;&$(n,\mu) \in \linkset$  \cr
           U_{n,\mu}    \,,\;&$(n,\mu) \notin \linkset$ \cr}
   \label{eq:505}
\end{equation}
Thus, the gauge links for which the corresponding interaction phase is not 
well defined will not get changed under the transformation. One can easily see
that $\cM^D\in \Umaps$ thus inducing a valid deformation of the gauge
action. It is expected that the chiral ordering conjecture {\bf CI2} applies
and the Kolmogorov entropy of $S^{\cM^D}$ will not grow relative to $S$.
\medskip

\noindent{\em (ii)} The analysis of the repeated exact partial chiral ordering 
has the same ingredients as in the exact case. For arbitrary $D \in \cS^F$ we
can define the set of perfectly ordered configurations $\cU^{L,D}$ in complete
analogy with Definition 1 with Eq.~(\ref{eq:350}) required to be valid
only on some non-empty set $\linkset (U)$ of links. Direct analogues of 
conjectures {\bf C1} and {\bf C2} are expected to hold.
\medskip

\noindent{\em (iii)} The implementation of approximate partial chiral ordering
requires a non-trivial modification relative to the full case. Indeed, it is not
possible just to take e.g. the cases ({\bf I})--({\bf VI}) of 
Sec.~\ref{sec:aco_various} and restrict them to a subset of links. The point
is that for generic $D\in \cS^F$ the equation (\ref{eq:500}) will not be satisfied
exactly for any links, and thus the selection of $\linkset$ is part of 
the constructed map. The set $\linkset$ is configuration--dependent and the rule
specifying it must be chosen such that the resulting $\cM^D$ is local. Below 
we discuss few examples that conform to our requirements.
\medskip

\noindent {\bf (Ip)} Consider the map $\cM^D$ for case {\bf (IV)} of full 
approximate chiral ordering with arbitrary $D\in \cS^F$. For given $U\in \cU^L$ 
let $\epsilon_{n,\mu}(U)$ be the relative residues of Eq.~(\ref{eq:470}), and let 
$\delta > 0$. Then we define the map $\cM^{D}_\delta$ by
\begin{equation}
   U \,\longrightarrow\, \cM^{D}_\delta(U) \qquad\qquad 
   (\cM^{D}_\delta)_{n,\mu}(U) \,\equiv\, 
   \cases{\cM^D_{n,\mu}(U) \,,\;&$\epsilon_{n,\mu}(U) <\delta$  \cr
           U_{n,\mu}       \,,\;&$\epsilon_{n,\mu}(U) \ge \delta$ \cr}
   \label{eq:508}
\end{equation}
We emphasize that $\cM^{D}_\delta$ is local and thus the element of $\Umaps$.
Depending on the value of $\delta$ it will chirally order all the links
or some subset $\linkset$. 
\footnote{Note that $\delta$ is not to be viewed as a free parameter that one
can change at will when changing the configuration or the lattice spacing.
The map $\cM^{D}_\delta$ has to be fixed when investigating the evolution 
in the set of actions.}
We can obviously base the construction of  $\cM^{D}_\delta$ on other full
approximate chiral orderings in an analogous manner.
\medskip

\noindent {\bf (IIp)} Qualitatively different kind of prescription can be 
obtained as follows. Consider the matrix element $D_{n,n+\mu}(U)$ and the 
configuration $\tU(V)$ obtained from $U$ by replacing $U_{n,\mu}$ with 
$V\in \mbox{\rm SU(N)}$ while keeping all the other link variables the same i.e.
\begin{equation}
      \tU_{m,\nu}(V) \,\equiv\, \cases{V         \,,\;&$(m,\nu)=(n,\mu)$ \cr
                                       U_{m,\nu} \,,\;&$(m,\nu)\neq (n,\mu)$\cr}
      \label{eq:510}
\end{equation}
Furthermore, let ${\frak V}_{n,\mu}(U)$ denote the set of all 
$V\in \mbox{\rm SU(N)}$ such that $D_{n,n+\mu}(\tU(V))$ leads to an exact 
spin--color factorization of chiral ordering type, i.e.   
\begin{equation}
   {\frak V}_{n,\mu}(U) \,\equiv\, \{\, V\in \mbox{\rm SU(N)} \,:\,
    D_{n,n+\mu}\Bigl(\tU(V)\Bigr) \,=\, D^f_\mu \times \Ubar_{n,\mu}(V)\,,\;
    \Ubar_{n,\mu}(V) \in \mbox{\rm SU(N)} \,\}
    \label{eq:515}  
\end{equation}  
and let $V_0$ be the element of ${\frak V}_{n,\mu}$ closest to $U_{n,\mu}$, i.e.
such that
\begin{equation}
   \min_{{\frak V}_{n,\mu}} ||\, U_{n,\mu} - V \,|| \,=\, ||\, U_{n,\mu} - V_0 \,|| 
   \label{eq:520}
\end{equation}
In the last equation we have implicitly assumed that if ${\frak V}_{n,\mu}$ 
is non-empty then $V_0$ is unique generically. We then define 
\begin{equation}
  \cM^D_{n,\mu}(U) \,\equiv\, \cases{\Ubar_{n,\mu}(V_0) \,,\;& 
                                    if $\,{\frak V}_{n,\mu} \neq \emptyset$ \cr
                                    U_{n,\mu} \,,\;&
                                    if $\,{\frak V}_{n,\mu} = \emptyset$ \cr} 
  \label{eq:525} 
\end{equation}
The above equation defines the approximate chiral ordering $\cM^D$ (element of
$\Umaps$) with the property that it only modifies links for which the corresponding 
spin--color factorization is achievable in a prescribed manner. It is thus a partial 
chiral ordering prescription.
\medskip

\noindent {\bf (IIIp)} The disadvantage of the partial chiral ordering defined above
is that it is not clear at this point how to carry it out in practice. In other words,
there are no straightforward numerical ways to determine $V_0$, and they will have 
to be developed. We thus also advocate another possibility. Recall that one of the main
purposes of approximate chiral ordering is to determine the structure of perfectly 
chirally ordered configurations obtained by repeated application of map $\cM^D$. 
A natural way to proceed in this regard is to enforce ``local perfectness'' in 
the map $\cM^D$. By this we mean the following. Using the same notation as in 
case {\bf (IIp)} let ${\frak V}^p_{n,\mu}(U)$ denotes the set of all possible
``perfect links'' $\hU_{n,\mu}$ associated with fixed $(n,\mu)$, i.e.
\begin{equation}
   {\frak V}^p_{n,\mu}(U) \,\equiv\, \{\,\hU_{n,\mu} \in \mbox{\rm SU(N)} \,:\,
    D_{n,n+\mu}\Bigl(\tU(\hU_{n,\mu})\Bigr) 
    \,=\, D^f_\mu \times \hU_{n,\mu} \,\}
    \label{eq:530}  
\end{equation}  
We now select the element $\hU^0_{n,\mu}$ of ${\frak V}^p_{n,\mu}(U)$ (if any) 
closest to $U_{n,\mu}$, i.e. such that
\begin{equation}
   \min_{{\frak V}^p_{n,\mu}} ||\, U_{n,\mu} - \hU_{n,\mu} \,|| \,=\, 
                              ||\, U_{n,\mu} - \hU^0_{n,\mu} \,|| 
   \label{eq:535}
\end{equation}
and define the map $\cM^D$ via
\begin{equation}
  \cM^D_{n,\mu}(U) \,\equiv\, \cases{\hU_{n,\mu}^0 \,,\;& 
                                    if $\,{\frak V}^p_{n,\mu} \neq \emptyset$ \cr
                                    U_{n,\mu} \,,\;&
                                    if $\,{\frak V}^p_{n,\mu} = \emptyset$ \cr} 
  \label{eq:540} 
\end{equation}
Note that the important difference between the sets ${\frak V}_{n,\mu}(U)$   
and ${\frak V}^p_{n,\mu}(U)$ is that in the former case we have twice as many
variables to solve for defining condition than in the latter case 
(16 versus 8 in SU(3) case), and thus the set 
${\frak V}^p_{n,\mu}(U) \subset {\frak V}_{n,\mu}(U)$ is naively expected to be
quite small generically. In fact, one expects at most a finite set of 
solutions in the perfect case. However, the comparison to case {\bf (IIp)} 
can only be made by studying these solutions on specific backgrounds typical 
for the starting theory $S$. Practically relevant feature of the perfect case is 
that one can attempt to find the elements of ${\frak V}^p_{n,\mu}(U)$ in 
a straightforward manner iteratively.
\medskip

\noindent{\em (iv)} The content of conjecture {\bf C2a} carries over naturally 
to the partial chiral ordering case and the analogous statement is expected to be 
valid. One can also take a dual view and think in terms of evolving Dirac operator 
as in the full case. 
\medskip

Let us finally remark that the usefulness of the idea of partial chiral ordering
can only be decided by studying the above options (full and partial) explicitly.
Indeed, it is an open question whether QCD dynamics leads to the phenomenon of 
excluded space--time, i.e. whether the natural splitting of space--time into 
regions of well-defined and ill-defined chiral phase actually occurs in typical
configurations dominating the QCD path integral.

\section{Chiral Ordering and the Renormalization Group}
\label{sec:rg}

As discussed extensively in the previous parts of this paper, the search for the
{\em fundamental} QCD vacuum structure in the path integral formalism naturally 
leads to exploring the behavior of Kolmogorov entropy in the set of lattice QCD 
actions. Moreover, we proposed that a suitable tool to guide us in this search 
might be the iterated chiral ordering transformations defining certain trajectories 
in this set. Thus, at least superficially, there exist some similarities here with 
the Renormalization Group (RG) concepts and techniques. The aim of this section 
is to suggest, and make it plausible, that the connection might be more than 
superficial. 

In case of RG the important structure in the set of actions is represented by
renormalized trajectories associated with different RG transformations.
The lattice actions on renormalized trajectory are expected to yield exact 
predictions (no cutoff effects) for all spectral quantities (or any physical 
quantities if the operators are correspondingly improved)~\cite{Has94A}.
While not clear a priori at all, it is tempting to think that renormalized 
trajectories will also represent a relevant feature from the point of view of 
Kolmogorov entropy. 
Such expectation is tied to the fact that, in certain situations, it appears valid 
to intuitively associate the physical (continuum) content of the lattice theory 
with order in typical configurations (clearly defined collective variable $C$), 
while to view scaling violations as being caused by an excess of ``disorder'' 
superimposed on top of it. Indeed, consider the theory $S^G(a(\beta))\in \cS^G$ 
and follow the trajectory of actions as $a$ is being changed towards the continuum 
limit. As the lattice spacing is decreased, the scaling violations become 
smaller while the Kolmogorov entropy is expected to decrease since the associated
distribution of configurations 
is becoming more sharply peaked. In this sense the transition from lattice spacing
$a_1$ to lattice spacing $a_2  < a_1$ can be intuitively viewed as a 
``noise reduction'' in typical configurations of $S^G(a_1)$. Looking now in the 
orthogonal direction in the set of actions (lattice spacing fixed), the actions
on renormalized trajectories represent the solutions to the problem of
maximizing the information content of the lattice theory relative to the continuum. 
In other words, among the actions at the particular fixed cutoff, they carry the most 
information about the continuum limit. As such, the underlying order shaping
this physics is expected to be closely related to the one associated with the physics
of the continuum limit. Thus, in analogy to approaching the continuum limit, it is
natural to expect that departures from this specific order (moving away from 
renormalized trajectory) are caused by extra randomness present in the dynamics
defined by the corresponding actions. In other words, one tends to conclude that
actions on renormalized trajectories correspond to local minima of 
Kolmogorov entropy at fixed lattice spacing.

While the above considerations appear quite plausible, they are strictly heuristic.
At the same time, the conclusion on renormalized trajectories is rather non-trivial.
\footnote{We point out that in case of approaching the continuum limit in given  
lattice theory, one can support the conclusion of minimal Kolmogorov entropy with 
the fact that all correlation lengths increase sufficiently close to the limit. This 
is not the case when approaching the action on renormalized trajectory in which case 
the correlation lengths are merely adjusted to reproduce the continuum masses. This 
is one of the reasons why the conclusion of minimal Kolmogorov entropy is non-trivial 
in this case.} 
In this section we extend the concepts needed to introduce the principle of 
chiral ordering in a manner that brings the relation to the renormalization 
group ideas more to the forefront. We will then argue that, in this extended
framework, the above conclusion on renormalized trajectories suggests itself
naturally. In fact, we will propose that the iterated chiral ordering can drive
gauge actions close to renormalized trajectories of particular RG transformations.

\subsection{RG Transformations Based on Chiral Ordering}

Let $D \in \cS^F$ and $\cM^D$ be an arbitrary full approximate chiral ordering map. 
\footnote{Note that if it happens that $D\in \cS^{F,C}$ then all approximate chiral 
ordering prescriptions define the same map, namely the associated exact chiral 
ordering.} 
The action of $\cM^D$ on $U_{n,\mu}$ is determined by $D_{n,n+\mu}(U)$ and it 
represents the interaction matrix phase associated with ``elementary hopping'' of 
chiral fermion from $n+\mu$ to $n$. In the same manner, we can think of an 
{\em effective} interaction matrix phase associated with fermion moving from $n+2\mu$ 
to $n$. Such map will be based on $D_{n,n+2\mu}$ but otherwise be identical 
(in form) to $\cM^D$. To illustrate this explicitly, let us consider the approximate 
chiral ordering $\cM^D$ of case {\bf (IV)} (see Eqs.~(\ref{eq:440},\ref{eq:445})). 
The associated map $\cM^{D,2}$ will be given by
\begin{equation}
   \cM^{D,2}: \,\; U_{n,\mu} \,\longrightarrow\, \cM^{D,2}_{n,\mu}(U)
   \equiv \Ubar_{n,\mu} \qquad
   \mbox{\rm such that} \qquad  
   \min_{V \in SU(N)} \,F(V) \,=\, F(\Ubar_{n,\mu})
   \label{eq:545}
\end{equation}
where  
\begin{equation}
   F(V) \,\equiv\, ||\, D_{n,n+2\mu}(U) - D^f_{2\mu} \times V \, ||
   \label{eq:550}  
\end{equation}
In general, we associate with arbitrary approximate chiral ordering map $\cM^D$
the family of maps $\cM^{D,s}$ defined by the formal replacement
\begin{equation}
  \cM^D \,\longrightarrow\,  \cM^{D,s}  \qquad
  \mbox{\rm under} \qquad
  \cases{ D_{n,n+\mu}(U) & $\longrightarrow\;\, D_{n,n+s\mu}(U)$ \cr
          D^f_\mu     & $\longrightarrow\;\, D^f_{s\mu}$ \cr}
  \qquad
  s=1,2,\ldots
   \label{eq:555}  
\end{equation}
where obviously $\cM^{D,1}=\cM^D$.

It is now straightforward to define the associated family of RG transformations
with scaling factor $s$. For simplicity, we will work formally on the infinite
lattice. The discussion can be repeated for the finite lattice in a straightforward
manner if one lets the lattice sizes $L$ to be the appropriate factors of powers
of $s$. The RG transformation $\rgmap^{D,s}$ is defined by
\begin{equation} 
     U \;\longrightarrow\; \rgmap^{D,s}(U) \;:\quad
     U_{n,\mu} \;\longrightarrow\; \rgmap^{D,s}_{n,\mu}(U) \,\equiv\,
     \cM^{D,s}_{sn,\mu}(U)  
     \label{eq:560}
\end{equation}
where reverting to the original lattice is already performed. Note that
$\rgmap^{D,1}\equiv \cM^D$. Repeated application of $\rgmap^{D,s}$ on $U$
induces the sequence of configurations 
$U^{s,j} \equiv (\rgmap^{D,s})^j(U),\;j=0,1,2,\ldots$ with $U^{s,0} \equiv U$.
Starting from the gauge action $S\equiv S^{\rgmap^{D,s}}_{(0)} \in \cS^G$
represented by the corresponding ensemble, this repeated map defines the sequence 
of actions $S^{\rgmap^{D,s}}_{(j)},\,j=0,1,2,\ldots$ in $\cS^G$, namely 
the RG trajectory. In fact, because of the freedom in choosing $s$, we have
defined a family of RG trajectories with different scaling factors $s=2,3,\ldots$. 
Chiral ordering procedure ($s=1$) represents the limiting case in this family and we 
will refer to the corresponding trajectory as the {\em chiral ordering trajectory}.

To close this section, we need to discuss a particular aspect of the above 
transformations that one needs to keep in mind. In a generic case of RG transformation
with scaling factor $s$, one argues on intuitive physical grounds that the 
long-distance behavior in the induced theory of blocked variables is exactly the same 
as in the original theory. This is expressed in the fact that the correlation lengths 
(before reverting to the original lattice) remain exactly the same.
As a result we find (after reverting to the original lattice) that 
$\xi_i \rightarrow \xi_i/s$
in the transformed theory for all correlation lengths $\xi_i$. Consequently,
the lattice spacing (being based on one of the correlation lengths) gets magnified
as $a\rightarrow s a$. The conclusion that interactions on renormalized trajectory 
specify perfect actions (no cutoff effects) relies heavily on the fact that
the above scaling of correlation lengths is {\em exact}. This is difficult 
to prove rigorously for generic non-linear RG transformations. Moreover, even
in case of linear transformations (e.g. for scalar fields) the argument
assumes that the transformations are strictly ultralocal, i.e. that the new variable
does not depend at all on old variables outside the considered block. At the same 
time, it is not difficult to see that if the dependence on variables outside the
block is non--ultralocal but local, i.e. that the influence of variables outside 
the block is bounded by the exponential decay with characteristic lattice length 
$\alpha$, then the correlation lengths of blocked variables can in principle change
at most by $\alpha$. If such change occurs, it will certainly make little difference
close to the critical surface ($g=a=0$) when the correlation lengths are very 
large. However, in principle it can make a difference when transforming a system
with lattice correlation lengths of order $\alpha$. 

The RG transformations proposed here are based on Ginsparg--Wilson fermionic kernels
that are necessarily non-ultralocal in fermionic variables~\cite{nonultr} and expected
to be non--ultralocal also in gauge variables which is of relevance here. Consequently, 
we will not claim that actions on the corresponding renormalized trajectories are perfect 
at arbitrary lattice spacing (correlation length). Rather, we will implicitly understand
that they are close to perfect down to correlation lengths comparable to $\alpha$. 
\footnote{We emphasize that we do not claim that non-ultralocality necessarily 
forces the small change of correlation length in the transformed system. We only
say that this is possible with non-ultralocal transformations. Also, if one wants to be 
more precise, our underlying assumption here is that performing the blocking operation
arbitrarily many times (without reverting to original lattice) leads asymptotically
only to a finite total change of correlation length. Then the concept of perfect action
is meaningful only up to correlation length comparable to this total change which is
expected to be of order $\alpha$.}
Since there are Ginsparg--Wilson actions with $\alpha \approx 1$ 
(see e.g. Ref.~\cite{ov_loc}), this is not an essential restriction. Moreover, as we 
emphasized all along, the nature of 
the RG transformation based on chiral ordering is expected to be such that the physics 
of the gauge field remains largely preserved even at small lattice distances. We thus 
expect that the correlation lengths will only change by the amount smaller than $\alpha$.
\footnote{It is worth mentioning here that standard RG transformations in momentum
space correspond to non-ultralocal (and sometimes even non--local) transformations in 
position space. Nevertheless, they are widely used albeit for analyzing different
questions than those relevant here.} 
The main purpose of the above remarks is to convey two messages. (1) Even though the
RG transformations proposed here are non-ultralocal, we expect that ``perfectness'' on 
the renormalized trajectory is still a valid concept. (2) Because of non--ultralocality 
we admit the possibility that the correlation lengths of blocked variables can in 
principle change by very small amounts.

\subsection{Renormalized Trajectory and the Line of Perfect Chiral Order}

\begin{figure}[t]
\begin{center}
   \vskip -0.20in
   \centerline{
   \hskip  0.00in
   \includegraphics[width=7.9truecm,angle=0]{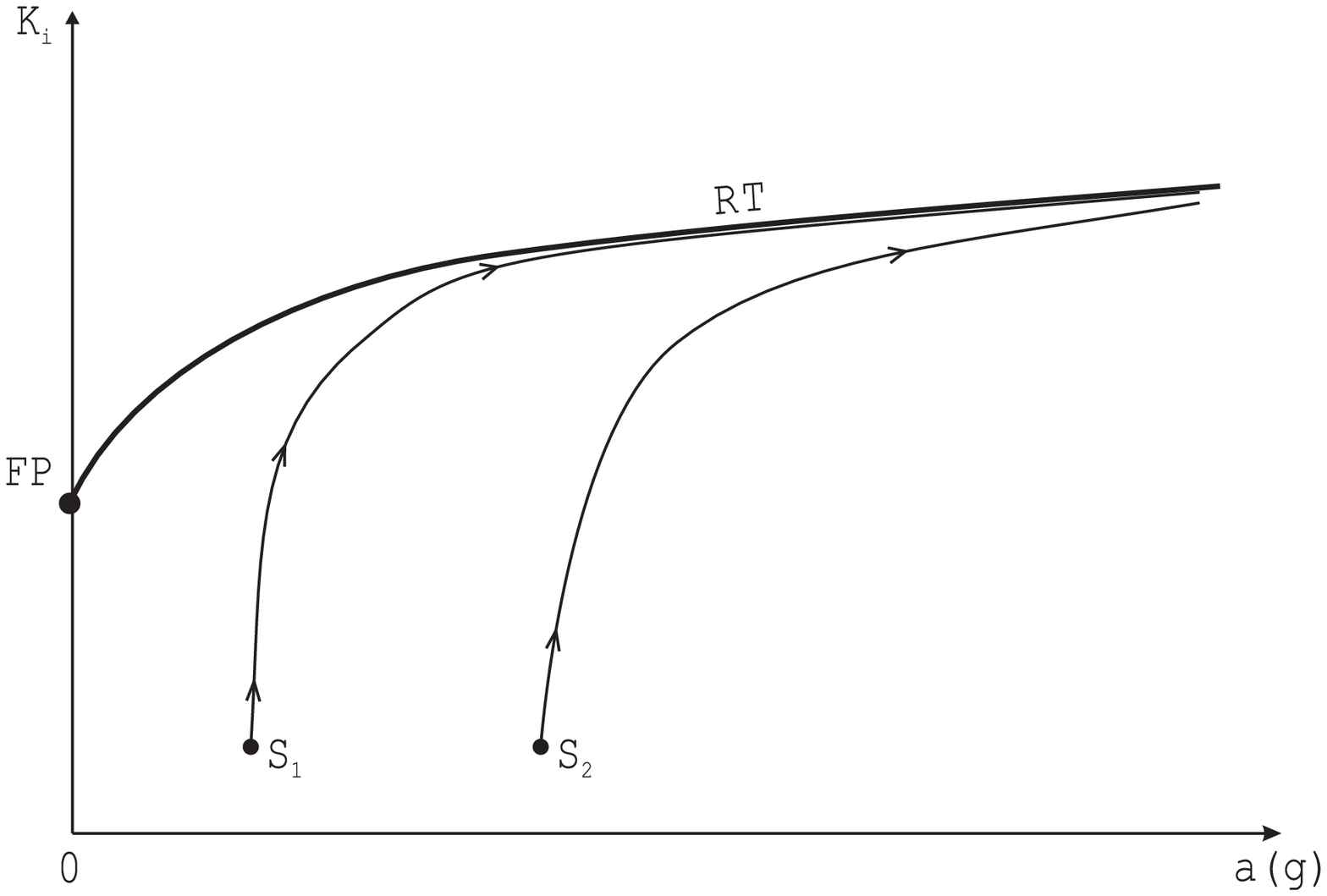}
   \hskip  0.20in
   \includegraphics[width=7.9truecm,angle=0]{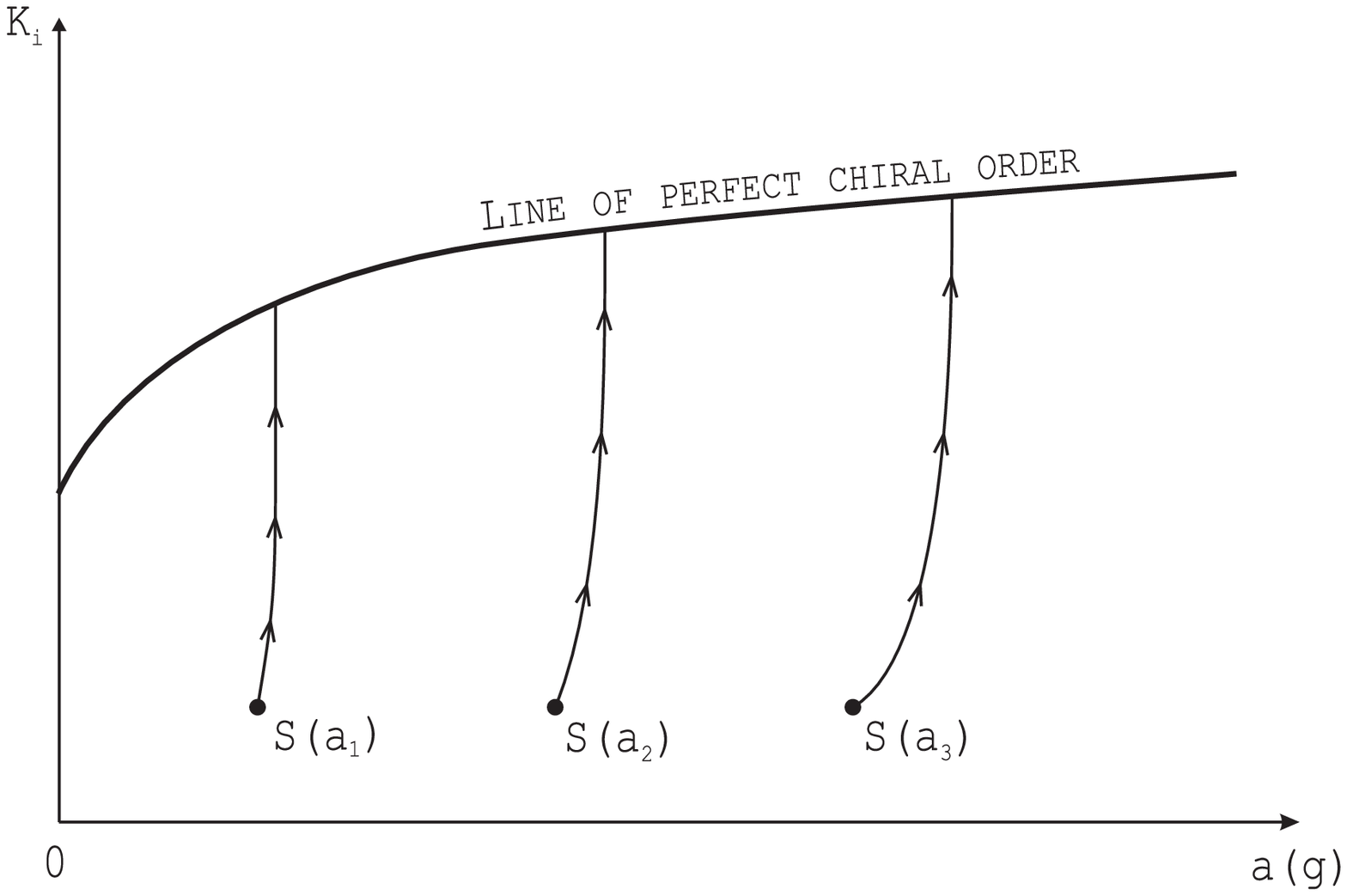}
   }
   \caption{The expected flow under transformations $\rgmap^{D,s}$ for $s>1$ 
            (left) and $s=1$ (right). $K_i$ collectively denotes the set of 
            couplings other than the gauge coupling. The line of perfect chiral
            order plays the role analogous to that of the renormalized trajectory.}
   \vskip -0.4in 
   \label{flows:fig}
\end{center}
\end{figure}

While the chiral ordering transformation $\cM^D$ and the RG transformations 
of Eq.~(\ref{eq:560}) have been put on the same footing via the family of maps 
$\rgmap^{D,s},\, s=1,2,\ldots$, we would conventionally tend to think of them 
in rather different terms. Starting with RG transformations
$\rgmap^{D,s},\, s=2,3,\ldots$, we accept the standard picture of the RG flow
for Yang--Mills theory in the set of actions. In other words, for given $s$ we
assume that there exists a unique fixed point (FP) of $\rgmap^{D,s}$ on the critical 
surface ($g=a=0$), and that there is a single weakly relevant coupling associated with 
this FP, namely the gauge coupling. Starting with arbitrary action $S(a(g)) \in \cS^G$ 
close to critical surface ($g$ very small), the RG trajectory will first run towards
the position of the FP and then away towards larger couplings (larger lattice spacings).  
The limiting procedure where the bare gauge coupling of the starting action $S(a)$
approaches zero (critical surface) defines a trajectory running from FP towards 
larger couplings, namely the renormalized trajectory corresponding to $\rgmap^{D,s}$.
The generic flow of actions (starting from $S$ not necessarily very close to 
critical surface) in relation to renormalized trajectory is shown schematically 
in Fig. \ref{flows:fig} (left).

Turning now to the chiral ordering map $\cM^D$, one could naively view it as nothing 
more than a very sophisticated way of smoothing the gauge field. Generic procedures
of this type are sometimes distrusted if used inappropriately. However, as we argued 
extensively, our suggestion here is that chiral ordering is considerably more 
fine--tuned than generic smoothing 
\footnote{By ``generic'' smoothing we mean any map that tends to average the gauge 
fields locally. Examples of this could be the APE smearing~\cite{APE} or the
``stout link'' operation~\cite{stout}.} 
since it tends to preserve the defining feature of the gauge field, namely its 
influence on the charged particle. 
This point of view is expressed by the principle of chiral ordering and, in more 
definite terms,
by conjectures {\bf C1}, {\bf C2} and {\bf C2a}. These imply that the evolution
in the set of actions under $\cM^D$ could actually be very constrained, and 
the tendency to preserve physics leads to the fact that the actions 
$S^{\cM^D}_{(j)} \equiv S^{\rgmap^{D,1}}_{(j)}, j=0,1,2,\ldots$ converge to 
the limiting action ${\hat S}(S,\cM^D)$ (fixed point) under repeated application 
of $\cM^D$. Fixing the starting lattice theory $S(a)$ parametrized by its lattice
spacing, we obtain a one--parameter line of fixed points ${\hat S}(a,\cM^D)$.
We will refer to this line as the {\em line of perfect chiral order}.
Note that the lattice spacing of ${\hat S}(a,\cM^D)$ is not expected to be
exactly $a$, but rather ${\hat a}(a)$ since correlation lengths can change 
by small amounts upon the repeated chiral ordering. However, because of tendency
to preserve physics, we expect that ${\hat a} \approx a$ with increased accuracy 
as $a\to 0$. We should also mention that by writing ${\hat S}(a,\cM^D)$
we implicitly assumed that the line of perfect chiral order only depends on 
transformation $\cM^D$ and not on the starting theory chosen (analogue of the 
uniqueness of the fixed point in the RG case). We thus propose a flow pattern
for $\cM^D \equiv \rgmap^{D,1}$ as shown in Fig. \ref{flows:fig} (right).

Given two arbitrary and unrelated RG transformations, it is not expected that
the positions of their corresponding fixed points and RG trajectories are 
highly correlated. However, it is clear that various transformations in the family 
$\rgmap^{D,s}$ are strongly related by the fact that they are all defined from 
a single and highly ``fine--tuned'' object -- the chirally symmetric lattice Dirac 
operator. Moreover, all these transformations are based upon the same underlying
principle: they express the effective matrix phase associated with the propagating
chiral fermion. Thus, for example, one would expect that the evolution of 
the action under a single $s=4$ transformation $\rgmap^{D,4}$ is quite similar to 
repeating two $s=2$ transformations $\rgmap^{D,2}$. Consequently, one would also 
expect that the corresponding fixed points and RG trajectories for different scaling
factors are closely positioned in the set of actions. At the same time, as made 
clear by the discussion in this section, RG trajectories for $s=2,3,\ldots$ are 
the direct analogs of the line of perfect chiral order in the $s=1$ case. 
Consequently, it is natural to conclude that the line of perfect chiral order 
runs in the vicinity of RG trajectories corresponding to transformations 
$\rgmap^{D,s}$. 
\smallskip

\noindent Accepting the above scenario entails two noteworthy implications:
\medskip

\noindent {\em (i)} Repeated chiral ordering brings lattice gauge theory close
to particular RG trajectories and thus, due to the conjecture {\bf CI2}, the 
regions around such RT--s are special in that their Kolmogorov complexity is very 
low. Thus, the framework of chiral ordering is compatible with the intuitive
reasoning put forward at the beginning of this section.
\medskip

\noindent{\em (ii)} To the extent that the arguments put forward here will 
turn out to be valid, one can also expect that chiral ordering might be capable of 
not only ``preserving the physical content'' as stipulated by 
{\em \underline{Goal 2}}, but also to {\em improve} the scaling behavior. 
Indeed, since theories in the immediate vicinity of RT are expected to have small 
cutoff effects, it can happen that the various correlation lengths will readjust
themselves in order to mimic more closely the continuum behavior.
\medskip

Finally, let us mention that we carried out this discussion in the context of 
{\ em full} chiral ordering. Adaptation to the case of partial chiral ordering is 
possible but depends to a large extent on which particular ordering is 
used. We thus postpone the detailed discussion for this case to dedicated 
publications.

\section{Effective Lattice QCD I.}
\label{sec:eff_lqcd}

The final topic we will deal with in this first paper of the series is the
construction of the framework for studying the {\em effective} QCD vacuum
structure and its scale dependence (see Sec.~\ref{sec:intro_effective} and 
points {\bf A3} and {\bf A4}). More precisely, here we will only discuss a 
specific issue (in fact a particular version of it) relevant for this problem. 
The topic will be discussed further in the second paper, and in its full generality 
in the third paper of this series. 
 
We will examine the problem of defining the theory $S_{\Lambda_F}$ associated 
with $S\in \cS^G$, for arbitrary fermionic response scale $\Lambda_F$. 
The particular approach proposed here is based on the approximate chiral 
ordering, and will not allow us to do it for arbitrary $S \in \cS^G$. 
However, with every $S\in \cS^G$ we will associate another valid lattice action(s) 
with the same physics content (in the spirit of the principle of chiral ordering), 
for which it can be done. Viewed alternatively, the construction applies for arbitrary 
$S \in \cS^G$ but with restricted set of observables, i.e. we will not be able to 
define the mean values of all lattice operators $O(U)$ in $S_{\Lambda_F}$, but only 
a subset of it. However, this subset is sufficiently rich so that this is in fact 
a very mild restriction. 

Let $S \in \cS^G$, $D\in \cS^F$ and let $\cM^D \in \Umaps$ be an arbitrary 
approximate chiral ordering based on $D$. As discussed in 
Secs.~\ref{sec:deformation},\ref{sec:pco_sub}, the map $\cM^D$ can be used
to associate with $S$ another gauge theory $S^{\cM^D}$ via the transformation 
of the ensemble   
\begin{equation}
   \cE_S =\{\ldots U^{i-1},U^{i},U^{i+1}, \ldots \} 
          \; \longrightarrow \;
          \{\ldots \cM^D(U^{i-1}),\cM^D(U^{i}),\cM^D(U^{i+1}), \ldots \} 
          \equiv \cE_{S^{\cM^D}} 
   \label{eq:565}  
\end{equation}
To define the action $S^{\cM^D}_{\Lambda_F}$ of {\em effective} lattice QCD at 
fermionic response scale $\Lambda_F$ via its ensemble, we first define the chiral 
ordering map $\cM^{D,\lambda_F}$ at the corresponding scale in lattice units 
($\lambda_F = \Lambda_F a$). To do that, we will proceed in the same way 
as in definition of the effective topological charge density in 
Refs.~\cite{Hor02B,Hor05A}, i.e. we will perform the eigenmode expansion of the
corresponding matrix element of $D$. Indeed, the map $\cM^D$ depends on the 
collection of spin--color matrices $D_{n,n+\mu}$ and we can write
\begin{equation}
   D_{n,n+\mu} \,=\,
   \sum_j \psi_n^j \; \lambda_j \;(\psi_{n+\mu}^j)^\dagger
   \qquad\quad
   D_{n,n+\mu}^{\lambda_F} \,\equiv\, 
   \sum_{j:|\lambda_j| \le \lambda_F} 
   \psi_n^j \; \lambda_j \; (\psi_{n+\mu}^j)^\dagger
   \label{eq:570}  
\end{equation}
where $\psi^j$ is an eigenmode of $D$ with eigenvalue $\lambda_j$ and,
in what follows, we will always assume that the labeling of eigenmodes
respects the ordering of eigenvalues by magnitude, i.e. if $j<k$ then
$|\lambda_j| \le |\lambda_k|$.
\footnote{Note that the expansion in the form (\ref{eq:570}) applies when $D$ is 
a normal matrix, e.g. for the overlap Dirac operator. If $D$ is not normal
then we can still use the expansion in the bi--orthogonal left--right 
eigensystem.} 
The map $\cM^{D,\lambda_F}$ is then defined via a replacement 
\begin{equation}
   \cM^{D}_{n,\mu} \;\longrightarrow \cM^{D,\lambda_F}_{n,\mu}
   \qquad \mbox{\rm under} \qquad
   D_{n,n+\mu} \;\longrightarrow\; D_{n,n+\mu}^{\lambda_F} \qquad 
   D_\mu^f \;\longrightarrow\; D_\mu^{f,\lambda_F}  
   \label{eq:580}  
\end{equation}
Note that in the above equations $\lambda_F\in [\,0,|\lambda|_{\max}\,]$, 
where $|\lambda|_{\max}$ is the maximal possible magnitude of an eigenvalue. 
The effective lattice QCD $S^{\cM^D}_{\Lambda_F}$ associated with $S^{\cM^D}$ 
is then specified by its ensemble
\begin{eqnarray}
   \cE_S & = & \{\ldots U^{i-1},U^{i},U^{i+1}, \ldots \}  
             \;\longrightarrow \; \nonumber \\
         & \longrightarrow &
          \{\ldots \cM^{D,\abar\Lambda_F}(U^{i-1}),\cM^{D,\abar\Lambda_F}(U^{i}),
                   \cM^{D,\abar\Lambda_F}(U^{i+1}), \ldots \}  
         \,\equiv\, \cE_{S^{\cM^D}_{\Lambda_F}}  
         \label{eq:585}  
\end{eqnarray}
with the new probability distribution (action $S^{\cM^D}_{\Lambda_F}(U))$ given
by equations (\ref{eq:295}, \ref{eq:300}). Note that $\abar=\abar(a)$ is the lattice
spacing of $S^{\cM^D}$. According to the principle of chiral ordering we expect that 
$\abar(a)\approx a$.  
\medskip

\noindent Several important points regarding the definition and meaning of effective 
lattice QCD now need to be discussed. 
\medskip

\noindent {\em (i)} It needs to be stressed that, at this point, we have not yet 
given the precise definition of $D_\mu^{f,\lambda_F}$ (i.e. of the effective free 
matrix element) in Eq.~(\ref{eq:580}). At the same time, the definition of 
$\cM^{D,\lambda_F}_{n,\mu}$ requires both $D_{n,n+\mu}^{\lambda_F}$ and
$D_\mu^{f,\lambda_F}$ for all prescriptions of approximate chiral ordering we 
discussed except for the case {\bf (VI)}. We will rectify this in the following
subsection devoted to this point. 
\medskip

\noindent {\em (ii)} We wish to reiterate that the map $\cM^{D,\lambda_F}$ associates
with arbitrary $U \in \cU^L$ an {\em effective} chirally ordered configuration 
$\Ubar^{\lambda_F} \in \cU^L$ since 
\begin{equation}
     \lim_{\lambda_F \to |\lambda|_{\max}} \Ubar^{\lambda_F} \,=\,
     \lim_{\lambda_F \to |\lambda|_{\max}} \cM^{D,\lambda_F}(U) \,=\,
     \cM^D(U) \,=\, \Ubar
     \label{eq:590}  
\end{equation}
i.e. $\Ubar^{\lambda_F}$ is an expansion of $\Ubar$ (not $U$). This is, of course, 
why $S^{\cM^D}_{\Lambda_F}$ given by (\ref{eq:585}) defines an effective description
of $S^{\cM^D}$ rather than that of $S$. However, at the same time, following 
the reasoning in Sec.~\ref{sec:deformation}, we can view the map $\cM^D$ as the 
means of associating with arbitrary local operator $O_\alpha(U)$, that we wish 
to evaluate in $S$, a new local operator $O_\alpha^{\cM^D}(U)\equiv O_\alpha(\cM^D(U))$ 
with the same classical limit. The map $\cM^{D,\lambda_F}$ then defines an effective 
operator $O_\alpha^{\cM^D,\Lambda_F}$ (to be valuated in $S$) via
\begin{equation}
     O_\alpha^{\cM^D,\Lambda_F}(U)  \,\equiv\, 
     O_\alpha\Bigl( \cM^{D,a \Lambda_F}(U) \Bigr)
     \label{eq:595}   
\end{equation}
Thus, from the point of view of action $S$, we can define the effective description 
at fermionic response scale $\Lambda_F$ for all operators that can be obtained
as $O_\alpha^{\cM^D}(U)$ for some $O_\alpha$.
\medskip

\noindent {\em (iii)} It is tempting to think that for the definition 
of $D_{n,n+\mu}^{\lambda_F}$ (and thus of $\cM^{D,\lambda_F}$) it would be 
advantageous to utilize the trick used in~\cite{Hor02B,Hor05A} to define the 
effective topological density i.e. to eigenmode--expand 
$(D-|\lambda|_{\max} \identity)_{n,n+\mu}=D_{n,n+\mu}$ rather than $D_{n,n+\mu}$. 
itself. This would seemingly give the low--lying modes the largest weight. However, 
contrary to the case of topological density, it turns out that $\gamma_5$--Hermiticity 
forces the two expansions be identical.
\medskip

\noindent {\em (iv)} We emphasize that the purpose and the meaning of 
effective lattice QCD is quite different from that of the common notion of 
effective field theory. In the latter, one constructs new degrees of freedom and
couples them via effective interaction designed so that it takes into account
the influence of high-frequency fluctuations that have been eliminated from 
the theory. However, in case of effective lattice QCD the nature of field variables
is preserved (they are still the same gauge fields), and the interaction is changed
only to {\em filter out} the high-frequency fluctuations in a meaningful manner. 
This effective interaction is non--local with the corresponding range being 
controlled by the scale $\Lambda_F$. The continuum limit of the theory 
$S^{\cM^D,\Lambda_F}(\abar(a))$ is achieved via bringing the driving theory
$S(a)$ to the continuum limit ($a\rightarrow 0$) while keeping 
$\Lambda_F$ fixed. The values of measurable quantities in the continuum 
theory $S^{QCD}_{\Lambda_F}$ will differ from those of the full continuum
theory $S^{QCD}$, and this difference must then be directly ascribed to the influence
of high--frequency fluctuations in the gauge field. In the language of the 
typical vacuum configurations, the space--time vacuum structure acquires a 
{\em scale--dependence} via $\Lambda_F$, and the relation of this structure to
the aforementioned differences will teach us how is vacuum structure at different 
scales related to various physical phenomena. In other words, the purpose of effective 
lattice QCD is to provide us with the {\em scale--dependent} picture of QCD vacuum. 
\medskip

\noindent {\em (v)} While we have carried out the discussion of effective lattice QCD 
in the context of pure--glue QCD, it can be clearly repeated without any change for 
full QCD. In fact, for given full lattice QCD defined by $S^G\in \cS^G$ and 
$S^F \in \cS^F$, there is an obvious natural choice of Dirac operator on which
we base our chiral ordering transformation $\cM^D$, namely the $D$ that specifies
$S^F$.
\medskip

\noindent {\em (vi)} At the technical level, our way of defining effective QCD 
is similar to the recently discussed procedure of Laplacian filtering~\cite{Bru05A}, 
wherein the authors define a ``filtered link'' via performing the eigenmode expansion 
of relevant matrix element of the covariant Laplacian.

\subsection{Expansion of the Free Matrix Element}
\label{sec:eff_free}

In this section we come back to the issue of specifying  $D_\mu^{f,\lambda_F}$,
which is necessary for definition of $\cM^{D,\lambda_F}$ with approximate 
chiral orderings we discussed (except case {\bf (VI)}). Our goal is to preserve
the physical meaning of chiral ordering procedure in the effective case. In other 
words, we want that the map $\cM^{D,\lambda_F}$ can still be interpreted as 
extracting the matrix phase acquired by fermion when hopping from $n+\mu$ to $n$ 
relative to the free case. The issue is what exactly do we mean by ``free case'' 
in effective theory. There are several ways to proceed and we will describe them 
in turn.
\medskip

\noindent $(\alpha)$ The most straightforward possibility is to expand the free
operator $D^f$ (no color indices in this case)
\footnote{Note that since $D^f$ is translation invariant we frequently used $D^f_m$ 
(single index) as a shorthand for $D^f_{n,n+m}$, $\forall n$. The meaning of 
single and double--indexing is thus clear.}
in its own eigenmodes $\chi^j$. The relevant matrix element is
\begin{equation}
   D^{f,\lambda_F}_\mu \,\equiv\, D^{f,\lambda_F}_{n,n+\mu} \,\equiv\, 
   \sum_{j:|\omega_j| \le \lambda_F} 
   \chi_n^j \; \omega_j \; (\chi_{n+\mu}^j)^\dagger
   \qquad\quad
   D^f \,\chi^j \,=\, \omega_j \,\chi^j
   \label{eq:600}  
\end{equation}
While this is a sensible definition satisfying our requirement, it might not be 
suitable if we are interested in (or restricted to) studying the low--energy behavior 
in finite physical volume $V^p=(La)^4$ (fixed as continuum limit is approached). 
Indeed, with antiperiodic boundary conditions in time direction, there will be 
a non--zero lower bound of the free spectrum $|\omega|_{\min}(L)$ such that 
$|\omega|_{\min}(L)/a \propto 1/(La)$ converges to a finite physical 
value $\Lambda_F^{\min}$ as $a\rightarrow 0$, when $V^p$ is held fixed. Thus, in this 
case we will not be able to define the map $\cM^{D,a\Lambda_F}$ for 
$\Lambda_F < \Lambda_F^{\min}$.
In other words, the effective lattice theory $S^{\cM^D}_{\Lambda_F}$ will not be 
defined for $\Lambda_F < \Lambda_F^{\min}$.
\footnote{Note that this is not necessarily a bad thing since $1/(La)$ is a natural
infrared scale for this system.}
\medskip

\noindent $(\beta)$ The possible issue with definition $(\alpha)$ is that for
$\Lambda_F$ close to $\Lambda_F^{\min}$ there will be an increasingly larger mismatch 
between the number of eigenmodes contributing to $D^{f,\lambda_F}_\mu$ and 
$D_{n,n+\mu}^{\lambda_F}$ as the continuum limit is approached. This comes about 
as a result of the fact that we force the scale (lattice spacing $a$) determined in 
interacting theory also on the free dynamics. This is not natural. Indeed, one 
should rather match the free theory to the interacting one by the ``number of degrees
of freedom'', which are measured in this case by the number of eigenmodes included
in the expansions. In other words, it is meaningful to ask what is an effective matrix
phase acquired by interacting fermion described by $N$ eigenmodes relative to the
dynamics described by $N$ free eigenmodes. To implement this properly, let us recall 
that, due to $\gamma_5$--Hermiticity, the eigenmodes of $D$ with complex eigenvalues 
are naturally paired, i.e. for each such eigenmode $\psi^j$ of $D$ there is a 
mode $\psi^{k}=\gfive\psi^j$ with complex--conjugated eigenvalue 
($\lambda_k=\lambda_j^\star$).
The unpaired eigenmodes have real eigenvalues and reflect the topology of the 
underlying gauge field. The eigenmodes of $D^f$ are all paired via the $\gamma_5$
operation. For given fixed $U$, let $N_0$ be the number of zeromodes of $D(U)$,
and let $N_p(\alpha)$ is the number of paired modes such that 
$|\lambda_j| \le \alpha >0$. By definition, $N_p(\alpha)$ is an even 
number. We then have
\begin{equation}
   D^{f,\lambda_F}_\mu \,\equiv\, D^{f,\lambda_F}_{n,n+\mu} \,\equiv\, 
   \sum_{j \le N_p(\lambda_F)} 
   \chi_n^{j} \; \omega_j \; (\chi_{n+\mu}^{j})^\dagger
   \label{eq:605}  
\end{equation}
Two points need to be emphasized with regard to the above definition.
\medskip

\noindent {\em (i)} Since there can be degeneracies in the free spectrum,
we implicitly assume that the ordering of the eigenmodes within the 
degenerate subspace is fixed by some arbitrary rule, and that the 
$\gamma_5$--conjugated modes follow one another when the sequence 
$\{ (\omega_j,\chi^j), j=1,2,\ldots \}$ is formed.
\medskip

\noindent {\em (ii)} Note that there is still seemingly a small mismatch 
between the number of modes contributing to $D_{n,n+\mu}^{\lambda_F}$
(equal to $N_0+N_p(\lambda_F)$) and the number of modes contributing
to $D^{f,\lambda_F}_\mu$ (equal to $N_p(\lambda_F)$). However, one can
easily convince himself that the real modes in fact do not contribute
to $D_{n,n+\mu}^{\lambda_F}$.

Using the above construction, the map $\cM^{D,\lambda_F}$ (and thus action
$S^{\cM^D}_{\Lambda_F}$) is defined by equations (\ref{eq:570},\ref{eq:580})
and (\ref{eq:605}). To see more explicitly how is this done for some basic
forms of approximate chiral ordering, consider the case {\bf (I)}.
We have $\cM^D = \cM^{(3)} \circ \cM^{(2)} \circ \cM^{(1)}$, where only
the map $\cM^{(1)}$ (projection to color space) will need modification when
transiting from $\cM^D$ to $\cM^{D,\lambda_F}$. In particular, in place of
Eq.~(\ref{eq:370}) we will have
\begin{eqnarray}
   (\,U_{n,\mu}\,)_{a,b} \,\longrightarrow\, M_{a,b}^{\lambda_F} 
   & = &
   \frac{1}{4} \mbox{\rm tr}^s \Bigl[\,( \, D^{f,\lambda_F}_\mu \times \identity^c \,)^{-1}
   \, D_{n,n+\mu}^{\lambda_F}(U) \,\Bigr]_{a,b} \nonumber \\
   & = &
   \frac{1}{4} \sum_{j:|\lambda_j|\le \lambda_F} \lambda_j \; 
   ( \psi_{n+\mu}^j )^\dagger_b 
   \; \Bigl( D^{f,\lambda_F}_\mu \Bigr)^{-1} \;
   ( \psi_n^j )_a  
   \label{eq:610}
\end{eqnarray}
where $( \psi_n^j )_a$ is the 4-component object (color index fixed to $a$) and
$D^{f,\lambda_F}_\mu$ is given by (\ref{eq:605}). For approximate chiral ordering
of case {\bf (II)} we will have in a similar manner
\begin{eqnarray}
   (\, U_{n,\mu} \,)_{a,b} \,\longrightarrow\, M_{a,b}^{\lambda_F} 
   & = &
   \frac{1}{4B_{\mu\mu}^{\lambda_F}} \mbox{\rm tr}^s \Bigl[\, \gamma_\mu
   \, D_{n,n+\mu}^{\lambda_F}(U) \,\Bigr]_{a,b} \nonumber \\
   & = &
   \frac{1}{4B_{\mu\mu}^{\lambda_F}} \sum_{j:|\lambda_j|\le \lambda_F} \lambda_j \; 
   ( \psi_{n+\mu}^j )^\dagger_b 
   \; \gamma_\mu \;
   ( \psi_n^j )_a  
   \label{eq:615}
\end{eqnarray}
where $B_{\mu\mu}^{\lambda_F}$ is the $\gamma_\mu$ component in the Clifford
decomposition of $D^{f,\lambda_F}_\mu$. We proceed in a straightforward manner
also in the remaining cases of the approximate chiral ordering. 
\medskip

\noindent $(\gamma)$ The final possibility that we wish to discuss treats
$D_{n,n+\mu}(U)$ and $D_{n,n+\mu}(\identity) = D^f_\mu \times \identity^c\,$ 
in a symmetric manner with respect to the eigenmode expansions performed. 
In particular, we will expand both operators in interacting eigenmodes 
$\psi^j$. To do this, let $\chihat^j$ denotes 
the eigenmode of $D(\identity)=D^f \times \identity^c$ with eigenvalue
$\omega_j$, i.e. $\chihat^j$ has both the spin and color indices and is related 
to eigenmodes of $D^f$ in an obvious manner. We can write $D(\identity)$ as
\begin{equation}
    D(\identity) \,=\, 
    \sum_j \,\chihat^j \,\omega_j\, (\chihat^j)^\dagger \,=\,
    \sum_{k_1,k_2} \, c_{k_1,k_2} \, \psi^{k_1} \, (\psi^{k_2})^\dagger  
    \qquad\qquad   
    c_{k_1,k_2} \equiv  (\psi^{k_1})^\dagger\, D(\identity) \,\psi^{k_2}
    \label{eq:625}
\end{equation}
which can be truncated to define
\begin{equation}
    D^{\lambda_F}_{n,n+\mu}(\identity) \,\equiv\, 
    \sum_{\scriptstyle k_1,k_2 \atop 
           \scriptstyle |\lambda_{k_1}|,|\lambda_{k_2}| \le \lambda_F}
    c_{k_1,k_2} \;\, \psi^{k_1}_n \, (\psi^{k_2})^\dagger_{n+\mu}  
    \label{eq:630}
\end{equation}
The aspect that needs to be discussed with regard to the above definition is that,
contrary to cases $(\alpha)$, $(\beta)$ where spin and color
variables are a priori separated, the matrix $D^{\lambda_F}_{n,n+\mu}(\identity)$ 
is not expected to be exactly expressible as a direct product in the corresponding 
subspaces. If this was possible and we could write it as
$D^{f,\lambda_F}_\mu \times \identity^c$, then we would use this correspondence as 
a definition of $D^{f,\lambda_F}_\mu$ in this case. However, as it stands, we need
to take an additional step. While there are other ways to proceed,
\footnote{One immediate possibility would be to define $D^{f,\lambda_F}_\mu$ 
by minimizing the norm 
$||D^{\lambda_F}_{n,n+\mu}(\identity) - D^{f,\lambda_F}_\mu \times \identity^c ||$ 
with respect to $D^{f,\lambda_F}_\mu$.}
we wish to explicitly discuss the approach where, instead of defining 
$D^{f,\lambda_F}_\mu$ explicitly and proceeding as in cases $(\alpha)$, $(\beta)$,
we simply replace the factors $D^{f,\lambda_F}_\mu \times \identity^c$ in various
definitions of approximate chiral orderings with $D^{\lambda_F}_{n,n+\mu}(\identity)$.
To see this explicitly, let us consider some interesting cases of approximate chiral 
ordering. In case {\bf (I)} we can replace the projection to color space $\cM^{(1)}$
(see Eq.~(\ref{eq:370})) directly with 
\begin{equation}
   U_{n,\mu} \,\longrightarrow\, M^{\lambda_F} \,\equiv\,
   \frac{1}{4} \mbox{\rm tr}^s 
   \Bigl[\,( \, D^{\lambda_F}_{n,n+\mu}(\identity) \,)^{-1}
   \, D^{\lambda_F}_{n,n+\mu}(U) \,\Bigr] 
    \label{eq:635} 
\end{equation}
thus defining the map $\cM^{D,\lambda_F}$ by composing it with $\cM^{(2)}$
and $\cM^{(3)}$. For case {\bf (IV)} we define the map $\cM^{D,\lambda_F}$
via the following minimization problem.
\begin{equation}
   \cM^{D,\lambda_F}: \,\; U_{n,\mu} \,\longrightarrow\, \Ubar^{\lambda_F}_{n,\mu} 
   \qquad
   \mbox{\rm such that} \qquad  
   \min_{V \in SU(N)} \,F(V) \,=\, F(\Ubar^{\lambda_F}_{n,\mu})
   \label{eq:640}
\end{equation}
where  
\begin{equation}
   F(V) \,\equiv\, ||\, D^{\lambda_F}_{n,n+\mu}(U) - 
                        D^{\lambda_F}_{n,n+\mu}(\identity) \,
                        ( \identity^s \times V ) \, ||
   \label{eq:645}  
\end{equation}
where $\identity^s$ is the identity in spinor space.

Finally, we emphasize that practical utility of cases $(\alpha)-(\gamma)$ is to
be decided numerically by studying which expansion leads to the most accurately
satisfied chiral ordering condition. 
 
\section{Summary}

The purpose of this series of articles is to discuss in detail the set of
ideas that, we think, can form a basis of a consistent framework for systematic
study of QCD vacuum structure in the path integral formalism. While the history 
of using the lattice definition of QCD to pursue these issues is rather long, it has 
not matured into a well-defined subject with accepted scope, goals, and a standard 
collection of tools. Thus, while in the areas such as hadron spectroscopy one walks 
on the firm ground of well--defined quantitative notions, the realm of QCD vacuum 
structure is typically associated with constant stumbling over (at best) intuitive 
concepts such as ``typical configuration'', ``object in the vacuum'', ``space--time
structure'', etc.. More importantly, the ultimate goal of the effort 
(the analog of obtaining the mass) is the part of the research question here. 
In other words, solving the problem of QCD vacuum means different things 
to different researchers, even though (vaguely) stated goals such as 
``understanding confinement'' and ``understanding chiral symmetry breaking'' 
are of universal interest.

Nevertheless, the common aspect of various efforts can be summed up rather 
succinctly as follows. We are trying to replace the QCD path integral 
ensemble $\cE^{QCD}$ (defined via some regularized limiting procedure, 
e.g.\ using lattice gauge theory) with the ensemble $\cE^{STR}$ which is defined
in terms of some collective variables $C$ that arise naturally as a result 
of strong dynamics. The analysis of strong interactions in terms of $C$ 
is expected to be very transparent, and the space--time nature of $C$ will 
represent the space--time structure of the QCD vacuum in the path integral
formalism. Stating the problem in this way, there are two kinds of issues
that need to be addressed for there to be a chance that a systematic framework 
emerges. {\em (i)} First, it needs to be clarified what is the relation
of $\cE^{QCD}$ (full theory of strong interactions) to $\cE^{STR}$ 
that we seek to find (see {\bf P1}). {\em (ii)} Secondly, once the underlying 
goal of the effort is specified, it is necessary to have a set of
guidelines/tools that will make the inquiries into specific questions
possible and streamlined (see {\bf P2}). Our suggestions with regard to 
{\em (i)} and {\em (ii)} can be schematically described as follows.
\medskip

\begin{description}

\item[{\em (i)}] Given the existence of the fundamental topological structure
observable in regularized QCD ensembles~\cite{Hor03A},
we propose that the ensemble $\cE^{STR}[C]$ can be regarded as fully
equivalent to $\cE^{QCD}$ for all questions of physical interest. In other
words, we can basically view the transition from $\cE^{QCD}[A]$ to $\cE^{STR}[C]$
as a complicated change of variables in the QCD path integral. Analogously
to the case of topological vacuum, we refer to the structure entailed
by variables $C$ as a {\em fundamental QCD vacuum structure}.
\footnote{Note that discovering the nature of variables $C$ does not 
necessarily imply the analytic solution of QCD. However, if QCD can be 
solved analytically, then specifying $C$ would clearly mean an important 
step in that direction.} 
Constructing $\cE^{STR}$ will not provide us with direct understanding
of various phenomena in strong interactions because of scale dependence
inherent in field--theory description. To achieve such understanding, 
we propose that it is also necessary to study the {\em effective vacuum 
structure} which captures the dependence of vacuum properties on the scale 
of fluctuating elementary fields~\cite{Hor02B,Hor05A}. The effective 
structure is specified by the ensemble 
$\cE^{STR}_\Lambda \equiv \cE^{STR}[C(\Lambda),\Lambda]$ describing the
{\em effective QCD} at scale $\Lambda$.\ \footnote{For more precise meaning
of ``effective'' in this case see comment {\em (iv)} of 
Sec.~\ref{sec:eff_lqcd}. More detailed discussion will be given in the
third paper of this series.}
Thus, in summary, the ultimate goal
in our approach is to construct the ensembles $\cE^{STR}$ and 
$\cE^{STR}_\Lambda$ for all $\Lambda$. 
\medskip

\item[{\em (ii)}] The above goal can be systematically pursued using the 
lattice definition of QCD and, in particular, by studying the space--time 
structure in numerically--generated ensembles of finite systems in a 
{\em Bottom--Up} manner (see section \ref{sec:intro_bottomup}). This means 
that we will attempt to arrive at the definition of  
$\cE^{STR}$ and $\cE^{STR}_\Lambda$ inductively, using the observed structure 
in the regularized ensembles as the only input (see {\bf A1}--{\bf A4}). 
In other words, the criterion for selecting the appropriate collective 
variables $C$ (or $C(\Lambda)$) is identified in this approach with actual 
{\em observability} of space--time attributes associated with $C$ in 
the regularized path--integral ensembles.

\end{description}
  
\medskip

Having the basic overall scheme described above in mind, the content of this 
series of papers can be described as an attempt to elevate it to the state where 
its various elements are properly defined to the largest extent, and where 
it is possible to practically pursue its goals using the currently available 
methods of lattice gauge theory. The present work addresses three topics related
to these issues.
\medskip

\noindent {\bf (I)} When searching for fundamental vacuum structure in 
a Bottom--Up manner, we can take advantage of a freedom that is inherent in the
process of regularization. In particular, the continuum theory can be defined in many 
ways using different lattice regularizations (theories in the set of valid lattice 
actions). Since the underlying goal is to identify the space--time patterns in 
configurations dominating the regularized path integral, we naturally prefer lattice 
theories for which these configurations exhibit a high degree of space--time order. 
However, such reasoning assumes that there exists a well--defined notion for 
``degree of space--time order'' in an arbitrary configuration. We argued that 
Kolmogorov entropy (Kolmogorov complexity) of binary strings describing 
coarse--grained configurations provides the appropriate quantitative measure. 
The ensemble average of Kolmogorov entropy associated with any theory then defines 
the ranking of regularizations (at a particular fixed cutoff) by the degree of 
space--time order the corresponding lattice interaction generates. We wish to make 
a few comments regarding these issues.
\medskip

\noindent $(\alpha)\;$ The Kolmogorov entropy measure is highly universal, 
capturing the space--time order in essentially any form. There is a price 
that we have to pay for this generality. Indeed, if one wishes to consider 
Kolmogorov entropy as a practical tool (rather than a conceptual construct),
then one needs to emphasize that the entropy $\ken(U,k)$, while well defined, 
is not computable in general. Also, the use of pseudo--random (deterministic) 
numbers in actual simulations will most likely lead to a systematic error in 
determining the average value $\ken[S,k]$ for theory $S$. Thus, if pseudo--random 
numbers are used for generating actual ensembles, then we implicitly assume
that our conclusions are not affected by this error. 
\medskip

\noindent $(\beta)\;$ We included a rather detailed discussion of the information 
theory aspects in Sec. \ref{sec:kentropy}. This is not just to introduce 
the Kolmogorov entropy, but also because we believe that the language and 
methods of information theory are both appropriate and fruitful for this area of 
research. Indeed, there are many relevant questions regarding the vacuum structure 
that can be both properly formulated and potentially solved by using the framework 
of information theory. For instance, regarding the issue at hand, one can attempt 
to formulate less general but computable measures (based on the quantity of 
information) that would mimic the role of Kolmogorov entropy. 
\medskip

\noindent $(\gamma)\;$ It is also worth mentioning that Kolmogorov entropy and 
the information viewpoint are quite useful for providing intuitive understanding 
in situations that are otherwise difficult to grasp. For example, Kolmogorov
entropy of the Wilson gauge action at currently studied couplings is probably
rather high since no observable structure could be identified in its typical
configurations when standard ultralocal operators for composite fields were used. 
From the information theory viewpoint it is clear that in order to reach the theory 
with significantly lower Kolmogorov entropy, it is necessary that it is related to 
Wilson theory by a highly complex map. In other words, this theory must be defined 
by an action that is difficult to work with computationally. Thus, the actions 
possessing low Kolmogorov entropy are expected to be very complicated. It is for 
related reasons that we are able to observe a well-defined structure in the 
configurations of overlap--based topological density (which is difficult to compute) 
but not in the ultralocal definitions which are comparably ``simple''.
\medskip 

\noindent {\bf (II)} In order to identify the space--time nature of collective 
variables $C$ in regularized configurations most easily, one would prefer to work 
with lattice theories that maximize the degree of space--time order. However, it 
is not sufficient (even as a conceptual goal) to simply minimize the Kolmogorov 
entropy in the set of actions (see {\em Goal 1} and {\em Goal 2} of 
Sec. \ref{sec:pco_main}). Indeed, one should rather perform a minimization
constrained by the requirement that the physics content of the theory remains 
largely preserved in the process. As a tool for achieving this in practice, 
we propose the configuration--based deformation of the action defined via 
{\em chiral ordering transformation}. 
Chiral ordering transformation associates with an arbitrary gauge configuration $U$ 
a new configuration $\Ubar$ such that $\Ubar_{n,\mu}$ represents an effective 
SU(N) matrix phase associated with hopping of chiral fermion from $n+\mu$ to $n$.
Since the transformation is constructed to represent the local physical meaning of 
the gauge field (local influence on a charged particle), it is expected that both 
short and long distance (in lattice units) properties of the theory will change very 
little upon the deformation ({\em principle of chiral ordering}). The repeated 
application of chiral ordering transformation on a given ensemble induces the evolution 
in the set of actions. We propose that there exist nontrivial configurations on a finite 
lattice that are stable under the chiral ordering transformation (perfectly chirally 
ordered configurations), and that the evolution in the set of actions can lead to 
theories where such configurations play a dominant role. We wish to mention 
the following points related to these issues.
\medskip

\noindent $(\alpha)\;$ Two basic forms of chiral ordering transformations were
discussed. The {\em full ordering} where all links of the lattice are forced 
to acquire their effective value dictated by chiral fermion, and 
the {\em partial ordering} where only a subset of all links is changed.
The second option was introduced in order to incorporate the possibility 
that the QCD path integral is dominated by configurations containing regions
that are not accessible to fermion propagation. Indeed, if that was the case
then the effective interaction phase would not have to be well defined for links 
over which the propagation is prohibited. Detailed numerical studies 
will be necessary to determine which option is closer to that being realized by 
QCD dynamics.
\medskip

\noindent $(\beta)\;$ Connections to RG ideas suggest themselves rather 
naturally in our framework, and they were discussed in Sec. \ref{sec:rg}. 
We defined the family of RG transformations with scale factor $s$ via 
chiral orderings extended over straight paths containing $s$ links.
The original chiral ordering can then be viewed as a limiting case of these
transformations with $s=1$. We proposed that actions on renormalized 
trajectories corresponding to generic RG transformations represent
important features in the landscape of Kolmogorov entropy over the set of
actions. The particular suggestion is that there are deep local minima of 
Kolmogorov entropy in the vicinity of renormalized trajectories. Assuming that 
chiral ordering evolution indeed leads to fixed points belonging to the set 
of actions, we argued that this conclusion can be made plausible at least for 
renormalized trajectories corresponding to RG transformations based on chiral 
ordering.
\medskip

\noindent {\bf (III)} In the Bottom--Up approach, the construction of 
$\cE^{STR}_\Lambda$ (effective structure) is to be guided by the space--time 
behavior in {\em effective QCD} ($S^{QCD}_\Lambda$), which can be defined 
by its ensemble $\cE^{QCD}_\Lambda$. The ensemble $\cE^{QCD}_\Lambda$ is
in turn to be defined as a continuum limit of lattice-regularized ensembles 
$\cE^{LQCD}_\Lambda$ with $\Lambda$ fixed. However, such regularizations have 
not been defined yet. Here we suggested a particular form of the first step toward 
the general definition of this kind. Guided by the fruitfulness of bringing in 
the chiral fermion structure into definition of lattice topological field, we 
proposed lattice theories defined at a given fermionic response scale $\Lambda_F$. 
This definition is based on the chiral ordering transformation discussed
here and, in this regard, we wish to briefly mention the following two points.
\medskip

\noindent $(\alpha)$ It needs to be emphasized that in our framework the 
search for the fundamental and effective structure is completely unified and 
streamlined under the umbrella of the principle of chiral ordering. Indeed, 
in a nutshell we are contemplating the following chain of logic. Consider
an arbitrary lattice action $S \in \cS^G$. If the goal is to examine its ensemble
to obtain an information on the fundamental structure, we suggest that it 
is beneficial to make a transition to the action $S^{\cM^D}$ representing a 
configuration--based deformation of $S$ constructed via chiral ordering map 
$\cM^D$, since this is expected to lower the Kolmogorov entropy. 
Next, we propose to consider the effective theories
$S^{\cM^D}_{\Lambda_F}$ of $S^{\cM^D}$, whose definition is made possible by 
the fact that theory became a function of Dirac kernel $D$. This will provide
information on the effective structure in a manner fully consistent
with fundamental structure since
$\lim_{\Lambda_F\to \Lambda_F^{\max}} S^{\cM^D}_{\Lambda_F} = S^{\cM^D}$.  
\medskip

\noindent $(\beta)$ As is quite clear from our discussion, the framework 
proposed here is particularly natural in the context of full QCD. Indeed,
in this case the choice of $D\in \cS^F$ in chiral ordering transformations 
is fixed by the fact that $D$ is part of the definition of the theory
itself. 

\bigskip\medskip
\noindent
{\bf Acknowledgments:}
I am indebted to Andrei Alexandru and Jianbo Zhang for participating in 
the preliminary numerical work testing some of the core ideas presented here. 
Many thanks to Andrei for spending long patient hours of listening (and 
sometimes pointing me in better directions) to my attempts to sort out 
the issues discussed in this manuscript. Conversations with Terry Draper, 
Keh-Fei Liu, Nilmani Mathur, Ganpathy Murthy, Thomas Streuer and Sonali 
Tamhankar are also gratefully acknowledged. This work was supported
by the U.S. Department of Energy under the grant DE-FG05-84ER40154.

\vfill\eject


\begin{thebibliography}{99}

\bibitem{Wil74A}
  K.~Wilson, Phys.~Rev.~{\bf D10}, 2445 (1974).

\bibitem{Cr79A} M.~Creutz, Phys.~Rev.~Lett. {\bf 43} (1979) 553.
                Erratum ibid., {\bf 43} (1979) 890.

\bibitem{Cr79B} M.~Creutz, L.~Jacobs, C.~Rebbi, Phys.~Rev.~Lett. {\bf 42} (1979) 1390.

\bibitem{Hor02B} I.~Horv\'ath, S.J.~Dong, T.~Draper, F.X.~Lee, K.F.~Liu, H.B.~Thacker, J.B.~Zhang, 
                 Phys.~Rev. {\bf D67}, 011501(R) (2003).

\bibitem{Hor03A} I.~Horv\'ath, S.J.~Dong, T.~Draper, F.X.~Lee, K.F.~Liu, N.~Mathur, H.B.~Thacker, 
                 J.B.~Zhang, Phys.~Rev. {\bf D68}, 114505 (2003).

\bibitem{Hor05A} I.~Horv\'ath, A.~Alexandru, J.B.~Zhang, S.J.~Dong, T.~Draper, F.X.~Lee, K.F.~Liu, 
                 N.~Mathur, H.B.~Thacker,
                 Phys. Lett. {\bf B612} (2005) 21.

\bibitem{Ale05A} A.~Alexandru, I.~Horv\'ath, J.B.~Zhang,
                 Phys. Rev. {\bf D72} (2005) 034506.

\bibitem{Has98A}
   P. Hasenfratz, V. Laliena, F. Niedermayer,
   Phys. Lett. {\bf B427} (1998) 125.

\bibitem{NarNeu95} R.~Narayanan and H.~Neuberger, Nucl.~Phys. {\bf B443}, 305 (1995).


\bibitem{Gin82A} P.H.~Ginsparg and K.G.~Wilson, Phys.~Rev. {\bf D25}, 2649 (1982).


\bibitem{Hor04A} I.~Horv\'ath, Nucl.~Phys. {\bf B710} (2005) 464;
                 Erratum-ibid. {\bf B714} (2005) 175.

\bibitem{Neu98BA}
   H.~Neuberger, Phys.~Lett {\bf B417} (1998) 141; 
                 Phys.~Lett. {\bf B427} (1998) 353.

\bibitem{fermion_glob} 
   J.~Smit, J.~Vink, Nucl.~Phys.~{\bf B284}, 234 (1987);
                     Nucl.~Phys.~{\bf B298}, 557 (1988);
   M.~Laursen, J.~Smit, J.~Vink, Nucl.~Phys.~{\bf B343}, 522 (1990).   

\bibitem{Hor01A} 
   I.~Horv\'ath, N.~Isgur, J.~McCune, H.B.~Thacker,
       Phys.~Rev.~{\bf D65}, 014502 (2002).  

\bibitem{deG01A} T.~deGrand, A.~Hasenfratz, Phys.~Rev. {\bf D64}, 034512 (2001).        

\bibitem{Followup} 
   T. DeGrand, A. Hasenfratz, Phys. Rev. {\bf D65}, 014503 (2002);
   I. Hip et al., Phys. Rev. {\bf D65}, 014506 (2002);
   R. Edwards, U. Heller, Phys. Rev. {\bf D65}, 014505 (2002);
   T. Blum et al., Phys. Rev. {\bf D65}, 014504 (2002);
   C.~Gattringer et al., Nucl.~Phys. {\bf B618} (2001) 205; 
                         Nucl.~Phys. {\bf B617} (2001) 101.

\bibitem{Hor02A} 
   I. Horv\'ath \emph{et al}., Phys. Rev. {\bf D66}, 034501 (2002).

\bibitem{Gat02A} C.~Gattringer, Phys. Rev. Lett. {\bf 88} 221601 (2002);

\bibitem{Other} N.~Cundy, M.~Teper, U.~Wenger, Phys.~Rev. {\bf D66}, 094505 (2002);
                P.~Hasenfratz \emph{et al.}, Nucl. Phys. {\bf B643} (2002) 280;
                C.~Gattringer, Phys.~Rev. {\bf D67}, 034507 (2003);
                T.~Draper \emph{et al.},
                   Nucl. Phys. {\bf B} (Proc. Suppl.) 140, 623 (2005);  
                C.~Aubin \emph{et al.} [MILC],
                   Nucl. Phys. {\bf B} (Proc. Suppl.) 140, 626 (2005); 
                J.~Gattnar \emph{et al.}, Nucl. Phys. {\bf B716} (2005) 105;
                F.V.~ Gubarev, S.M.~Morozov, M.I.~Polikarpov, V.I.~Zakharov,
                {\tt hep-lat/0505016};
                A.~Gonzalez-Arroyo, R.~Kirchner, JHEP 0601, 029 (2006);
                A.M.~Garcia-Garcia, J.C.~Osborn {\tt hep-lat/0512025}.

\bibitem{Gre05A} J. Greensite, {\v S}. Olejn\'{\i}k, M.I. Polikarpov, S.N. Syritsyn, 
                 V.I. Zakharov, Phys. Rev. {\bf D71}, 114507 (2005).

\bibitem{Bru05A} F. Bruckmann, E.-M. Ilgenfritz, Phys. Rev. {\bf D72}, 114502 (2005).

\bibitem{Tha05A} S. Ahmad, J.T. Lenaghan, H.B. Thacker, 
                 Phys. Rev. {\bf D72}, 114511 (2005).

\bibitem{follow_q} V. Weinberg \emph{et al}, PoS LAT2005 (2005) 171; 
                   E.-M. Ilgenfritz \emph{et al}, 
                   Nucl. Phys. {\bf B} (Proc. Suppl.) 153, 328 (2006);
                   Y. Koma \emph{et al}, PoS LAT2005 (2005) 300. 

\bibitem{quaternion} F.V. Gubarev, S.M. Morozov, Phys. Rev. {\bf D72}, 076008 (2005);
                     Phys. Rev. {\bf D73} 014512 (2006); {\tt hep-lat/0602001}.  

\bibitem{Lus98A}
M.~L\"{u}scher, Phys.~Lett. {\bf B428} (1998) 342.

\bibitem{Sha48} C.E.~Shannon, Bell System Tech. J. {\bf 27}, 379--423, 623--656, 1948.

\bibitem{Sol64} R.J.~Solomonoff, Inform. Contr. {\bf 7}, 1--22, 224-254, 1964. 

\bibitem{Kol65} A.N.~Kolmogorov, Problems Inform. Transmission {\bf 1(1)}, 1-7, 1965.

\bibitem{Cha69} G.J.~Chaitin, Journal of the ACM {\bf 16}, 145, 1969.

\bibitem{Vitanyi} M.~Li, P.~Vit\'anyi, {\em An Introduction to Kolmogorov Complexity
                  and its Applications.} Springer--Verlag, 1997.  

\bibitem{Zur89} W.H.~Zurek, Phys.~Rev. {\bf A40}, 4731 (1989).

\bibitem{Bron88} J.B.~Bronzan, Phys.~Rev. {\bf D38}, 1994 (1988).

\bibitem{RG} K.~Wilson, Phys.~Rev. {\bf 140}, B445 (1965);
                        Phys.~Rev. {\bf D2}, 1438 (1970);
                        Phys.~Rev. {\bf D3}, 1818 (1971);
                        Phys.~Rev. {\bf B4}, 3174 (1971);  
                        Phys.~Rev. {\bf B4}, 3184 (1971).  

\bibitem{nonultr}
   I.~Horv\'ath, Phys.~Rev.~Lett. {\bf 81} (1998) 4063;
   I.~Horv\'ath, Phys.~Rev. {\bf D60} (1999) 034510;
   I.~Horv\'ath, C.T.~Balwe, R.~Mendris, Nucl.~Phys. {\bf B599}, 283 (2001).


\bibitem{Cre83a} M.~Creutz, {\em Quarks, Gluons and Lattices}, 
         Cambridge University Press (1983).

\bibitem{Yan75a} C.N.~Yang, Phys. Rev. Lett. {\bf 33}, 445 (1975).

\bibitem{Cre01a} M.~Creutz, I.~Horv\'ath and H.~Neuberger,
                 Nucl. Phys. {\bf B} (Proc. Suppl.) 106 (2002) 760.

\bibitem{Lia92A} Y. Liang, K.F. Liu, B.A. Li, S.J. Dong, K. Ishikawa,
                 Phys. Lett. {\bf B307} (1993) 375.

\bibitem{Has94A} P. Hasenfratz, F.Niedermayer, Nucl.~Phys. {\bf B414}, 785 (1994).

\bibitem{ov_loc} P.~Hern\'andez, K.~Jansen, M.~L\"uscher, 
                 Nucl.~Phys. {\bf B552}, 363 (1999).

\bibitem{APE} 
   M.~Falcioni, M.L.~Paciello, G.~Parisi, B.~Taglienti, Nucl.~Phys. {\bf B251} (1985) 624;
   M.~Albanese et al, Phys.~Lett. {\bf B192} (1987) 163.

\bibitem{stout}
   C.~Morningstar, M.~Peardon, Phys. Rev. {\bf D69}, 054501 (2004).


\end{thebibliography}
\end{document}
\bye